\documentclass[aps,twocolumn,groupedaddress,nofootinbib,english,nobalancelastpage,floatfix,english,nofootinbib]{revtex4-1}

\usepackage[latin1]{inputenc}
\pdfoutput=1
\usepackage{bm}
\usepackage{multirow, amsmath, amssymb, amsfonts, array}
\usepackage{color}
\usepackage{bm,graphicx,cancel}
\usepackage{calc}
\usepackage{dcolumn}
\usepackage{mathrsfs}
\usepackage{url}

\newcommand{\rmd}{\rm d}

\begin{document}

\title{Complementarity of direct and indirect Dark Matter detection experiments}

\author{Chiara Arina}
\author{Gianfranco Bertone} 
\author{Hamish Silverwood}

\affiliation{GRAPPA Institute, University of Amsterdam, Science Park 904, 1090 GL Amsterdam, The Netherlands}

\begin{abstract}
We investigate the prospects for reconstructing the mass, spin-independent and spin-dependent cross-sections of Dark Matter particles with a combination of future direct detection experiments such as XENON1T, and the IceCube neutrino telescope in the 86-string configuration including the DeepCore array. We quantify the degree of complementarity between the two experiments by adopting realistic values for their exposure, energy threshold and resolution. Starting from benchmark models arising from a supersymmetric model with 25 free parameters, we show that despite the stringent constraints set by the run with 79 strings, IceCube can help break the degeneracies in the Dark Matter cross-section parameter space, even in the unfortunate case where it fails to discover high energy neutrinos from the Sun. We also discuss how the reconstruction of the Dark Matter particle parameters from the combined data sets is affected by the uncertainties associated with the nuclear structure of the target material in case of spin-dependent scattering and those associated with astrophysical quantities such as the Dark Matter density and velocity distribution. 
\end{abstract}

\maketitle

\section{Introduction}

The identification of the nature of Dark Matter (DM) particles is one of the central unresolved questions in modern particle physics and cosmology~\cite{BertoneBook}. Weakly interacting massive particles (WIMPs) are among the best motivated candidates for Cold Dark Matter, as they are thermally produced and naturally lead to the correct relic abundance in the early Universe, and they naturally arise from well-motivated extensions of the Standard Model ~\cite{Jungman,Munoz:2003gx,Bergstrom,Bertone:2004pz}. 

DM can be searched for with a variety of detection strategies. In particular, WIMPs can be detected {\it directly}, i.e. through their scattering off nuclei in underground detectors~\cite{Goodman:1984dc,CerdenoGreen}. Possible hints of detection of a light WIMP have emerged from data obtained by CDMS II \cite{Agnese:2013mga}, CRESST-II~\cite{Angloher:2011uu}, CoGeNT~\cite{Aalseth:2011wp} and DAMA/LIBRA~\cite{Bernabei:2010mq}, but the interpretation of these events as due to spin-independent (SI) scattering of a 10 GeV DM particle has been challenged~\cite{Arina:2012dr,Kopp:2011yr} by several other experiments such as XENON100~\cite{Aprile:2012nq}, CRESST commissioning run~\cite{Brown:2011dp} and CDMS~\cite{Ahmed:2010wy,Ahmed:2012vq}. 

Even if new particles are convincingly discovered with direct detection experiments, reconstructing the properties of DM particles, such as the mass and the WIMP-nucleon cross-sections, will be  a complex task. First, the astrophysical parameters describing the DM distribution in the Galactic halo and in the solar neighborhood are affected by large uncertainties. It is known for instance that the DM velocity distribution may substantially deviate from the idealized Maxwell-Boltzmann (MB) distribution which is usually assumed~\cite{Ling:2009eh,Kuhlen:2009vh,Vogelsberger:2008qb}, affecting reconstruction of the WIMP parameters~\cite{Green:2010gw,McCabe:2010zh}. A realistic reconstruction of the DM parameters requires marginalizing\footnote{By marginalizing we mean to integrate over all other parameters than the DM mass and cross-sections.} over astrophysical uncertainties ~\cite{Pato:2012fw,Strigari:2009zb,Pato:2010zk}, or integrating them out~\cite{Fox:2010bz}. 

Second, the spectrum of nuclear recoils is insensitive to the WIMP mass when the latter is much larger than that of the nuclei of the target material, making a mass determination impossible for WIMPs heavier than approximately 100 GeV ({\it e.g.}~\cite{Strege:2012kv,Kavanagh:2012nr}). Third, as recently pointed out in Ref.~\cite{Cerdeno:2012ix}, not only it is impossible to disentangle SI and spin-dependent (SD) couplings with a single direct detection experiment, but the large uncertainties associated with the nuclear structure function might lead to an error of about one order of magnitude on the reconstructed SD coupling~\cite{Cerdeno:2012ix,:2013uj}. 

Uncertainties and degeneracies in the parameter space can fortunately be reduced with a careful combination with other DM searches, {\it e.g.} combining the information arising from direct detection experiments with different targets ~\cite{Bertone:2007xj,Pato:2010zk,Pato:2011de,Cerdeno:2013gqa}, or combining direct detection with accelerator searches or indirect searches~\cite{Bertone:2010rv,Bernal:2008zk}. Here we explore the degree of complementarity between DM direct searches and an indirect detection strategy that consists of searching for high energy neutrinos produced by the capture and annihilation of WIMPs in the Sun's core with the IceCube neutrino telescope at the South Pole~\cite{Aartsen:2012kia} (see also Refs. \cite{Hooper:2006wv,PhysRevD.80.095019} for a similar analysis in the framework of supersymmetric models).

In this paper, we quantify the degree of complementarity of direct and indirect DM searches with a combination of a future direct detection experiments such as XENON1T, and the IceCube neutrino telescope in the 86-string configuration including the DeepCore array. We start by assessing the reconstruction capabilities of XENON1T~\cite{Aprile:2012zx} for 3 benchmark DM candidates within reach during the next 5 years: XENON1T is expected to reach $10^{-47} {\rm cm^2}$ in sensitivity for the SI interaction, and $10^{-42} {\rm cm^2}$ for the SD interaction (the same sensitivity to SD interaction as is expected for IceCube in the same time period~\cite{IceCube:2011aj}). We will demonstrate that even if one of the detectors does not see a signal the reconstruction of the physical parameters is still improved by utilizing both experiments.  We will then assess the impact of uncertainties from astrophysics and from the nuclear structure functions in the reconstruction of the WIMP parameters. Our final results will be obtained by marginalizing over all these nuisance parameters\,\footnote{A nuisance parameter is any parameter which is not of immediate interest but which must be accounted for in the analysis of those parameters which are of interest.}, assessing the degradation of the reconstruction of physical properties, such as the DM mass and the WIMP-nuclei cross-sections with respect to the scenario with fixed nuisance parameters. 

The paper is organized as follows. We first introduce the theoretical expected signal for DM scattering off nuclei in underground detectors, Sec.~\ref{sec:theoxe}, and for the neutrino flux from the Sun, Sec.~\ref{sec:theoic}. Section~\ref{sec:stat} describes the phenomenological approach we use in studying the WIMP signal as well as the statistical framework. Section~\ref{sec:Xe} describes the setup of the future XENON1T experiment and its sensitivity for detecting WIMP signals. On the same lines, in Sec.~\ref{sec:IC} we describe the prospect for detection with IceCube and its sensitivity in reconstructing WIMP parameters. Section~\ref{sec:comb} illustrates the effectiveness of combining different search strategies for reconstructing the benchmark models. We subsequently discuss the uncertainties that affect the signal reconstruction in Sec.~\ref{sec:unc}: we first  describe the impact of astrophysical uncertainties over the Galactic parameters, then the effect of undetermined nuclear structure functions, and lastly comment on the velocity distribution parametrization. Finally, we summarize our findings in Sec.~\ref{sec:concl}.

\section{Predicted signals from WIMPs}\label{sec:theo}
In this section we review the theoretical predictions for the direct detection rate and for the neutrino flux arising from annihilation of DM particles in the Sun. The scope of this brief summary is to introduce the key model parameters of our phenomenological analysis and underline the (different) dependence of the expected rates in direct and indirect detection experiments.

\subsection{Theoretical rate for direct detection}\label{sec:theoxe}

Direct detection experiments aim to detect nuclear recoils arising from the scattering of WIMPs off target nuclei. The differential spectrum of a DM particle recoiling off a nucleus, in units of  events per time per detector mass per energy, 
has the form
\begin{equation}
\label{eq:diffrate}
\frac{{\rm d} R}{{\rmd} E} = \frac{\rho_{\odot}}{m_{\rm DM} m_{\cal N}}  \int_{v>v_{\rm min}} {\rm d}^3v  \, \frac{\rm d\sigma}{{\rm d} E}(E,v) \, v \, f (\vec{v}(t))\,,
\end{equation}
where $E$ is the energy transferred during the collision,  $\rho_{\odot} \equiv \rho_{\rm DM}(R_{\odot})$ is the WIMP density in the solar neighborhood, $m_{\rm DM}$ is the WIMP mass, ${\rmd} \sigma/{\rmd} E$ is the differential cross-section for the scattering, and $f(\vec{v}(t))$ is the normalized WIMP velocity distribution in the Earth's rest frame. The integration in the differential rate is performed over all incident particles capable of depositing a recoil energy of $E$.
For elastic scalar interactions, this implies a lower integration limit of
\begin{equation}\label{eq:vmin}
 v_{\rm min} = \sqrt{\frac{M_{\cal N} E}{2 \mu}}\,,
 \end{equation}
 where $M_{\cal N}$ is the mass of the target nucleus, and $\mu=m_{\rm DM} M_{\cal N}/(m_{\rm DM}+M_{\cal N})$ is the WIMP-nucleus reduced mass. As for the velocity distribution $f(\vec{v}(t))$, we use the MB parametrization~\cite{Smith:1996fu,Savage:2008er} and neglect the time dependent modulation due to the Earth's motion around the Sun. We defer to Sec.~\ref{sec:astro} and~\ref{sec:results} the discussion about the role of astrophysical uncertainties.

The differential cross-section ${\rmd} \sigma/{\rmd} E$ encodes the particle and nuclear physics information and is in general 
separated into the SI and SD contributions as
\begin{equation}
\frac{\rmd \sigma}{{\rmd} E} = \frac{\rmd \sigma}{{\rmd} E}\Big|_{\rm SI} + \frac{\rmd \sigma}{{\rmd} E}\Big|_{\rm SD}\,. 
\end{equation}

\paragraph{Spin-independent interaction}
\begin{equation}
\label{eq:sipart}
\frac{\rmd \sigma}{{\rmd} E}\Big|_{\rm SI} = \frac{M_{\cal N} \sigma^{\rm SI}_n}{2 \mu^2_{n} v^2}\ \frac{\Big(f_p Z + (A-Z) f_n\Big)^2}{f_n^2} {\cal F}_{\rm SI}^2(E) \,  ,
\end{equation}
where $\mu_n=m_{\rm DM} m_n/(m_{\rm DM}+m_n)$ is the WIMP-nucleon reduced mass, $\sigma^{\rm SI}_n$ is the SI zero-momentum WIMP-nucleon cross-section, $Z$ ($A$) is the atomic (mass) number of the target nucleus used, and $f_p$ ($f_n$) is the WIMP effective coherent coupling to the proton (neutron). We assume the WIMP couples equally to the neutron and the proton ($f_n=f_p$), so that the differential cross-section ${\rmd} \sigma/{\rmd} E$ is sensitive only to $A^2$.  The nuclear form factor ${\cal F}_{\rm SI}(E)$ characterizes the loss of coherence for nonzero momentum transfer, and is well parametrized by the Helm form factor~\cite{Helm:1956zz,Lewin:1995rx} for all nuclei~\cite{Duda:2006uk}
\begin{equation}
{\cal F}_{\rm SI}(E)=3 e^{-q^2 s^2/2} \frac{\sin(qr)-qr \cos(qr)}{(qr)^3},
\end{equation}
where $s=1$~fm, $r=\sqrt{R^2-5 s^2}$, $R=1.2 \ A^{1/3}$~fm, and $q=\sqrt{2 M_{\cal N} E}$. 

\paragraph{Spin-dependent interaction}
\begin{equation}
\label{eq:sdpart}
\frac{\rmd \sigma}{{\rmd} E}\Big|_{\rm SD} = \frac{ 4 M_{\cal N} \sigma^{\rm SD}_n}{3 \mu^2_n\,  v^2}\frac{J+1}{J} (\langle S_p \rangle +a_n/a_p \langle S_n \rangle)^2 {\cal F}^2_{\rm SD}(E) \,,
\end{equation}
where $\sigma^{\rm SD}_n$ is the SD zero-momentum WIMP-nucleon cross-section, $a_p$ ($a_n$) are axial WIMP-proton (neutron) couplings, $J$ is the total spin of the nucleus and $\langle S_p \rangle \, (\langle S_n \rangle)$ is the proton (neutron) spin averaged over the nucleus. The nuclear form factor for SD is usually defined as
\begin{equation}
{\cal F}^2_{\rm SD}(E) = \frac{S(E)}{S(0)}\,,
\end{equation}
and 
\begin{equation}
S(q) = a^2_0 S_{00}(q) + a_0 a_1 S_{01}(q) + a^2_1 S_{11}(q)\,,
\end{equation}
with $a_0 = a_n+a_p$ ($a_1 = a_p - a_n$) being the isoscalar (isovector) coupling. Furthermore we assume equal coupling to neutron and proton ($a_n=a_p$), hence only the structure factor $S_{00}$ will be relevant for our analysis. This assumption is motivated by the fact that theoretical models of WIMPs typically predict a similar cross-section to proton and neutron~\cite{Cerdeno:2013gqa,Belanger:2008gy}.

For the xenon-based detector we will consider there are two isotopes that have a nonzero total spin because of the unpaired neutrons:  $^{129}\rm Xe$, with $J=1/2$ and abundance $26.44\%$, and $^{131}\rm Xe$, with $J=3/2$ and abundance $21.18\%$.

The structure functions $S_{00}(q)$ and $S_{11}(q)$ are related to the transverse electric and longitudinal projections of the axial current. These functions can be computed in a shell-model for the atomic nucleus, and the spin of the nucleus is computed by means of the wave functions of the unpaired nucleons. Assuming a particular interaction between nucleons, these are placed in energy levels according to the exclusion principle. As many excited levels as possible are included, making this kind of computation difficult. Finally the projected currents are computed by evaluating the matrix elements of the many-nucleon model. Nuclear shell models are more reliable for heavy nuclei, but even in this case there can be significant deviations at zero momentum transfer or at high momentum. To bracket the uncertainties in the case of the xenon nucleus, for both isotopes we consider two parametrizations for the structure functions: the first (NijmegenII hereafter) was computed in 1997 by Ressell and Dean~\cite{Ressell:1997kx}, while the second is based on a very recent computation using chiral effective field theory formalism and accounting for two body interactions~\cite{Menendez:2012tm} (from now on CEFT). In the first part of our discussion we assume nuclear structure functions from CEFT formalism, while in Sec.~\ref{sec:nucl} we discuss the effect of marginalizing over various realizations of the nuclear structure functions.

The total number of recoils expected, as a function of the DM parameters, in a detector of mass $M_{\det}$ in a given observed energy range $[E_{\rm min},E_{\rm max}]$ over an exposure time $T$ is obtained by integrating Eq.~(\ref{eq:diffrate}) over energy
\begin{equation}
\label{eq:totrate}
S_{\rm Xe}(m_{\rm DM}, \sigma^{\rm SI}_n, \sigma^{\rm SD}_n) = \epsilon \, M_{\rm det} \, T \, \int_{E_{\rm min}}^{E_{\rm max}}  {\rmd} E  \frac{{\rmd} R}{{\rmd} E} \,,
\end{equation}
where we have accounted for an energy independent efficiency factor $\epsilon$ and a finite energy resolution $\sigma(E)$ for the detector:
\begin{equation}
\frac{{\rm d}R}{{\rm d}E}   =   \int_{0}^{\infty} \frac{{\rmd} R}{{\rmd} E'}\,  \frac{\rm e^{\frac{-(E'-E)^2}{2 \sigma(E')^2}}}{\sqrt{2 \sigma(E')^2 \pi}}\, dE'\,.
\end{equation}
For this analysis we will focus on the future XENON1T experiment, the details of which are outlined in Sec.~\ref{sec:Xe}.

\subsection{Muon signal in neutrino telescopes}\label{sec:theoic}
Indirect detection of DM using neutrino telescopes involves four processes: capture of WIMPs by the Sun, annihilation of these WIMPs, production of neutrinos following the annihilation event, and finally detection of these neutrinos. The formalism described in this section is encoded in the \texttt{DarkSUSY 5.0.6} software package which we use in our analysis \cite{Gondolo2004}.

Capture of WIMPs in the Sun occurs when WIMPs elastically scatter off nuclei in the Sun and lose enough energy to reduce their velocity to below the solar escape velocity. Subsequent scattering events reduce the velocity of the captured WIMPs further, until they concentrate and thermalize in the core of the Sun. The capture rate is \cite{Gould1987}
\begin{equation}
\label{eq:CaptureRateTotalIntegral}
C_c = \frac{\rho_\odot}{m_{\rm{DM}}}  \int_0^R {\rm d}r \sum_i \frac{{\rm d}C_i}{{\rm d}V} 4\pi r^2,
\end{equation}
where
\begin{equation}
\label{eq:CaptureRateDifferentialPerVolume}
\frac{{\rm d}C_i}{{\rm d}V} = \int_0^{u_{\rm max}} {\rm d}u \frac{f(u)}{u} w \Omega^-_{v, i} (w)
\end{equation}
is the capture rate per unit shell volume, $R$ is the solar radius and index $i$ runs across nuclear species present in the Sun. The variable $w$ is the velocity of the WIMP at the shell, and $w = \sqrt{u^2 + v^2} $, where $u$ is the velocity at an infinite distance away from the shell ({\it i.e.} where the influence of the shell's gravitational potential is negligible) and $v$ is the escape velocity at the shell. The integration limit $u_{\rm{max}}$ is given by 
\begin{equation}
\label{eq:uMax}
u_{\rm{max}} = \frac{\sqrt{4 m_{\rm{DM}} M_{{\cal N}_i}}}{m_{\rm{DM}} - M_{{\cal N}_i}} v.
\end{equation}
The term $w \Omega^-_{v, i} (w)$ quantifies the probability that a WIMP will scatter to a velocity less than the escape velocity, and is proportional to $\sigma_i n_i,$ where $n_i$ is the number density of nuclei $i$ in the shell and $\sigma_i$ is the total interaction cross-section between the WIMP and nuclei $i$. This calculation, the default in \texttt{DarkSUSY 5.0.6}, does not include the effects of diffusion and planets as these have been shown to be minimal \cite{Sivertsson:2012qj}.

The WIMP-nuclei cross-section $\sigma_i$ can be expressed as~\cite{Jungman,Scott:2008ns}
\begin{equation}
\label{eg:SigmaFirstApproximation}
\sigma_i = \beta^2 \left[ \sigma^{\rm SI}_n A_i^2 + \sigma^{\rm SD}_n \frac{4(J_i + 1)}{3J_i} \left| \langle S_{\rm{p}, i} \rangle + \langle S_{\rm{n}, i} \rangle \right|^2 \right], 
\end{equation}
where 
\begin{equation}
\label{eq:SigmaBetaFactor}
\beta = \frac{M_{{\cal N}_i} (m_\chi + m_{\rm{p}})}{m_{\rm{p}} (m_\chi + m_i)}.
\end{equation}

The Sun is composed overwhelmingly of spin-$\frac{1}{2}$ hydrogen and spin-0 helium, with only small quantities of heavier elements. The increased atomic number of these heavier elements will compensate for their lower abundance via the $A_i^2$ term and give appreciable contributions to the SI part of $\sigma_i$. However no such enhancement occurs for the SD part, with total spin being related to the number of unpaired nucleons. Thus we can consider only a SD contribution from hydrogen nuclei, reducing Eq.~(\ref{eg:SigmaFirstApproximation}) to 
\begin{equation}
\label{eq:SigmaSecondApproximation}
\sigma_i = 
\begin{cases}
\sigma^{\rm SI}_n + \sigma^{\rm SD}_n & \text{for $i = 1$},\\
\beta^2 \sigma^{\rm SI}_n A_i^2 & \text{for $i \geq 2$}.
\end{cases}
\end{equation} 
While SI interactions are taken into account in the capture rate calculation, in practice the SD interaction is dominant. As the SD process occurs directly on protons, the theoretical rate is not affected by nuclear structure functions describing the coherence of the nucleus. 

As shown in Eq.~(\ref{eq:CaptureRateTotalIntegral}) the capture rate is dependent upon the density of WIMPs. However as WIMP capture is a continuous process it is sensitive not to the local density at the Sun's current position but instead samples the local density along the prior path of the Sun around the galaxy~\cite{Serpico:2010ae}. For this analysis, however, we assume a constant local DM density $\rho_{\odot}^{\rm obs}$, and we will discuss the uncertainties related to this assumption in Sec.~\ref{sec:astro}. 

Acting against the accumulation of WIMPs is the process of DM annihilation. We can describe the total population of WIMPs in the Sun $N(t)$~\cite{Griest1987} by the equation
\begin{equation}
\label{eq:WIMPpopDE}
\frac{\text{d} N(t)}{\text{d}t} = C_c - \Gamma_a (t)
\end{equation}
where $\Gamma_a (t) = \frac{1}{2} C_a N^2(t)$ is the annihilation rate. The parameter $C_a$ is dependent on the distribution of WIMPs in the Sun and $\langle \sigma_a v \rangle$, the zero velocity WIMP annihilation cross-section ~\cite{Jungman}.

Solving Eq.~(\ref{eq:WIMPpopDE}) gives us an expression for the annihilation rate:
\begin{equation}
\label{eq:AnnihilationRateSolution}
\Gamma_a (t) = \frac{C_c}{2} \tanh^2 \left(\frac{t}{\tau} \right)
\end{equation}
where $\tau = 1/{\sqrt{C_c C_a}}$ is the capture-annihilation equilibrium time scale. In the case of $t \gg \tau$, Eq.~(\ref{eq:AnnihilationRateSolution}) reduces to $\Gamma_a (t) = \frac{1}{2} C_c$, and equilibrium between capture and annihilation is reached. For this study we assume this \textit{steady state} scenario; it allows us to take a more model independent approach by eliminating the dependence on $\langle \sigma_a v \rangle$. 

WIMPs which have accumulated in the Sun can annihilate with each other and produce Standard Model particles. The majority of the decay products of these particles are absorbed almost immediately and without consequence in the core of the Sun. However, certain classes of WIMPs can decay directly into neutrinos, and in other cases the Standard Model decay products can themselves decay into neutrinos, which can escape from the Sun and potentially be detected on Earth. Detection of these neutrinos by neutrino telescopes occurs via the observation of the \v{C}erenkov radiation emitted by the particles produced following the weak force interactions between the neutrinos and the matter in and around the telescope. Of particular interest are muons created by the charged current interactions of muon neutrinos, as their range is such that they create long, relatively detectable tracks compared to other leptons~\cite{Halzen2006}. Optical detection of the \v{C}erenkov radiation in a transparent medium such as water or ice then allows the incoming neutrino's energy and origin to be reconstructed. For this analysis we will focus on the IceCube neutrino telescope, the details of which are outlined in Sec.~\ref{sec:IC}.

\section{Statistical framework and benchmark WIMP models}\label{sec:stat}

To illustrate the capabilities of reconstruction and complementarity of future direct detection experiments and neutrino telescopes we consider three benchmark models, described in Table \ref{tab:bench}. These WIMP models represent a phenomenological approach in the description of the theoretical parameters $\Theta=\{m_{\rm DM}, \sigma^{\rm SI}_n, \sigma^{\rm SD}_n\}$ we are interested in, and capture the relevant aspects of the analysis:
\begin{itemize}
\item[(i)] \textit{Benchmark A} is characterized by a light mass of 60 GeV and its cross-section on nucleons is dominated by the SD component. 
\item[(ii)] \textit{Benchmark B} has an intermediate DM mass of 100 GeV and sizable WIMP-nuclei cross-sections for both SI and SD interactions.
\item[(iii)] \textit{Benchmark C} is characterized by a heavy mass of 500 GeV and significant SI but negligible SD cross-sections.
\end{itemize}
We have checked that our conclusions are robust and hold for benchmarks with same behavior of the cross-sections but different masses as well. These benchmark WIMP models are representative of well-motivated neutralino configurations arising in scans of the MSSM25~\cite{Silverwood:2012tp} parameter space, which is a phenomenological parametrization of the minimally supersymmetric Standard Model (MSSM) with 25 free parameters defined at the electroweak scale. 

\begin{table*}[t!]
\caption{Below are listed the benchmark models used in the analysis, together with their mass, cross-sections, counts predicted in XENON1T and expected muon signal in IceCube (and the WIMP annihilation channel), as labeled. The first benchmark is characterized by a light DM mass and large SD contribution, the second has intermediate mass and both sizable SI and SD cross-sections, while the third has a large mass and dominant SI contribution. As for the SD contribution to XENON1T detector, we quote the predicted counts arising from both nuclear structure functions considered in this analysis (see text for further details). \label{tab:bench}}
\begin{center}
\begin{ruledtabular}
\begin{tabular}{c  c c c c c c  c}
& $m_{\rm DM}$ [GeV] & $\sigma^{SI}_n$ $\rm [cm^2]$ & $\sigma^{SD}_n$ $\rm [cm^2]$ &$S_{\rm Xe}^{\rm SI}(\Theta)$ & $S_{\rm Xe}^{SD}(\Theta)$ (CEFT)  & $S_{\rm Xe}^{SD}(\Theta)$  (NijmegenII) & $S_{\mu}(\Theta)$  \\
\hline
$A$ & 60 & $3.7 \times 10^{-49}$ & $2.0 \times 10^{-40}$ &$1.1$ & $422.8$ & $170.9$  & $24.9$ ($\tau^+\tau^-$) \\
$B$ & 100 & $8.8 \times 10^{-46}$ & $2.0 \times 10^{-40}$ & $252.8$ & $356.1$ & $122.3$ &  $66.0 $ ($W^+W^-$) \\
$C$ & 500 & $1.1 \times 10^{-45}$ & $9.6 \times 10^{-45}$ & $74.4$ & $4.4 \times 10^{-3}$ & $1.5 \times 10^{-3}$ &  $7.8$ ($\nu_\mu\bar{\nu}_\mu$)\\
\end{tabular}
\end{ruledtabular}
\end{center}
\end{table*}

For each benchmark model we generate mock data, using the experimental likelihoods $\mathcal{L}(\Theta)$, from the true model, {\it i.e.} without Poisson scatter~\cite{Cowan:2010js}. These theoretical signals expected in the detectors, which are the number of recoiling nuclei in XENON1T arising from both SI and SD part and the number of up-going muons $N_\mu$ in IceCube, are given in Table \ref{tab:bench} and will be described in detail in Secs.~\ref{sec:Xe} and~\ref{sec:IC}, respectively.  

For the sampling of the theoretical parameter space we adopt the Bayesian methodology. We employ the public code \texttt{MultiNest v2.12}~\cite{Feroz:2007kg,Feroz:2008xx}, which uses an ellipsoidal and multimodal nested-sampling algorithm to estimate the posterior probability over the full parameter space:
\begin{equation}
\mathcal{P}(\Theta | d) \propto \mathcal{L}(\Theta) \pi (\Theta)\,,
\end{equation}
where $\pi(\Theta)$ is the prior probability density function (pdf). The priors for the three theoretical parameters $m_{\rm DM}$, $\sigma^{\rm SI}_n$ and $\sigma^{\rm SD}_n$ are chosen to be flat on a logarithmic scale so as not to favor any particular order of magnitude, and are defined as follows: 
\begin{eqnarray}
& \log_{10}(m_{\rm DM}/\rm GeV)  :  1 \to 3\,,\nonumber\\
& \log_{10}(\sigma^{\rm SI}_n/\rm cm^2)  :   -60 \to -43\,,\nonumber\\ 
& \log_{10}(\sigma^{\rm SD}_n/\rm cm^2)  :   -55 \to -38\,.\nonumber
 \end{eqnarray} 
We set $n_{\rm live} = 25000$, and use an efficiency factor of $10^{-4}$ and a tolerance factor of 0.01~\cite{Feroz:2007kg}, which ensure that the sampling is accurate enough to have a parameter estimation similar to Markov chain Monte Carlo sampling methods~\cite{Feroz:2011bj}. The resulting chains are analyzed with an adapted version of the package \texttt{GetDist}, supplemented with \texttt{MATLAB} scripts from the package \texttt{SuperBayeS}~\cite{superbayes,Trotta:2008bp}. Two-dimensional posterior pdfs, $\mathcal{P}_{\rm marg}$, marginalized over the nuisance parameters and the remaining $n-2$ theoretical parameters, are obtained from the chains by dividing the relevant parameter subspace into bins and counting the number of samples per bin. An $x\%$ credible interval or region containing $x\%$ of the total volume of $\mathcal{P}_{\rm marg}$ is then constructed by demanding that $\mathcal{P}_{\rm mar}$ at any point inside the region be larger than at any point outside. The inferred pdfs are sensitive to the choice of the mass prior range, which we have checked by increasing the upper bound of the prior range to 100 TeV: the $x\%$ contours suffer from volume effects related to the behavior of the likelihood at very large mass, above 10 TeV, and we will comment upon this more in Sec.~\ref{sec:comb}. We however argue that these effects are not relevant for our analysis, since for these high masses the DM most likely does not meet the WIMP requirements anymore. Hence the mass prior range used here is driven by the standard mass range for WIMPs.

The astrophysical parameters describing the DM halo and the DM in the solar neighborhood are regarded as nuisance parameters. These parameters are the local standard of rest velocity $v_0$, the DM escape velocity in the halo $v_{\rm esc}$, and $\rho_\odot$. All these quantities have uncertainties that range from 20\% to a factor of two. Indeed the observed values with one standard deviation, from~\cite{Bovy:2012ba,Reid:2009nj,Gillessen:2008qv,Smith:2006ym,Dehnen:1997cq,Weber:2009pt,Salucci:2010qr,Bovy:2012tw}, are given by
\begin{eqnarray}
  v_0^{\rm obs} & = & 230 \pm 24.4 \ {\rm km \ s}^{-1}\,,\nonumber\\
  v_{\rm esc}^{\rm obs} & = & 544   \pm 39 \  {\rm km \ s}^{-1}\,,\nonumber \\
 \rho_{\odot}^{\rm obs} &  = & 0.4 \pm 0.2 \  {\rm GeV \ cm}^{-3}\,.\nonumber
 \end{eqnarray}
In the first part of our analysis we keep these nuisance parameters fixed to their observed value, while in Sec.~\ref{sec:astro} we present the final results marginalizing over them. The uncertainty over the nuclear structure function for SD interactions is also treated as a nuisance parameter in Sec.~\ref{sec:nucl}.

\section{Reconstruction with XENON1T}\label{sec:Xe}
XENON1T will be the ton scale continuation of the XENON100 detector, and will be constructed underground at the Laboratori Nazionali del Gran Sasso in Italy. The start of data taking is planned for 2015, with first results released in 2017. 

We consider the energy window for DM searches to be from 10 to 45 keV, divided into 7 bins. In this range, the XENON1T likelihood function is given by the product of independent Poisson likelihoods describing the probability of observing the predicted events for a given WIMP signal over the energy bins labeled by the index $i$:
\begin{equation}
\ln \mathcal{L}_{\rm XENON1T}(\Theta)   =   \sum_{i=1}^7 \ln P\left(N^{\rm obs}_i| S_i(\Theta)+\theta_{{\rm{BG}},i}\right)\,.
\end{equation}
$N^{\rm obs}_i$ is the observed number of events in each bin and $\theta_{{\rm{BG}},i}$ is the total background in each bin, described below. The detector energy resolution is $\sigma(E) = 0.6 \, {\rm keV} \sqrt{E/{\rm keV}}$. The effective exposure $\epsilon_{\rm eff} = 2\, \rm  ton \times year$ comes from $\epsilon \times \eta_{\rm cut} \times A_{\rm NR} = 5 \times 0.8 \times 0.5$, where $\epsilon$ is the total exposure, $\eta_{\rm cut}$ is the cut efficiency, and $A_{\rm NR}$ is the nuclear recoil acceptance. As XENON1T is expected to be almost background free~\cite{Aprile:2012zx}, we consider a background of $4 \times 10^{-8}$ counts/kev/day/kg, which accounts for a total of  $\theta_{{\rm{BG}},i} = 0.02$ events per bin after integrating over the exposure time. With this assumption we can safely consider it to be an energy independent background. 
\begin{figure*}[t]
\begin{minipage}[t]{0.32\textwidth}
\centering
\includegraphics[width=1.\columnwidth,trim=10mm 25mm 12mm 12mm, clip]{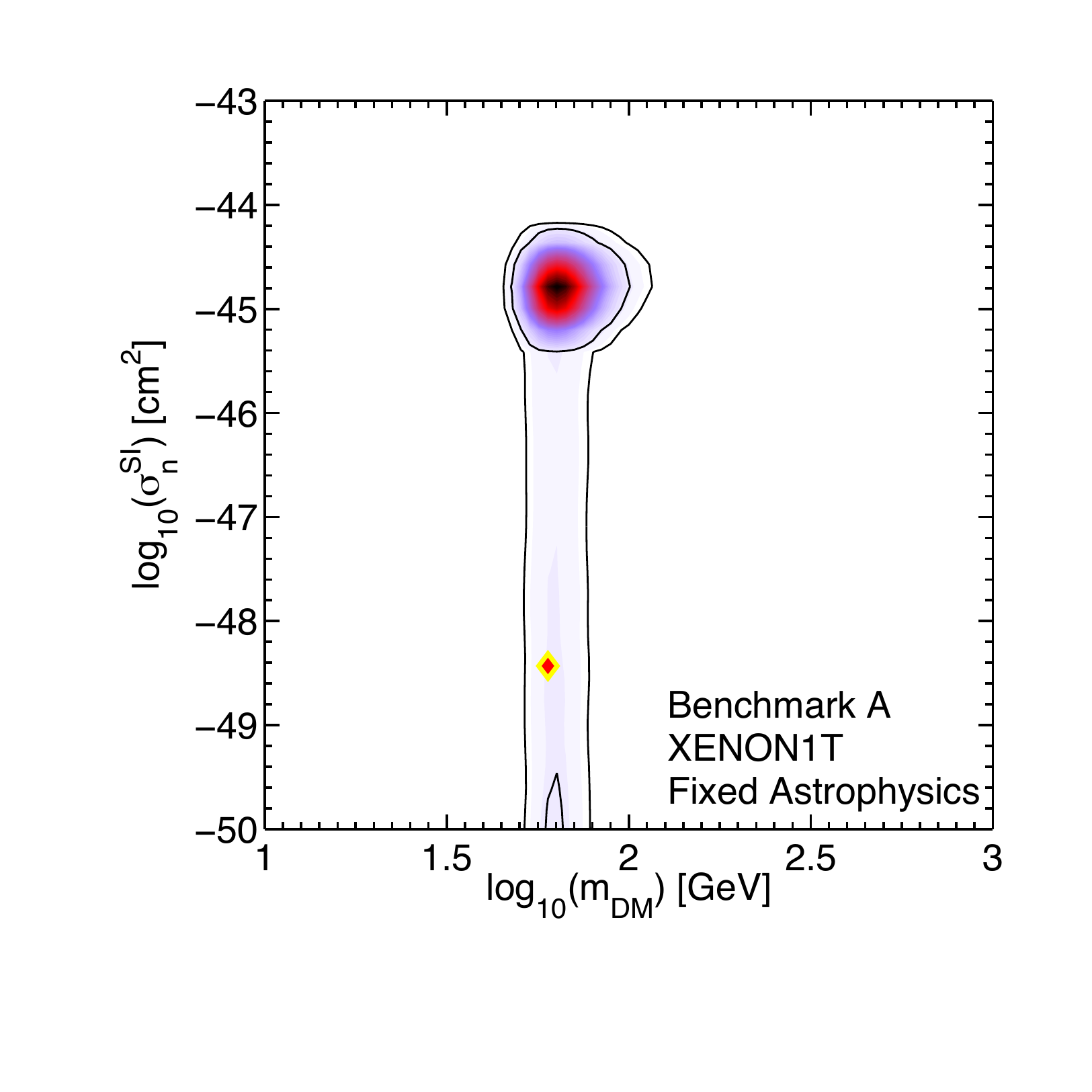}
\end{minipage}
\begin{minipage}[t]{0.32\textwidth}
\centering
\includegraphics[width=1.\columnwidth,trim=10mm 25mm 12mm 12mm, clip]{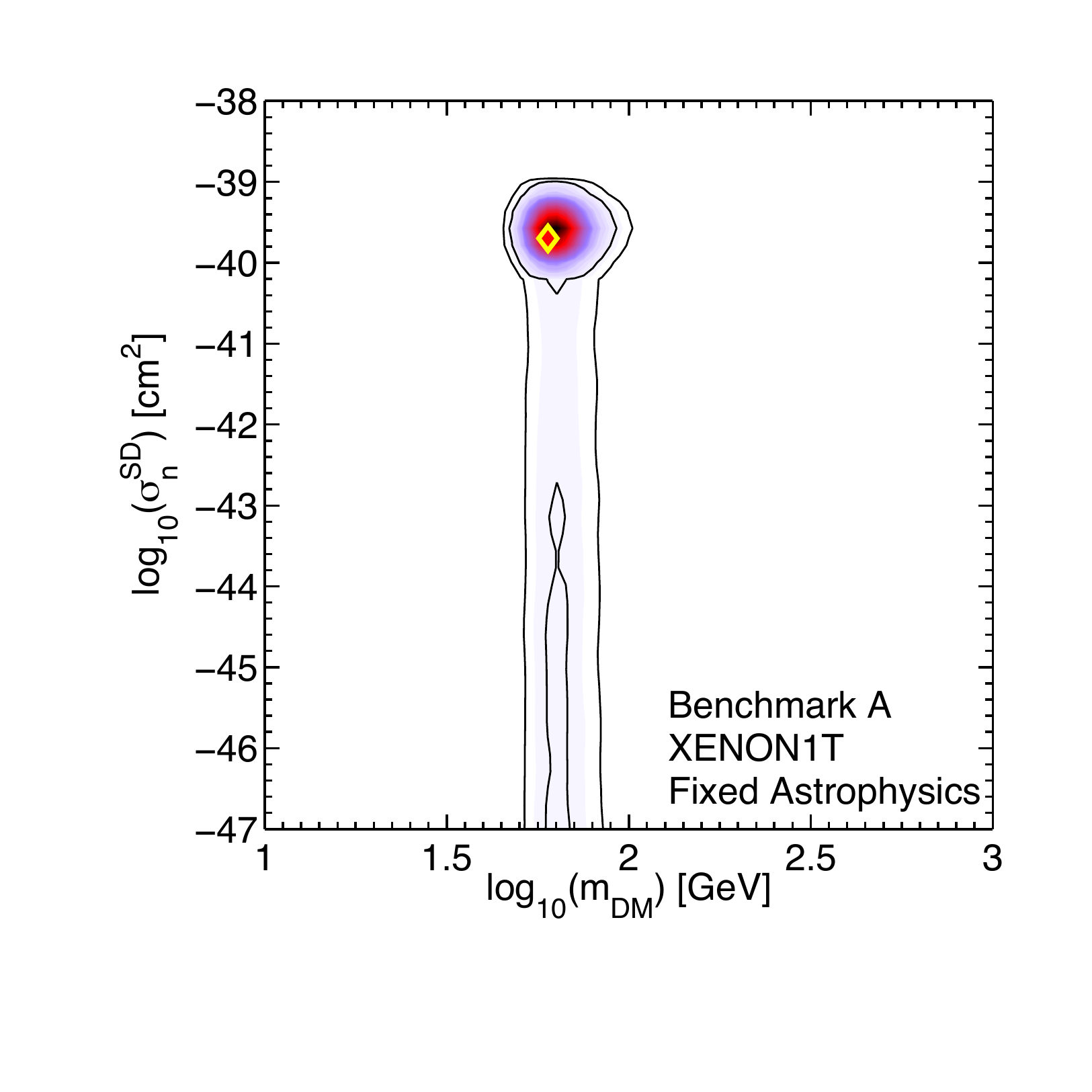}
\end{minipage}
\begin{minipage}[t]{0.32\textwidth}
\centering
\includegraphics[width=1.\columnwidth,trim=10mm 25mm 12mm 12mm, clip]{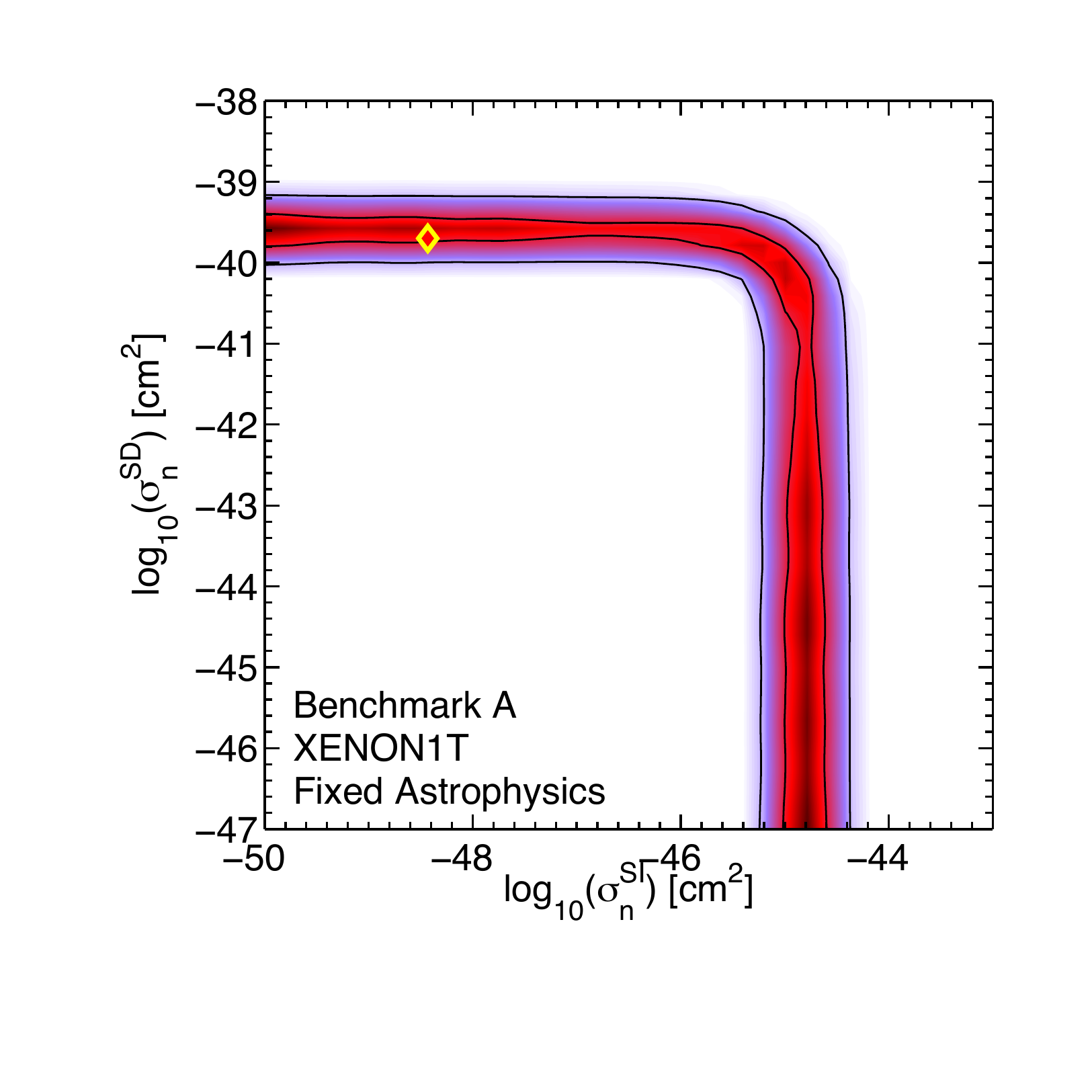}
\end{minipage}
\\
\begin{minipage}[t]{0.32\textwidth}
\centering
\includegraphics[width=1.\columnwidth,trim=10mm 25mm 12mm 12mm, clip]{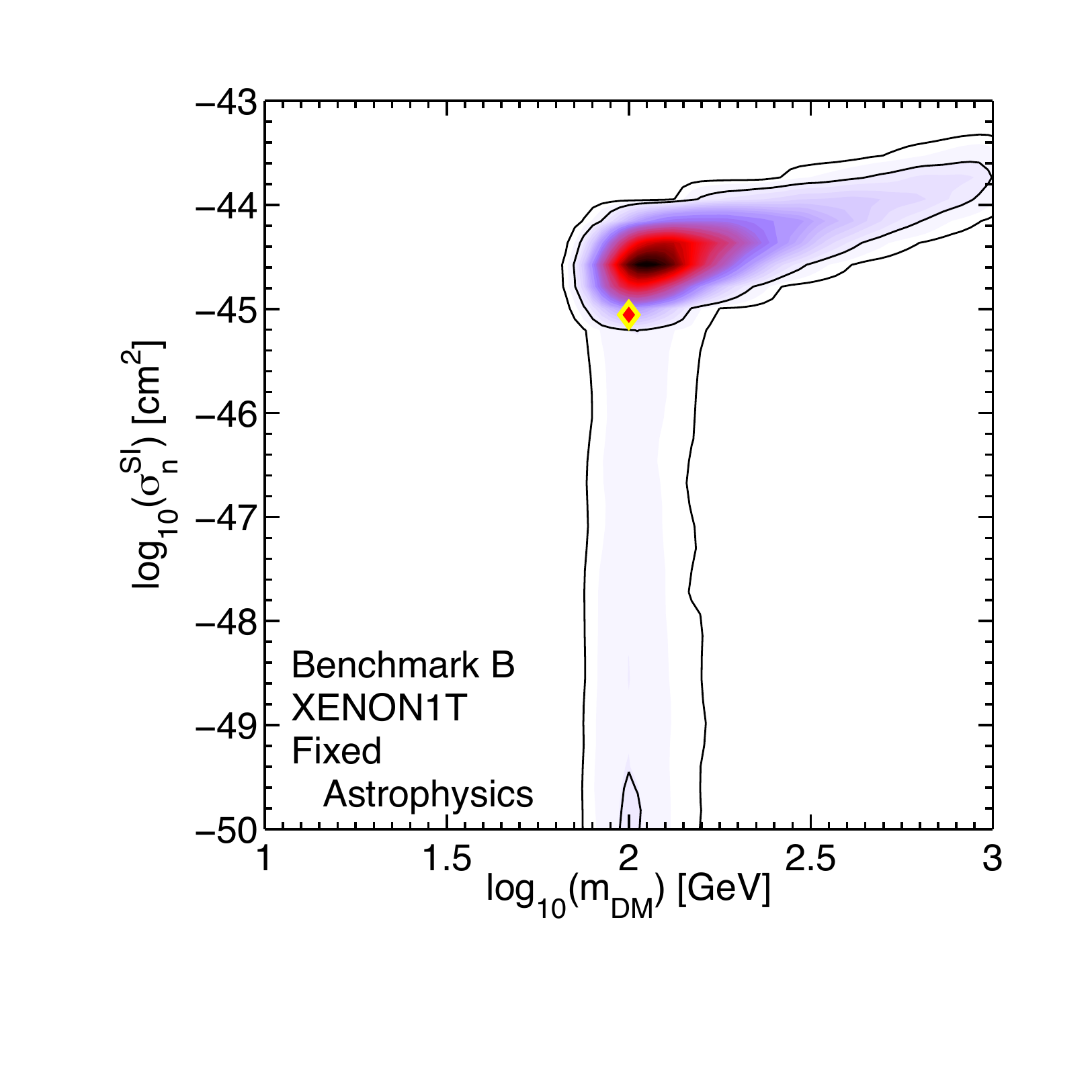}
\end{minipage}
\begin{minipage}[t]{0.32\textwidth}
\centering
\includegraphics[width=1.\columnwidth,trim=10mm 25mm 12mm 12mm, clip]{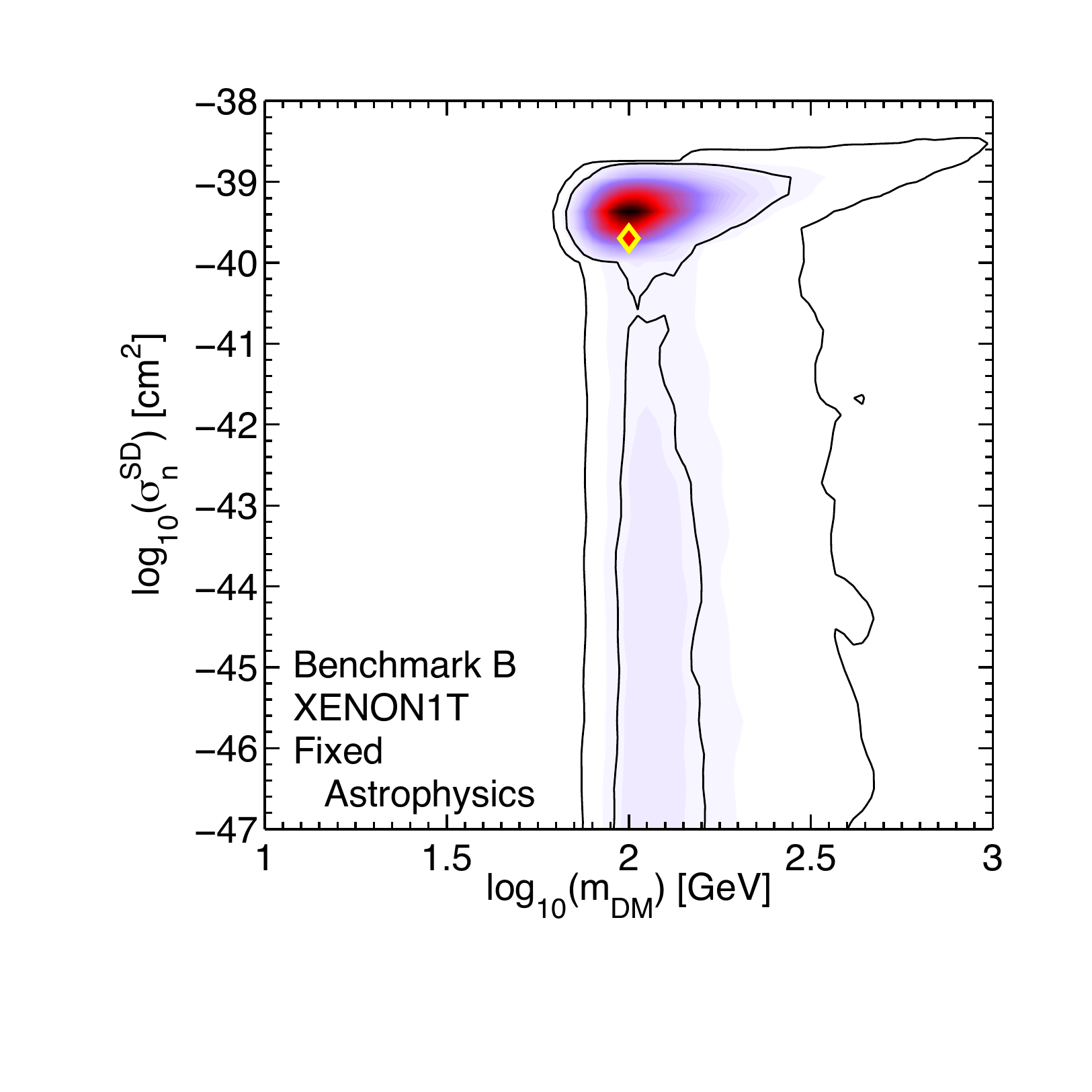}
\end{minipage}
\begin{minipage}[t]{0.32\textwidth}
\centering
\includegraphics[width=1.\columnwidth,trim=10mm 25mm 12mm 12mm, clip]{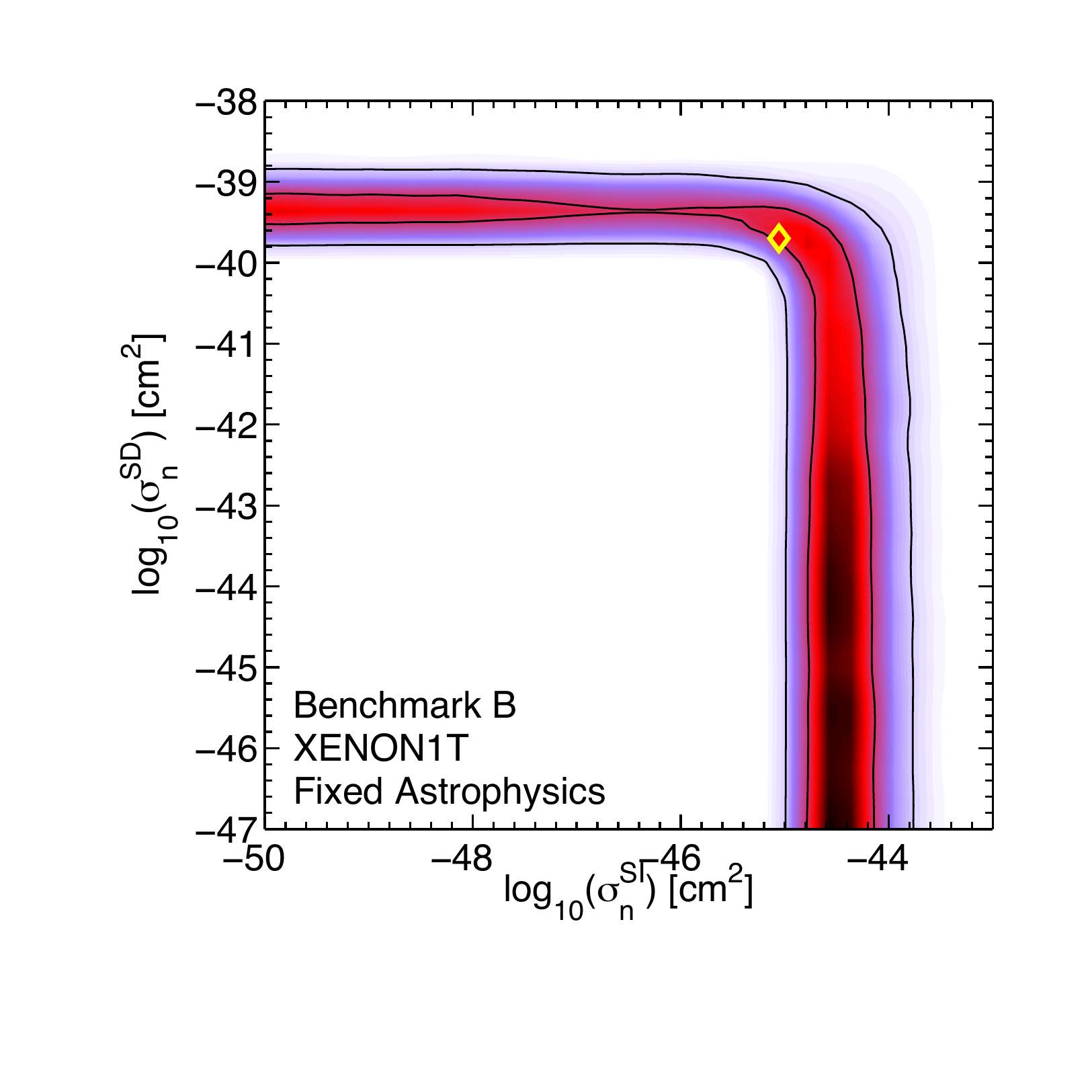}  
\end{minipage}
\\
\begin{minipage}[t]{0.32\textwidth}
\centering
\includegraphics[width=1.\columnwidth,trim=10mm 25mm 12mm 12mm, clip]{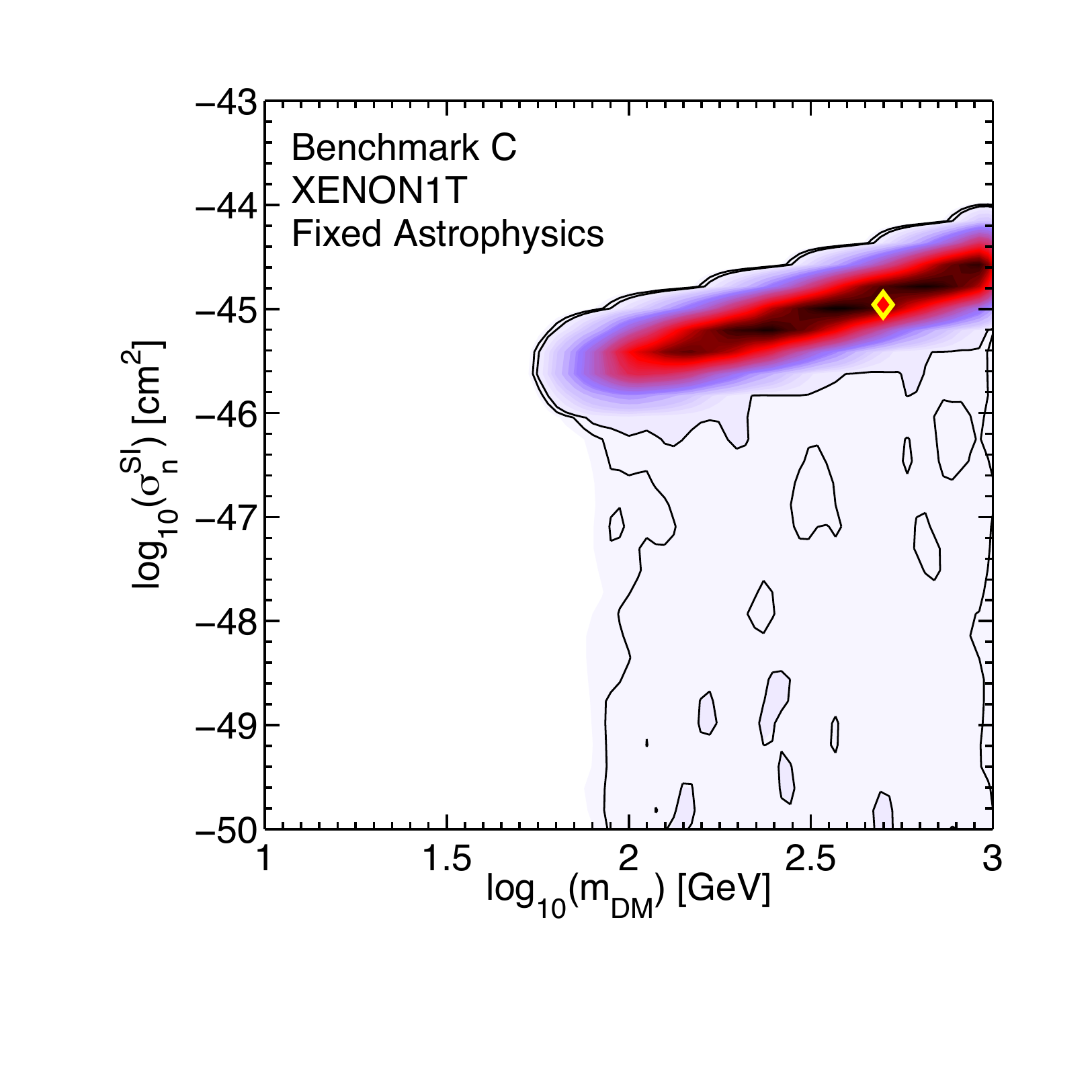}
\end{minipage}
\begin{minipage}[t]{0.32\textwidth}
\centering
\includegraphics[width=1.\columnwidth,trim=10mm 25mm 12mm 12mm, clip]{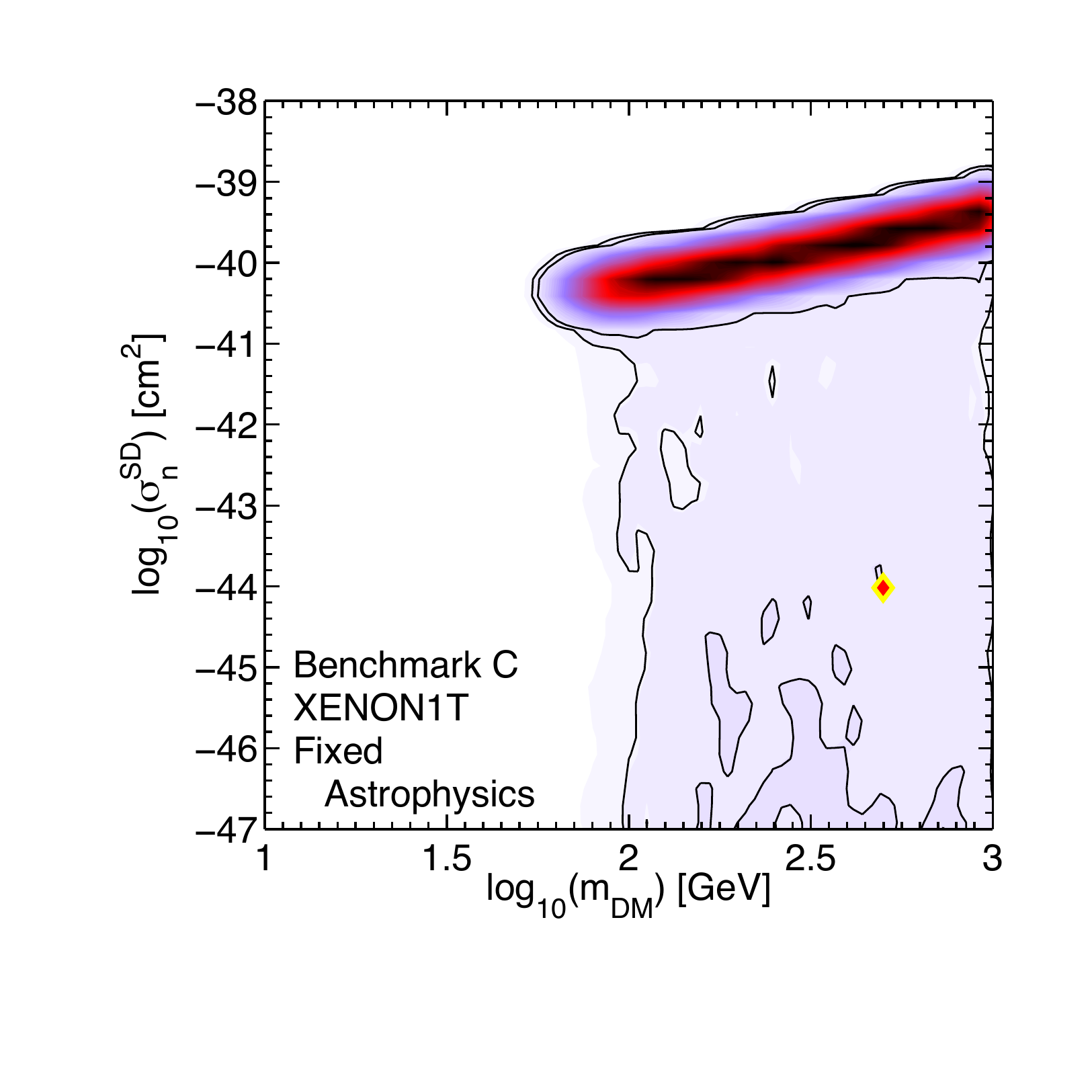}
\end{minipage}
\begin{minipage}[t]{0.32\textwidth}
\centering
\includegraphics[width=1.\columnwidth,trim=10mm 25mm 12mm 12mm, clip]{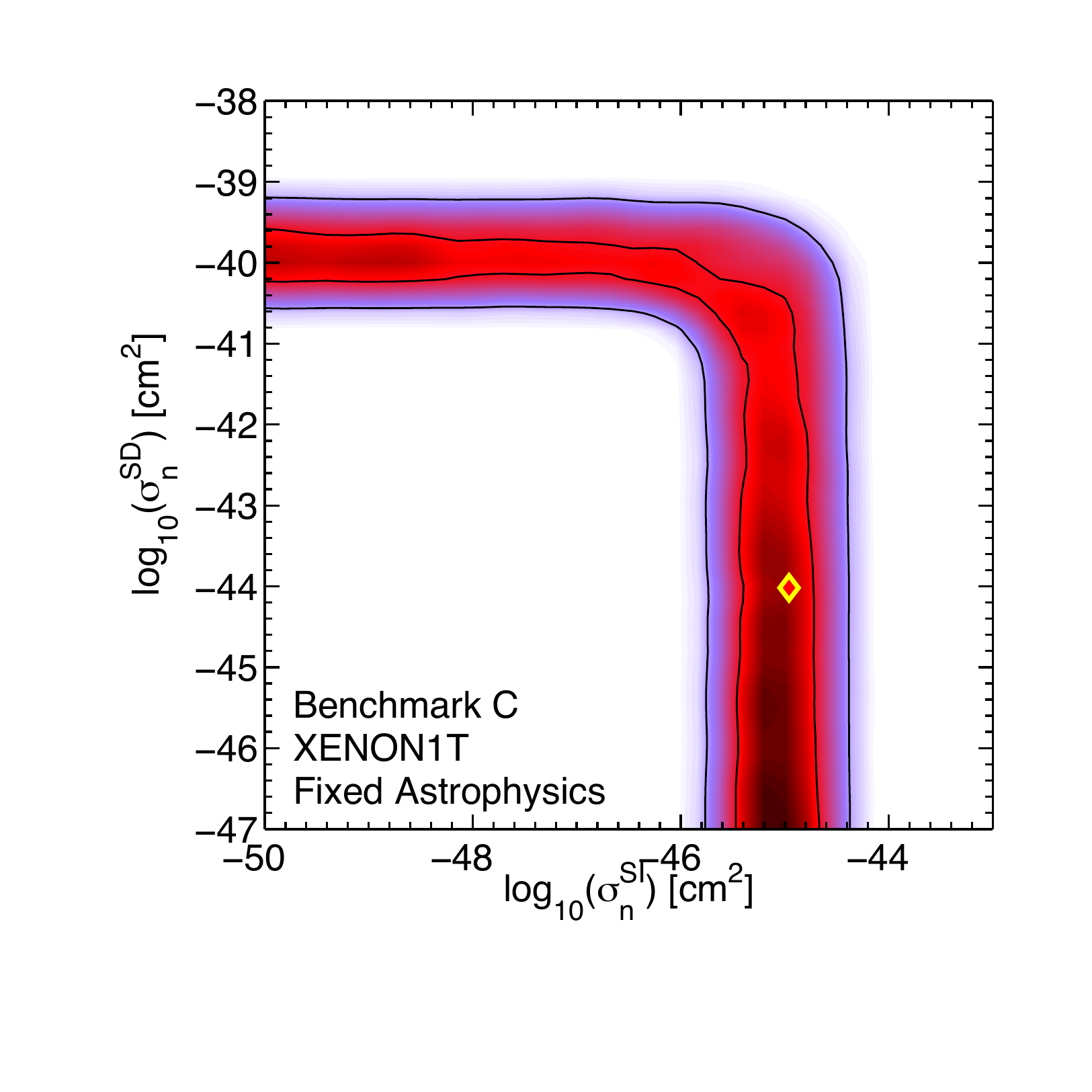}
\end{minipage}
\caption{The reconstruction with XENON1T of the $A$ (top), $B$ (middle) and $C$ (bottom) benchmark models with fixed astrophysics. No nuclear uncertainties are taken into account. The left, central and right plot shows the 2D marginal posterior pdf in the $\{m_{\rm DM}, \sigma^{\rm SI}_n\}$-plane, $\{m_{\rm DM}, \sigma^{\rm SD}_n\}$-plane and $\{\sigma^{\rm SI}_n,\sigma^{\rm SD}_n\}$-plane respectively. The contours denote the 68\% and 95\% credible regions and the diamond point shows the parameter values of the benchmark models.}
\label{fig:Xeonly}
\end{figure*}

The expected number of recoiling nuclei for our benchmark models are shown in Table \ref{tab:bench}. Benchmark $A$ has a negligible SI cross-section of about $10^{-49} {\rm cm^2}$, which is well below the expected sensitivity of XENON1T of $\sim 10^{-47} {\rm cm^2}$: the theoretical predictions are compatible with the background in this case. All the signal comes from the SD contribution, which can range from 200 to 400 events, depending on the nuclear structure function adopted for describing the xenon nuclei. In this case $\sigma^{\rm SD}_n = 2.0 \times 10^{-40} {\rm cm^2}$ is close to the upper bound of XENON100 for SD interaction on neutron~\cite{:2013uj} and in the sensitivity range of XENON1T for SD which is expected to be around $10^{-42} {\rm cm^2}$~\cite{Aprile:2012zx}. Benchmark $B$ is similar to $A$ as far as it concerns the SD contribution, with a number of events ranging from 100 to 350 depending on the structure function model. In contrast to Benchmark $A$ however, $B$ has a large contribution of 252 events from the SI interaction, which has a cross-section of $\sigma^{SI}_n = 8.8 \times 10^{-46} {\rm cm^2}$, just below the exclusion bound of XENON100~\cite{Aprile:2012nq}. The last model, Benchmark $C$, has a negligible contribution from the SD interaction, accounting for zero events, while the SI interaction predicts about 70 events.

Figure \ref{fig:Xeonly} shows the reconstruction of our benchmark models with XENON1T only, with the Galactic parameters fixed at their observed values and with the CEFT nuclear structure function in use. Hence no nuisance parameters are included in this part of our analysis. The left panels show the 2D marginal posterior pdf in the \{$m_{\rm DM}, \sigma_n^{\rm SI}$\}-subspace, the central panels show the 2D marginal posterior pdf in the \{$m_{\rm DM}, \sigma_n^{\rm SD}$\}-subspace, while the rightmost panels show the 2D marginal posterior pdf in the \{$ \sigma_n^{\rm SI}, \sigma_n^{\rm SD}$\}-subspace, for Benchmarks $A$, $B$ and $C$ in the first, second and third rows respectively. 

If we allowed the DM particle to have only one interaction, SI or SD, contributing to the observed events, we would expect the reconstruction of the WIMP-nucleon cross-section to be effective, namely to select only a closed region in the parameter space. However, as soon as both cross-sections are allowed to vary the marginal posterior pdf exhibits a tail towards very small values, since each parameter can compensate for the other to achieve the observed number of events. In other words, as the signal in XENON1T receives contributions from both SI and SD interactions a degeneracy can occur~\cite{Cerdeno:2012ix,Cerdeno:2013gqa}. This is apparent from all plots, where the 95\% credible region does not exhibit a closed contour. The effect is more pronounced for the SD interaction, while in the case of SI scattering the 68\% credible region is denoted by a closed contour around the Benchmark $A$ point, or always bounded from below in the $B$ and $C$ cases. This is due to the fact that XENON1T is more sensitive to SI interactions than SD interactions. In addition a low signal for one interaction will cause the corresponding true value to be poorly reconstructed: for example, for $A$ the SI interaction is very small and close to zero observed events, hence the signal in XENON1T at 68\% confidence level (CL) is attributed to the largest cross-section that accounts for such events, and the true value is contained only in the $95\%$ CL. This effect is mirrored for Benchmark $C$, where again the true value for the SD interaction is only contained in the $95\%$ credible region, which are as well upper bound for that interaction.

In the reconstruction of the DM mass parameter, we encounter the usual limitation of direct detection: after the WIMP mass matches and then exceeds the target nuclei mass, the mass reconstruction becomes progressively worse, as the slope of the differential cross-section becomes essentially insensitive to WIMP mass for $m_{\rm DM} \gg m_{\mathcal{N}}$. The degeneracy in the mass reconstruction can be seen as we move to the higher mass $B$ and $C$ points, as denoted in the left and central panel of Fig.~\ref{fig:Xeonly} by the tail extending towards large masses. The only exception is Benchmark $A$, whose posterior pdf does not exhibit this mass degeneracy and appears to be a closed contour around the benchmark model point at both 68\% and 95\% CLs. A viable solution to ameliorate the DM mass reconstruction is to combine different target materials, typically with different atomic number, as described in~\cite{Pato:2010zk,Cerdeno:2013gqa}. However heavy WIMPs still exhibit a tail in the posterior pdf extending towards high mass values. We will show in Sec.~\ref{sec:comb} that it is possible to break this degeneracy by complementing the signal seen in XENON1T with the signal of IceCube, or even with the nondetection of a signal.

It is worth commenting on the features appearing in the marginal posterior pdf contours at the 95\% CL, in particular affecting Benchmark $C$ of Fig.~\ref{fig:Xeonly}. These features are not physical, but are an indication that the posterior pdf is flat, which makes it a challenge to sample effectively. This also occurs with IceCube alone or when the likelihood is poorly constrained by the data.

\section{Reconstruction with IceCube}\label{sec:IC} 

The IceCube neutrino telescope consists of 5160 digital optical modules (DOMs) arranged vertically along 86 strings embedded in a cubic kilometer of extremely transparent natural ice below the South Pole~\cite{HalzenKlein2010}. Included in IceCube is the DeepCore subarray, which consists of six strings arranged at the center of IceCube with closer horizontal spacing and instrumented with DOMs of higher sensitivity and closer vertical spacing along the string. DeepCore has been designed to reduce the energy threshold of IceCube from 100 GeV down to 10 GeV~\cite{IceCubeDeepCore2012}.

The WIMP annihilations can produce a wide variety of final states, whose subsequent decays can produce high energy neutrinos, with the branching ratios to each state dependent on the properties of the DM particle and the underlying theory that generates it. In our phenomenological approach we consider three annihilation final states with branching ratio equal to one, as labeled in Table \ref{tab:bench}. For Benchmark $A$ we consider the $\tau^+\tau^-$ final state, which gives a harder neutrino signal than the $b\bar{b}$ channel. Benchmark $B$ annihilates into $W^+W^-$, which is the dominant channel for a common fermionic WIMP with a mass above the $W$ threshold. Finally we consider the $\nu_\mu\bar{\nu}_\mu$ channel for Benchmark $C$. The annihilation directly into neutrinos is heavily suppressed for Majorana particles ({\it e.g.} neutralinos)~\cite{Goldberg1983} but can occur for vector WIMPs, such as the Kaluza-Klein photon~\cite{HooperSilk2004,Bertone:2007xj}. 

For each benchmark model and its annihilation channel we first calculate the theoretical signal rate using the default routines in \texttt{DarkSUSY 5.0.6}~\cite{Gondolo2004,DarkSUSY2009} (which employ the methods outlined in Sec.~\ref{sec:theoic}) with a slight modification to enforce the {\it steady state} scenario. The MSSM25 models we build upon for our benchmark WIMP models (see Table \ref{tab:bench}) have equilibration times of the order of $1-4 \times 10^8$ years, \textit{e.g.} an order of magnitude smaller than the age of the Sun, making our assumption of a {\it steady state} scenario a valid one. We use an exposure time of five 180 day austral winter observing seasons ({\it e.g.} 900 days), and an angular cut around the solar position of $\phi_{\rm{cut}} = 3^\circ$. The predicted muon signals $S_\mu(\Theta)$ for each benchmark model and annihilation channel are given in Table \ref{tab:bench}. From the background files included in the \texttt{DarkSUSY 5.0.6} release we extracted the number of background events within the $3^\circ$ angular cut. For a 180 day season the number of background events was 41, yielding 205 background events for our 900 day exposure time. With this background estimation we have an estimate of the number of events needed for a detection within 5 seasons of data taking in IceCube, which is approximately 50 events, depending on the annihilation channel. This is compatible with the analysis released recently by the IceCube collaboration~\cite{Aartsen:2012kia}: by analyzing the data from one winter season with up-going muons plus one summer season with down-going muons, taken using 79 strings (including the six DeepCore strings), the strongest exclusion bounds are for a WIMP mass of 250 GeV annihilating into gauge bosons and with a spin dependent cross-section on proton of about $10^{-40} {\rm cm^2}$ producing $\sim 15$ events. The use of DeepCore allows the limit to be extended down to a mass of 20 GeV but the sensitivity rapidly diminishes, and at this mass only a SD cross-section of approximately $10^{-38}{\rm cm^2}$ can be excluded. For a WIMP with a mass of 50 GeV the current exclusion bound  is close to $10^{-40}{\rm cm^2}$, comparable with the bound obtained with XENON100. For our benchmark models the expected number of events range from $\sim 67$ muons for Benchmark $B$ to 7.8 muons produced by the $\nu_\mu\bar{\nu}_\mu$ final state for $C$. Benchmark A annihilating to the $\tau^+\tau^-$ final state is intermediate with 22 expected muons. Comparing these to Table I of~\cite{Aartsen:2012kia} we can see that all our benchmarks have signal rates compatible with the current exclusion limits. 

The likelihood function we use for the IceCube experiment, Eq.~(\ref{eq:IClikelihood}), is presented in~\cite{Scott:2012mq} and included in \texttt{DarkSUSY 5.0.6}. The likelihood of observing $N_{\rm{tot}}$ events given a background rate of $\theta_{\rm{BG}}$ and a theoretical model $\Theta$, which has a true expected signal $S_{\rm true}(\Theta) = \varepsilon S_\mu(\Theta)$ events, is 
\begin{widetext}
\begin{equation}
\label{eq:IClikelihood}
\mathcal{L}_{\rm IC86}(\Theta) \equiv \mathcal{L} \left(N_{\rm{tot}}|\theta_{\rm{BG}}, S_\mu \right) = \frac{1}{\sqrt{2\pi} \sigma_{\varepsilon}} \int_0^{\infty} \frac{\left( \theta_{\rm{BG}} + \varepsilon S_\mu(\Theta) \right)^{N_{\rm{tot}}}  e^{-(\theta_{\rm{BG}} + \varepsilon S_\mu(\Theta) )}}{N_{\rm{tot}}!} \rm{exp} \left[{- \frac{1}{2} \left(\frac{\rm{ln}\, \varepsilon}{\sigma_\varepsilon} \right)^2}\right] \text{d} \varepsilon.
\end{equation}
\end{widetext}
The variable $\varepsilon$ quantifies potential systematic errors between the true expected signal $S_{\rm true}(\Theta)$ and the nominally predicted signal $S_\mu(\Theta)$. The relative fractional error on $S_\mu(\Theta)$ is then $\varepsilon - 1$, and this is then marginalized over in a semi-Bayesian way, assuming a log-normal probability distribution and with standard deviation $\sigma_\varepsilon$. This likelihood takes into account only the total number of signal events. A likelihood function taking into account the energy of the signal muons is also presented in~\cite{Scott:2012mq} and encoded into \texttt{DarkSUSY 5.0.6}, but this requires an accurate description of the energy response function of IceCube, which is not publicly available for the 86 string configuration. 
\begin{figure*}[t]
\begin{minipage}[t]{0.32\textwidth}
\centering
\includegraphics[width=1.\columnwidth,trim=10mm 25mm 12mm 12mm, clip]{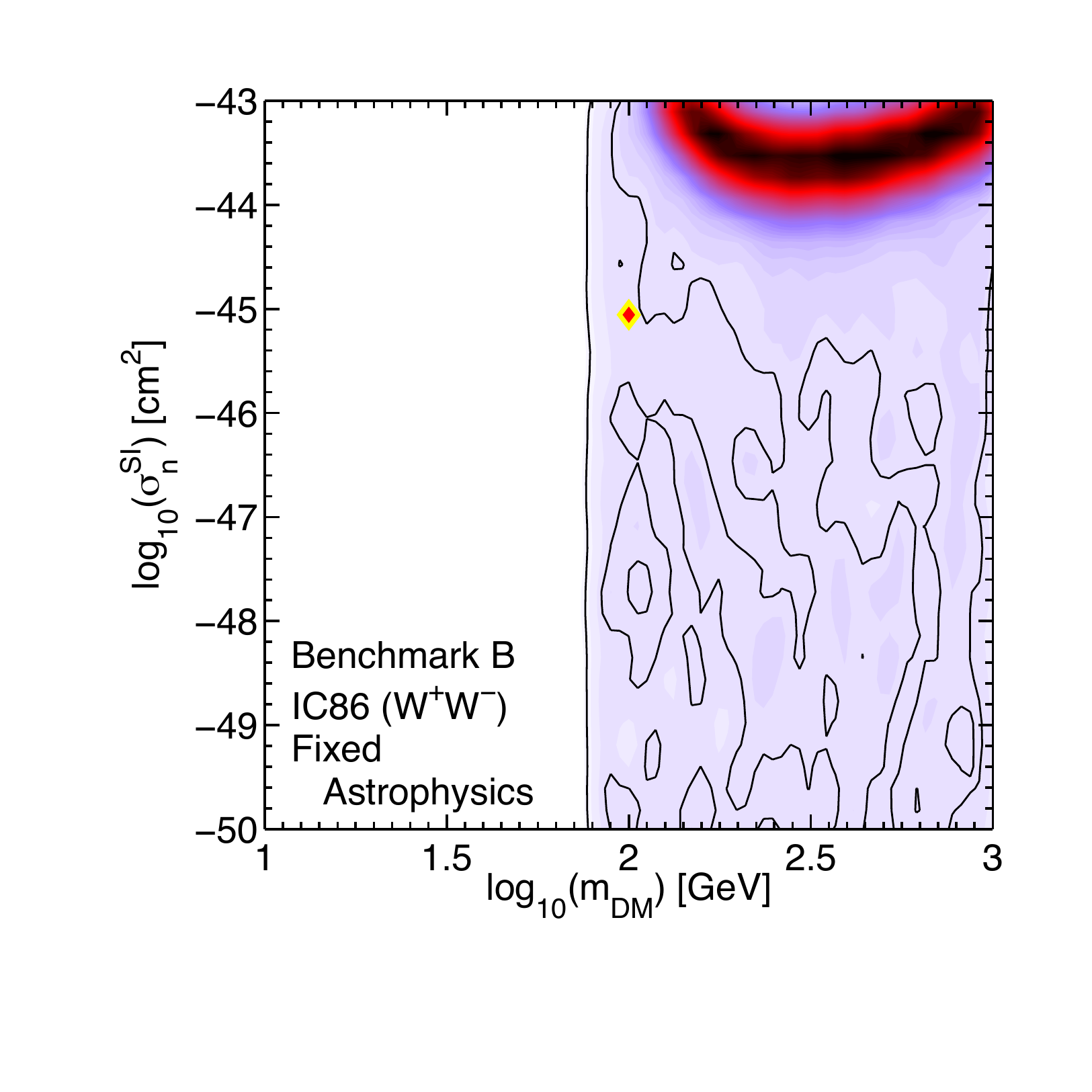}
\end{minipage}
\begin{minipage}[t]{0.32\textwidth}
\centering
\includegraphics[width=1.\columnwidth,trim=10mm 25mm 12mm 12mm, clip]{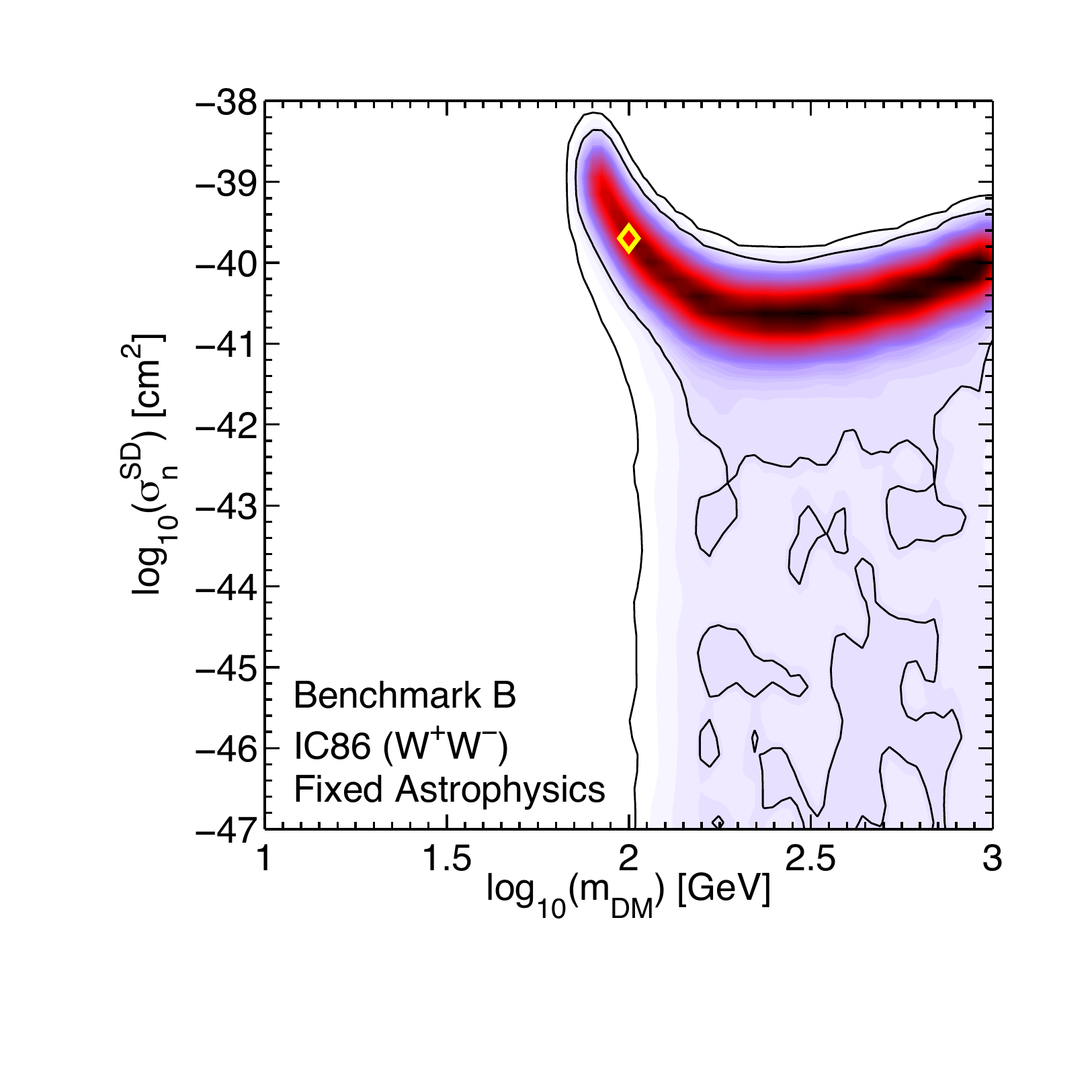}
\end{minipage}
\begin{minipage}[t]{0.32\textwidth}
\centering
\includegraphics[width=1.\columnwidth,trim=10mm 25mm 12mm 12mm, clip]{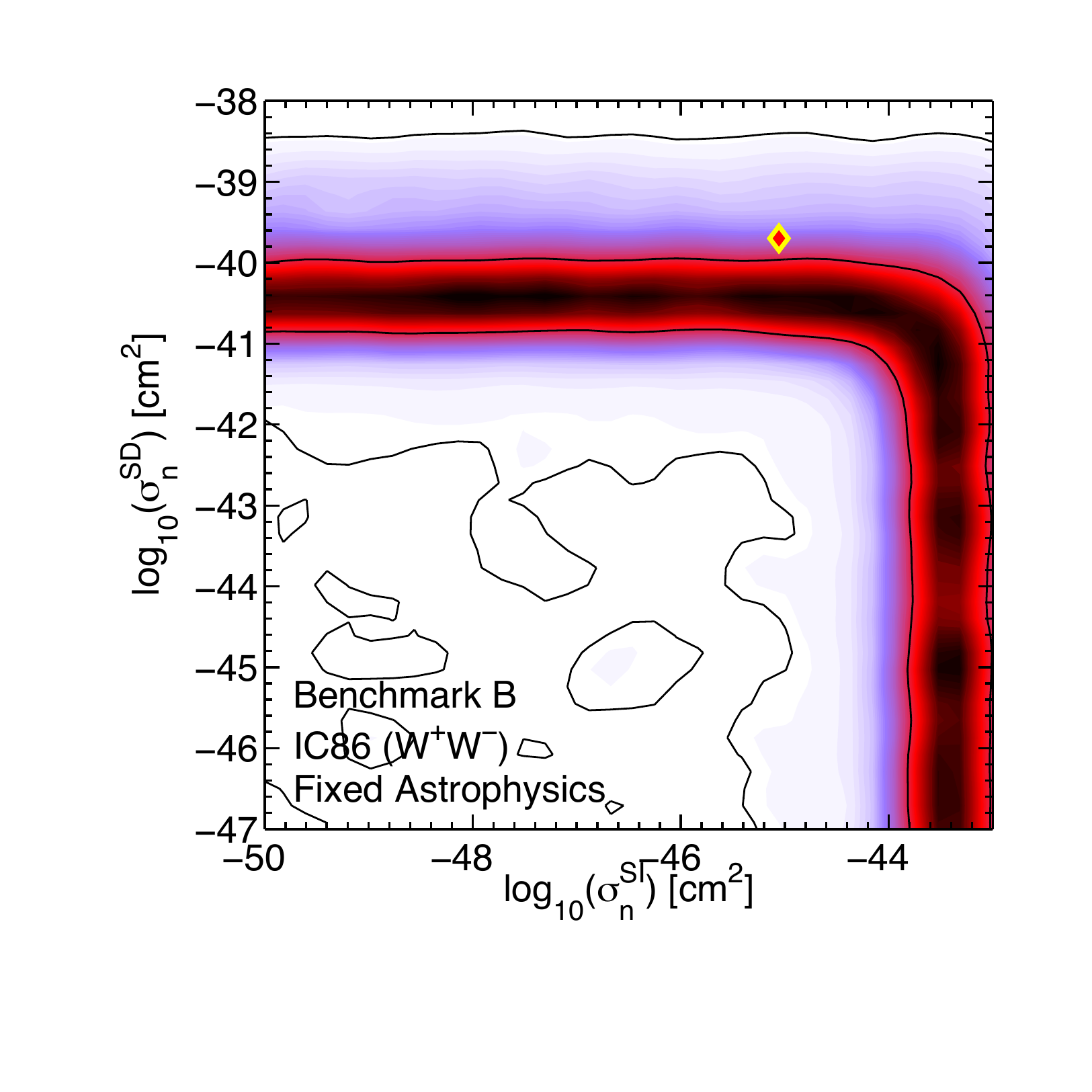}
\end{minipage}
\caption{Reconstruction using IceCube (denoted IC86), with fixed astrophysics, for benchmark model $B$ with annihilation channel $W^+W^-$. The left, central and right plot shows the 2D marginal posterior pdf in the $\{m_{\rm DM}, \sigma^{\rm SI}_n\}$-plane, $\{m_{\rm DM}, \sigma^{\rm SD}_n\}$-plane and $\{\sigma^{\rm SI}_n,\sigma^{\rm SD}_n\}$-plane respectively. The contours denote the 68\% and 95\% credible regions and the diamond point shows the parameter values of the benchmark model.}
\label{fig:ICfa}
\end{figure*}

There are several experimental aspects of IceCube that are salient to our discussions. First, the effective area of IceCube, which quantifies the sensitivity of the detector, increases with energy~\cite{Danninger2011}. Thus higher mass WIMPs will tend to produce stronger signals as they will, in general, produce higher energy neutrinos and thus higher energy muons. Second, as neutrino energy increases the mean angular error for the reconstructed muon track decreases~\cite{Danninger2011}. During calculation of signal in IceCube an angular cut ($\phi_{\rm{cut}}$) around the solar position is made. Thus lower mass WIMPs will produce lower signal, as the muons they produce can potentially have greater angular deviation from the solar position and so fewer will survive the angular cut. Conversely the higher energy muons will be more densely clustered around the solar position, and so more will pass the angular cut, increasing signal for the higher mass WIMPs.  

Given these limitations in the number likelihood of IceCube, there exists a degeneracy between WIMP mass and WIMP-nuclei cross-section $\sigma_{\rm{tot}} = \sigma^{\rm{SI}}_n + \sigma^{\rm{SD}}_n$. Consider two scenarios, one with a WIMP of high mass and low cross-section $\sigma_{\rm{tot}}$, and another with a WIMP of low mass but high $\sigma_{\rm{tot}}$. The first scenario will have a low capture rate, as this is dependent on $\sigma_{\rm{tot}}$ and inversely dependent on WIMP mass, Eq.~(\ref{eq:CaptureRateTotalIntegral}). Assuming capture-annihilation equilibrium, this scenario will feature a lower muon flux, but the muons will be of a higher energy and so hence produce higher signal due to the features of IceCube outlined above. The muons of the second scenario will be of lower energy, but they will have a higher flux as their capture rate is higher. This can compensate the lower muon energy and generate a high signal. As the energy spectrum, which could be used to determine the DM mass, is not defined in the number likelihood, these two scenarios cannot be distinguished from each other. As for direct detection signals, there is a degeneracy between $\sigma^{\rm{SI}}_n$ and $\sigma^{\rm{SD}}_n$. The solar capture rate, given in Eq.~(\ref{eq:CaptureRateTotalIntegral}), receives contributions from both SI and SD interactions, however the sensitivity of IceCube to the two type of interactions is different, as shown in~\cite{Aartsen:2012kia}: the SI is poorly constrained and the actual exclusion limits are one or two orders of magnitude less constraining than the XENON100 ones, while for the SD cross-section the exclusion limits are competitive with XENON100 at high masses. 

Figure \ref{fig:ICfa} shows the reconstruction of benchmark model $B$ with IceCube only, and exemplifies the discussion above. In the left panel, which shows the 2D marginal posterior pdf in the $\{m_{\rm DM}, \sigma^{\rm SI}_n\}$-plane, the signal clearly exhibits an almost independent behavior with respect to SI interaction. On the contrary, the SD cross-section as a function of the DM mass is both lower and upper bounded at 68\% CL (middle panel). The mass degeneracy is a result of the missing spectral information. The degeneracy between $\sigma^{\rm{SD}}_n$ and $\sigma^{\rm{SI}}_n$ is similar to the case of direct detection, as shown in the right panel and in Fig.~\ref{fig:Xeonly}. As far as it concerns Benchmarks $A$ and $C$, the muon signal produced by WIMP annihilations in the Sun cannot be disentangled from the background, leading to a flat and featureless marginalized posterior pdf and underlining that there is not detection in IceCube. Once again, the features in the pdfs have no physical meaning and are merely artifacts of a flat likelihood.

\section{Combined XENON1T and IceCube analysis}\label{sec:comb}
\begin{figure*}[t]
\begin{minipage}[t]{0.32\textwidth}
\centering
\includegraphics[width=1.\columnwidth,trim=10mm 25mm 12mm 12mm, clip]{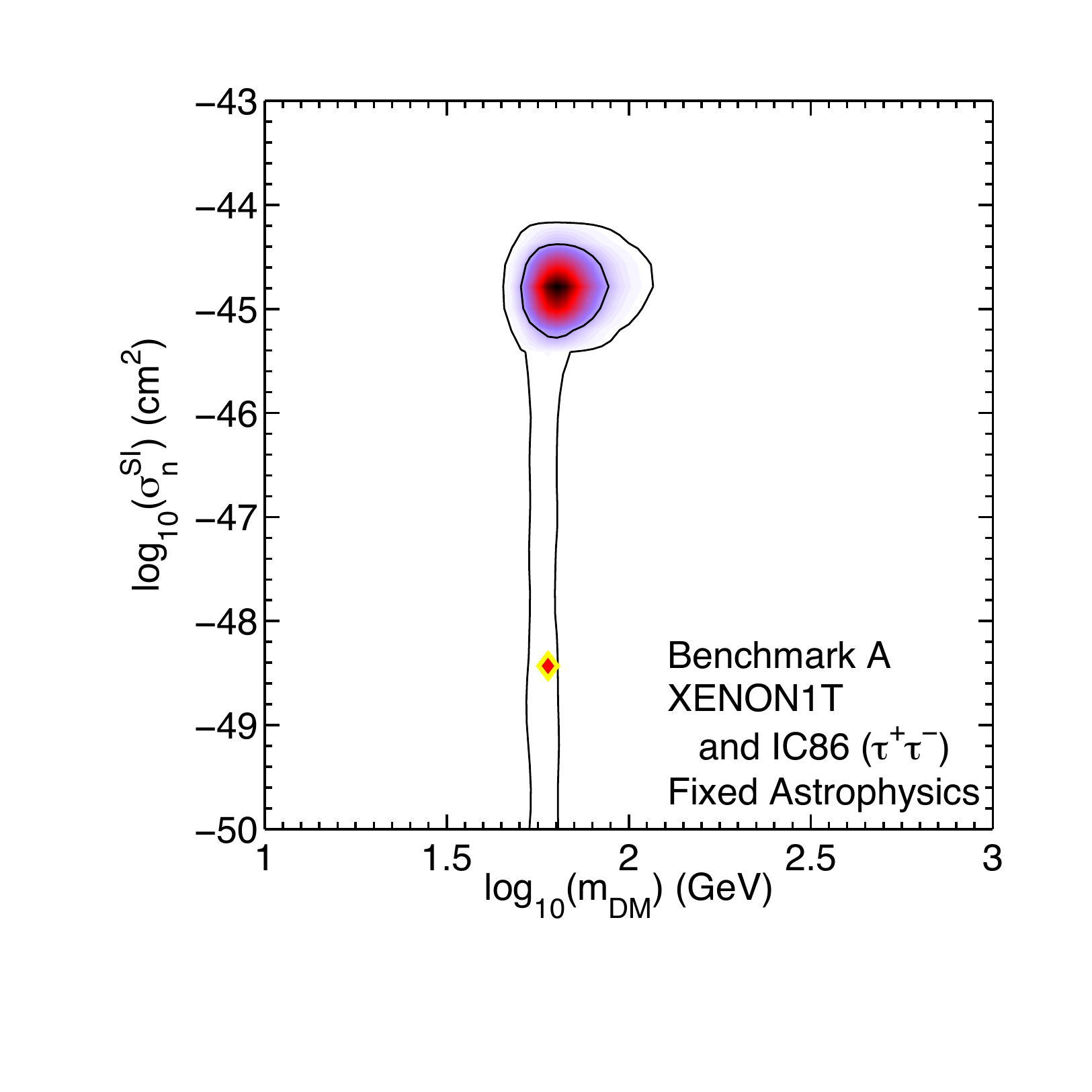}
\end{minipage}
\begin{minipage}[t]{0.32\textwidth}
\centering
\includegraphics[width=1.\columnwidth,trim=10mm 25mm 12mm 12mm, clip]{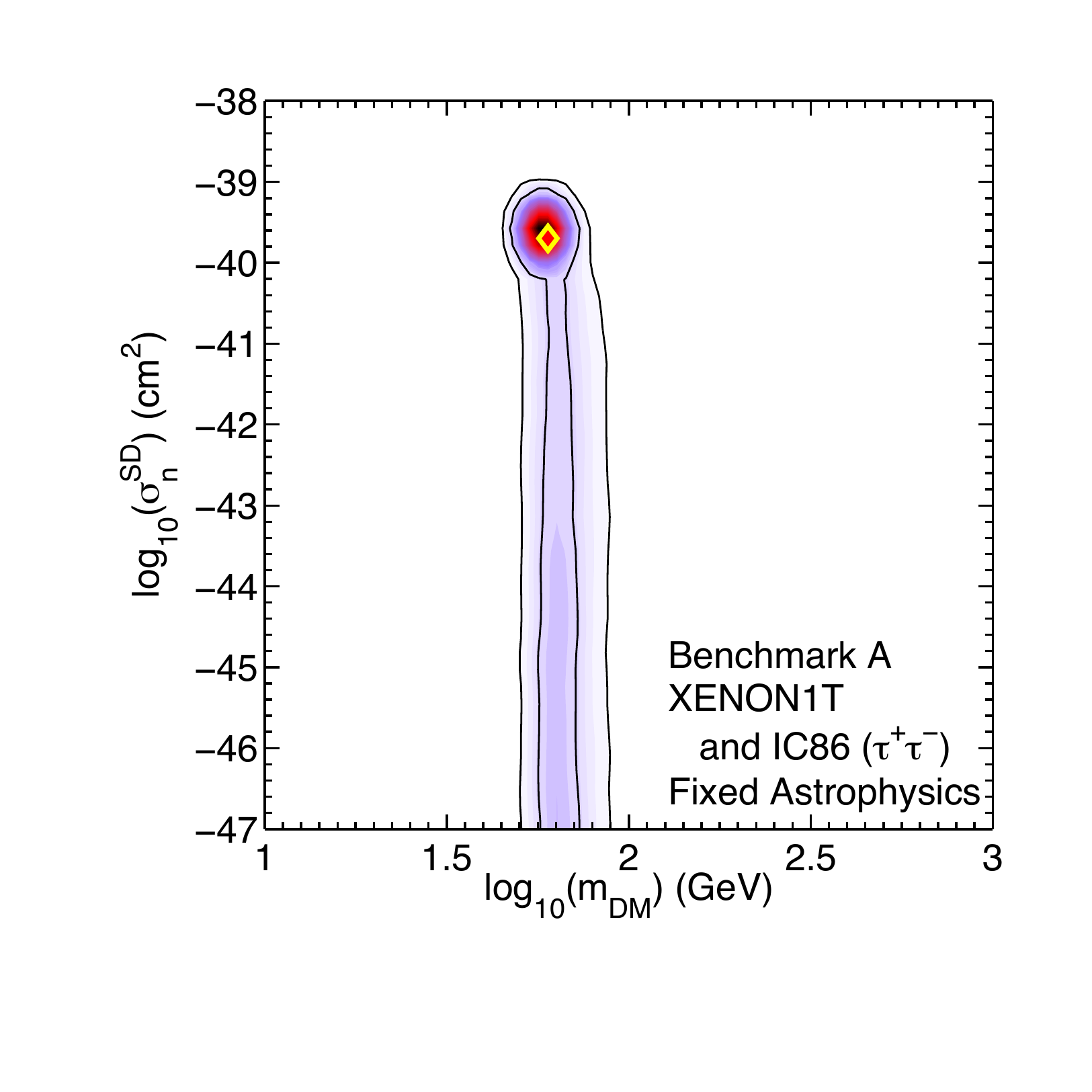}
\end{minipage}
\begin{minipage}[t]{0.32\textwidth}
\centering
\includegraphics[width=1.\columnwidth,trim=10mm 25mm 12mm 12mm, clip]{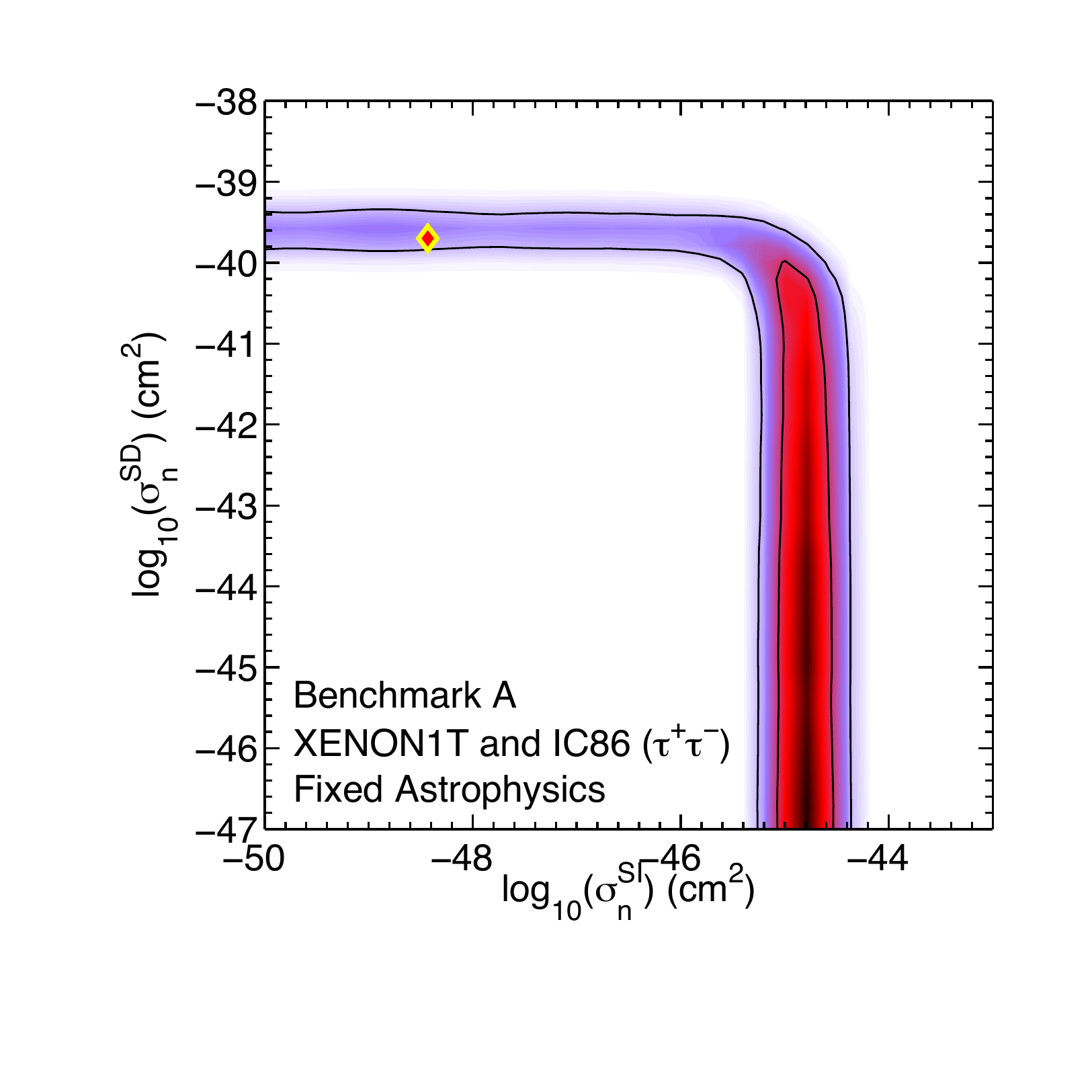}
\end{minipage}
\\
\begin{minipage}[t]{0.32\textwidth}
\centering
\includegraphics[width=1.\columnwidth,trim=10mm 25mm 12mm 12mm, clip]{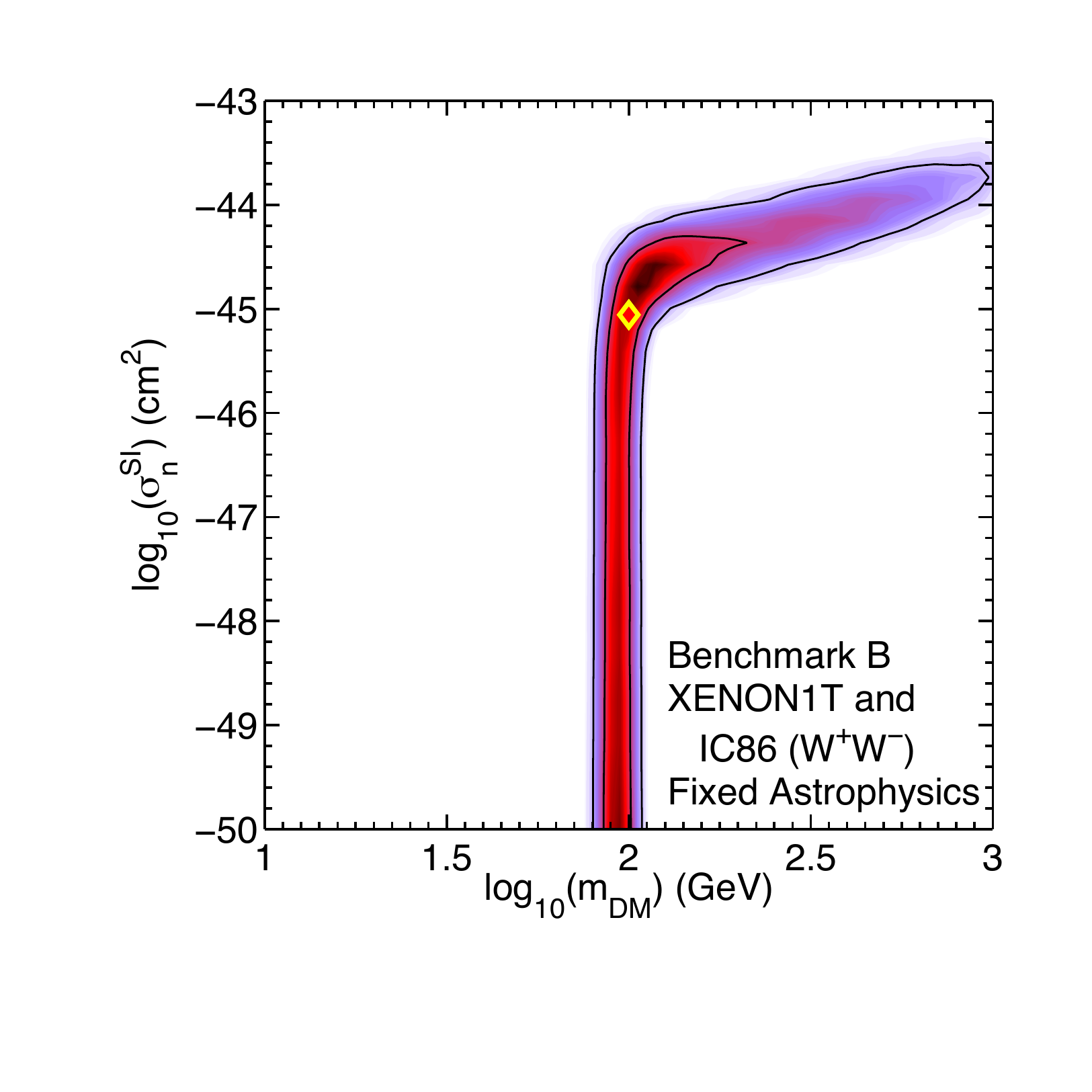}
\end{minipage}
\begin{minipage}[t]{0.32\textwidth}
\centering
\includegraphics[width=1.\columnwidth,trim=10mm 25mm 12mm 12mm, clip]{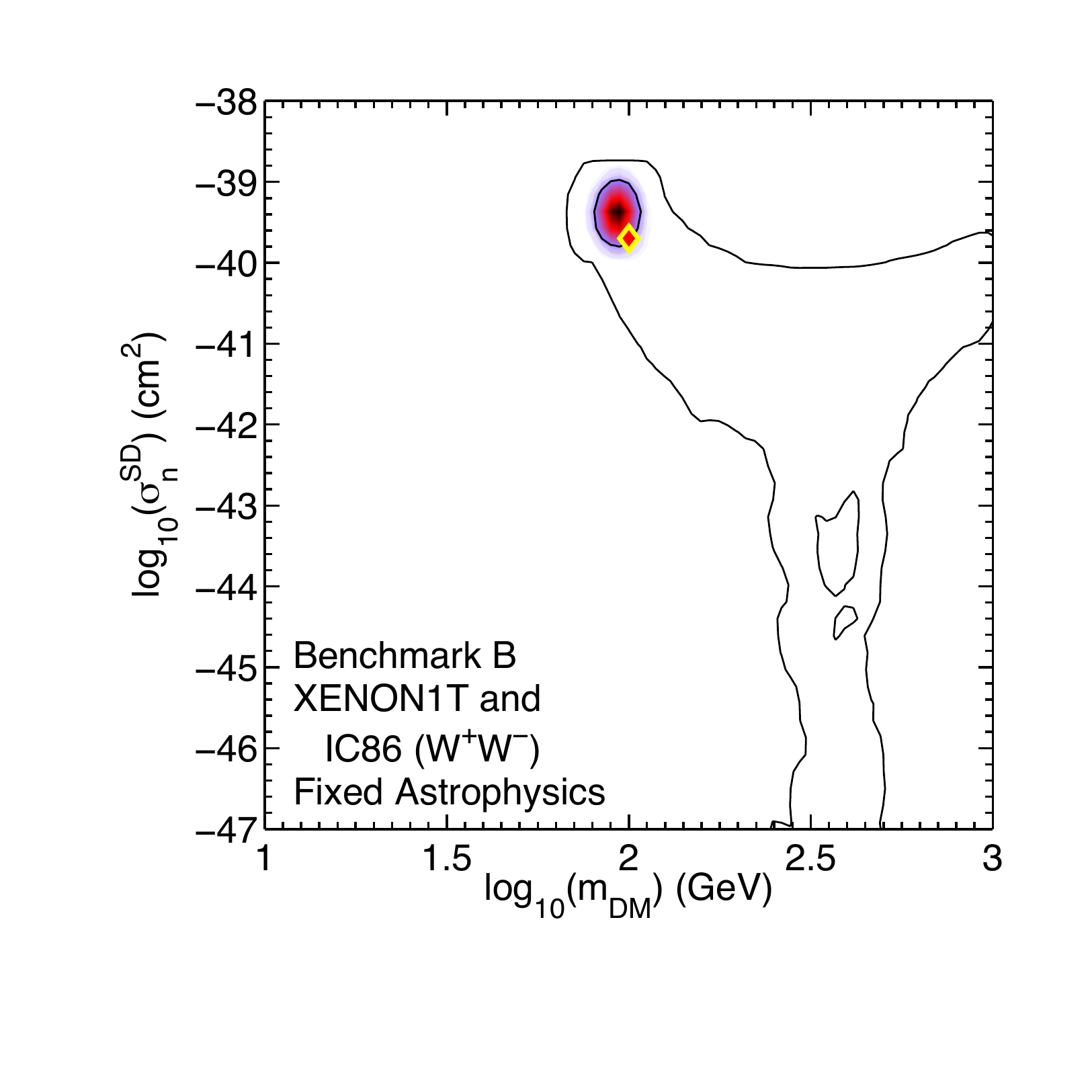}
\end{minipage}
\begin{minipage}[t]{0.32\textwidth}
\centering
\includegraphics[width=1.\columnwidth,trim=10mm 25mm 12mm 12mm, clip]{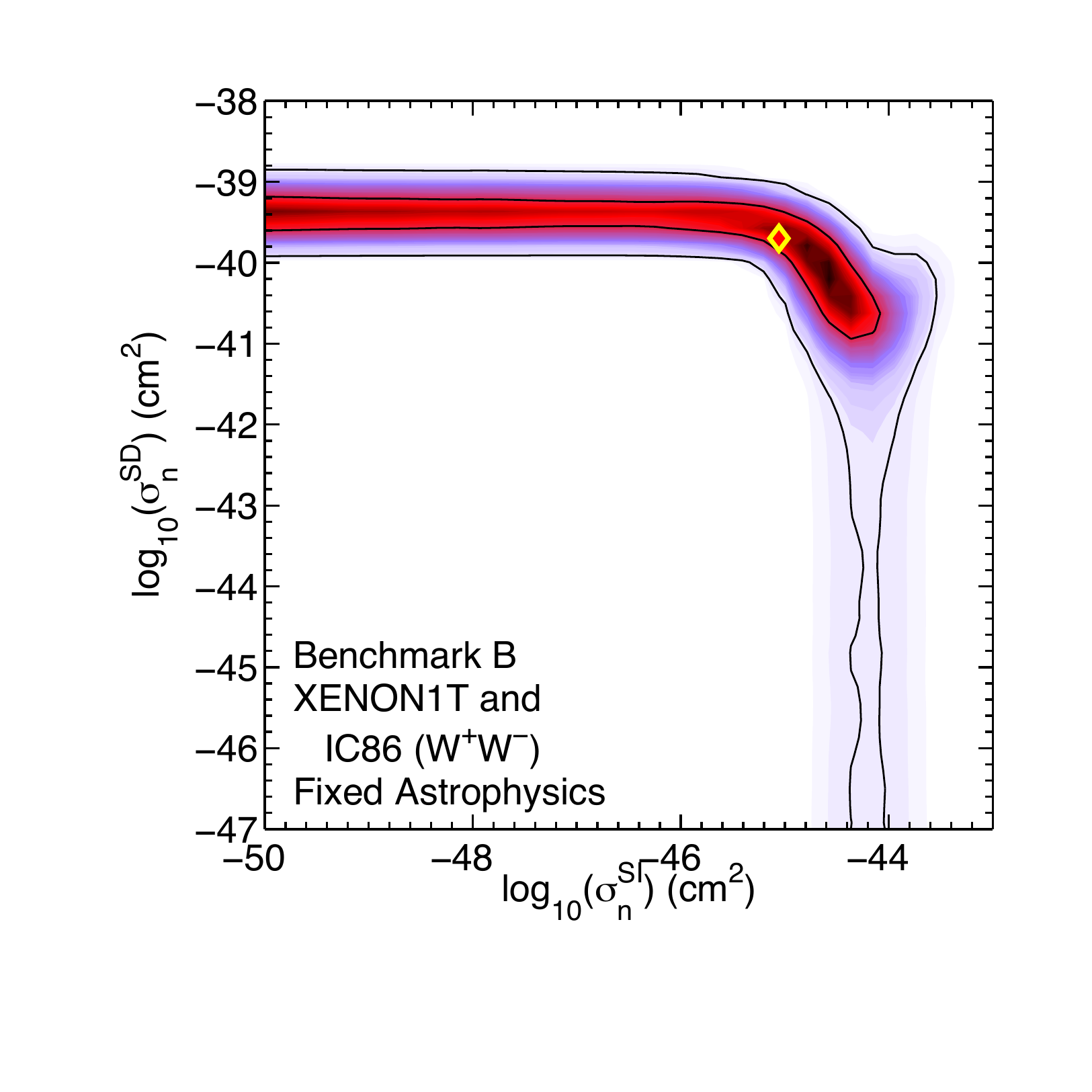}
\end{minipage}
\\
\begin{minipage}[t]{0.32\textwidth}
\centering
\includegraphics[width=1.\columnwidth,trim=10mm 25mm 12mm 12mm, clip]{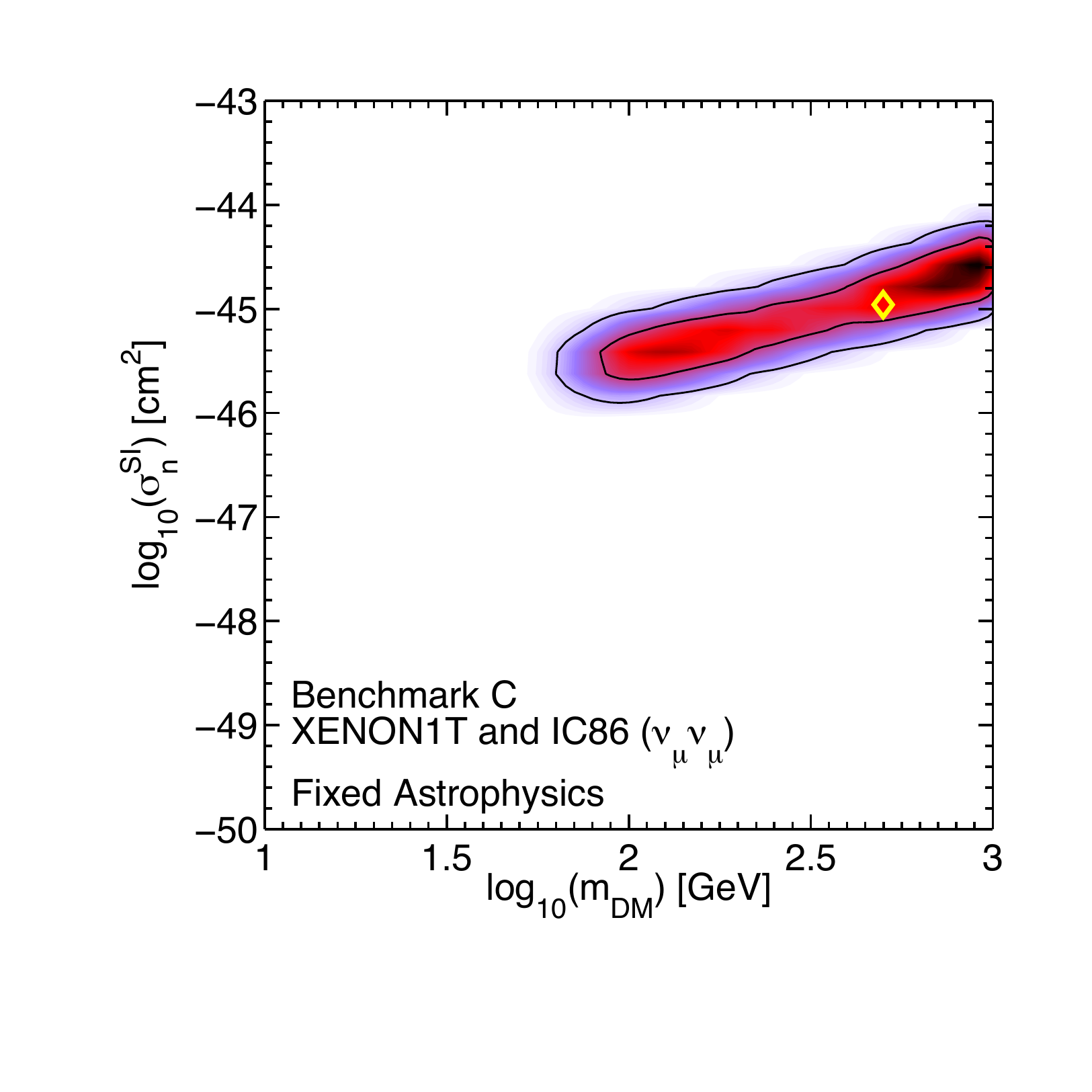}
\end{minipage}
\begin{minipage}[t]{0.32\textwidth}
\centering
\includegraphics[width=1.\columnwidth,trim=10mm 25mm 12mm 12mm, clip]{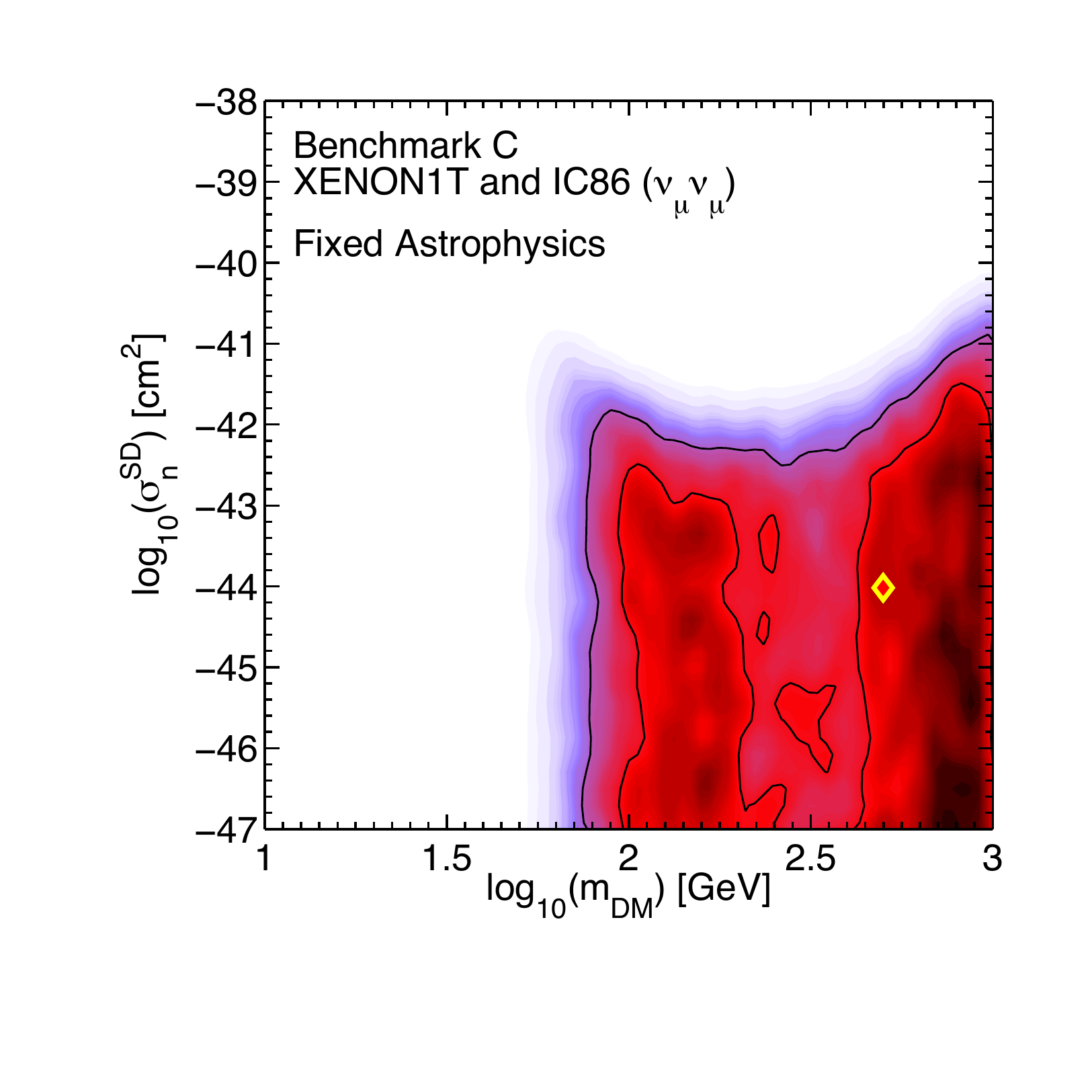}
\end{minipage}
\begin{minipage}[t]{0.32\textwidth}
\centering
\includegraphics[width=1.\columnwidth,trim=10mm 25mm 12mm 12mm, clip]{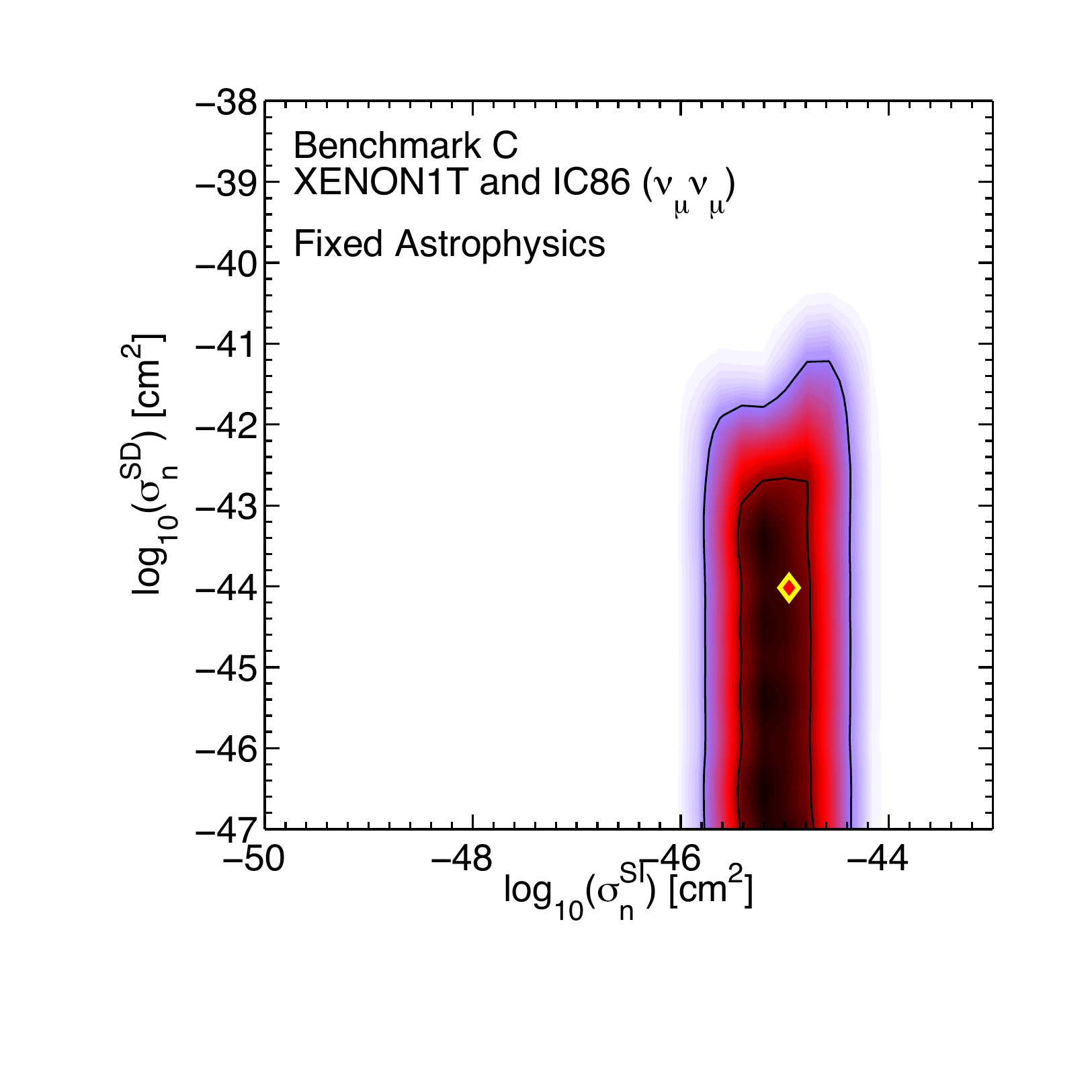}  
\end{minipage}
\caption{Reconstruction using XENON1T combined with IceCube (denoted IC86), with fixed astrophysics, for benchmark models $A$, $B$ and $C$ with annihilation channels $\tau^+\tau^-$ (top), $W^+W^-$ (centre) and $\nu_\mu\bar{\nu}_\mu$ (bottom). The left, central and right plot shows the 2D marginal posterior pdf in the $\{m_{\rm DM}, \sigma^{\rm SI}_n\}$-plane, $\{m_{\rm DM}, \sigma^{\rm SD}_n\}$-plane and $\{\sigma^{\rm SI}_n,\sigma^{\rm SD}_n\}$-plane respectively. The contours denote the 68\% and 95\% credible regions and the diamond point shows the parameter values of the benchmark model.}
\label{fig:XeICfa}
\end{figure*}

Figure~\ref{fig:XeICfa} shows how the future detection/constraint from IceCube complements a detection in XENON1T, for fixed astrophysical parameters. The top row of Fig.~\ref{fig:XeICfa} illustrates the combined reconstruction of Benchmark $A$. Even though this benchmark would produce only 12 events in IceCube, hence too small a signal to claim a detection, it still has the ability to improve the reconstruction. Comparing to the top row of Fig.~\ref{fig:Xeonly} the 68\% and 95\% CLs have shrunk.  The reconstruction of $\sigma^{\rm SI}_n$ and $\sigma^{\rm SD}_n$ exhibits the same trend as the case of XENON1T only, except that now the 68\% CL for SD cross-section in the right panel is upper bounded, because of  the IceCube upper bound. The mass determination remains unchanged, demonstrating the good reconstruction capability of XENON1T and the reduced sensitivity of IceCube.

Benchmark model $B$ (second row of Fig.~\ref{fig:XeICfa}) fully exploits the properties of complementarity between DM search strategies and is the principal illustration of the main point of our analysis, as it can be detected by both experiments. Comparing with Figs.~\ref{fig:Xeonly} and~\ref{fig:ICfa} we can see that the combination of the two experiments allows for the SD cross-section (middle panel) to be well determined at the 68\% CL. The mass degeneracy is also significantly reduced and it is well determined at the 68\% CL. The contours for the SI cross-section as a function of $m_{\rm DM}$ have been contracted sensibly with respect to the separate experiments, but at $68\%$ CL the SI cross-section is only upper bounded, as can be seen as well from the third panel: the SD cross-section is determined within one order of magnitude at most and contains the true value, while the SI cross-section cannot be larger than $10^{-44}{\rm cm^2}$. Therefore if a WIMP is detected by both experiments, meaning that it has sizable cross-sections, the SD cross-section and the mass can be well reconstructed, while a precise reconstruction of SI interaction would require the use of a further experiment; in this case a direct detection experiment with different target material would be the optimal choice~\cite{Cerdeno:2013gqa}.

Benchmark $C$ is illustrated in the last row of Fig.~\ref{fig:XeICfa} and shows remarkably the constraining power of IceCube, even with only 8 events from high energy neutrinos. The most dramatic improvement is in the determination of $\sigma^{\rm SI}_n$: where previously the 68\% and 95\% CLs extended from the upper bound all the way down to the bottom of our prior range, they are now bounded from below as well. As expected, only an upper limit can be set on $\sigma^{\rm SD}_n$\,: the 95\% CL is clearly bounded from above at values of $\sigma^{\rm SD}_n \approx 10^{-42} {\rm cm^2}$. The effect of the SD upper bound is much more striking than for Benchmark $A$, signaling the fact the IceCube is more likely to be sensitive to high WIMP masses. Note that in this analysis we assumed, in absence of published information on the energy response function of the 86-string configuration, that nothing can be said on the energy of the neutrinos measured by IceCube. It is known that the energy resolution of IceCube is poor; for a muon of energy $10^4 - 10^8$ GeV in the core of the detector there is an uncertainty of a factor of two~\cite{MontaruliIceCube2011}. Nonetheless, including the spectral information on observed events would certainly improve the mass determination, one of the main weaknesses of the reconstruction procedure from XENON1T data only, and substantially reduce the degeneracy in the DM particle parameter space. As the mass cannot be well reconstructed by either XENON1T alone or by adding the upper bound of IceCube, this last benchmark model is affected by the prior range choice on the DM mass: this will be visible in all two-dimensional projections of the dark matter parameter space as an increase of confidence levels, hence volume effect.

\section{Role of uncertainties in the reconstruction of WIMP parameters}\label{sec:unc}

In this section we first assess the effect in the combined reconstruction of benchmark models when DM Galactic parameters vary within their uncertainties (see Sec.~\ref{sec:stat}) and discuss the effective DM density probed by the capture rate (see Sec.~\ref{sec:theoic}). We then account as well for our lack of knowledge on the true form of the nuclear structure function (see Sec.~\ref{sec:theoxe}). At the end of this section, we return to the additional uncertainties induced by the shape and parametrization of the velocity distribution.

\subsection{Astrophysical uncertainties}\label{sec:astro}

A reconstruction of benchmark models fixing the astrophysical parameters to their observed value could hold for very well measured parameters, however, as mentioned in Sec.~\ref{sec:stat}, these are known only up to various degrees of precision. We therefore introduce these uncertainties into our analysis, treating the Galactic parameters as nuisances $\psi_{\rm astro}$ and marginalizing over them. The astrophysical likelihood governing these parameters follows a Gaussian distribution centered on the observed value. The total likelihood is then 
\begin{equation}
\label{eq:AstroExpLikelihood}
\mathcal{L} = \mathcal{L}(\psi_{\rm astro}) \times \mathcal{L}_{\rm XENON1T}(\Theta,\psi_{\rm astro}) \times \mathcal{L}_{\rm IC86}(\Theta,\psi_{\rm astro})\,.
\end{equation}
The results of our analysis are illustrated in Fig.~\ref{fig:XeICva}, which shows the 68\% CL for marginalized astrophysics denoted by the medium thickness red line, for $A$ (top), $B$ (center) and $C$ (bottom). The primary effect of marginalizing over the astrophysical parameters is to broaden the contour regions, which can be clearly seen by comparing the red contours to their fixed astrophysics counterparts given by the thick light green line. 

There is an additional uncertainty related to the local DM density and to the different way this parameter enters into the IceCube and XENON1T rates. The direct detection rate depends on $\rho_\odot$ at present time at the Sun's position, while the solar capture rate depends on the long term history of the Sun, which completes an orbit around the Galactic center in $\sim 2 \times 10^8$ years. If we drop the assumption that the DM Galactic halo is isotropic and smooth, throughout its journey the Sun will cross over-dense or under-dense regions with respect to an averaged density ($\langle \rho_\odot \rangle$), to which the capture rate is sensitive. This difference has been evaluated for a triaxial DM halo arising from N-body simulations~\cite{Pato:2010yq}. In case of simulation with baryons the difference is of the order of $30\%$, while it can reach a factor of 3 for pure DM simulations. The MSSM25 models which serve as the starting points for our benchmark models have small annihilation cross-section, close or below the thermal one, hence the equilibration time scales $\tau$ (see Sec.~\ref{sec:theoic}) ranges from $9 \times 10^7$ year for Benchmark $B$, up to $2-3 \times 10^9$ year for points $A$ and $C$. These values imply that the capture process is fully sensitive to this averaged DM density (case $B$ as well, since it spanned half of the solar period). Even if we do not know the initial position of the Sun we could assume the extreme scenario such that the local DM density is different from the average density and corresponds to an over-dense or under-dense region. In this case we would get a 30\% or even a factor of 3 bias in the theoretical predictions. In other words the signal in one detector could be potentially boosted or suppressed by the same factor with respect to the signal in the other detector, as both rates scale linearly with the local DM density, affecting the complementarity in either a positive or negative way.

\subsection{Nuclear uncertainties}\label{sec:nucl}
\begin{figure*}[t]
\begin{minipage}[t]{0.32\textwidth}
\centering
\includegraphics[width=1.\columnwidth,trim=10mm 25mm 12mm 12mm, clip]{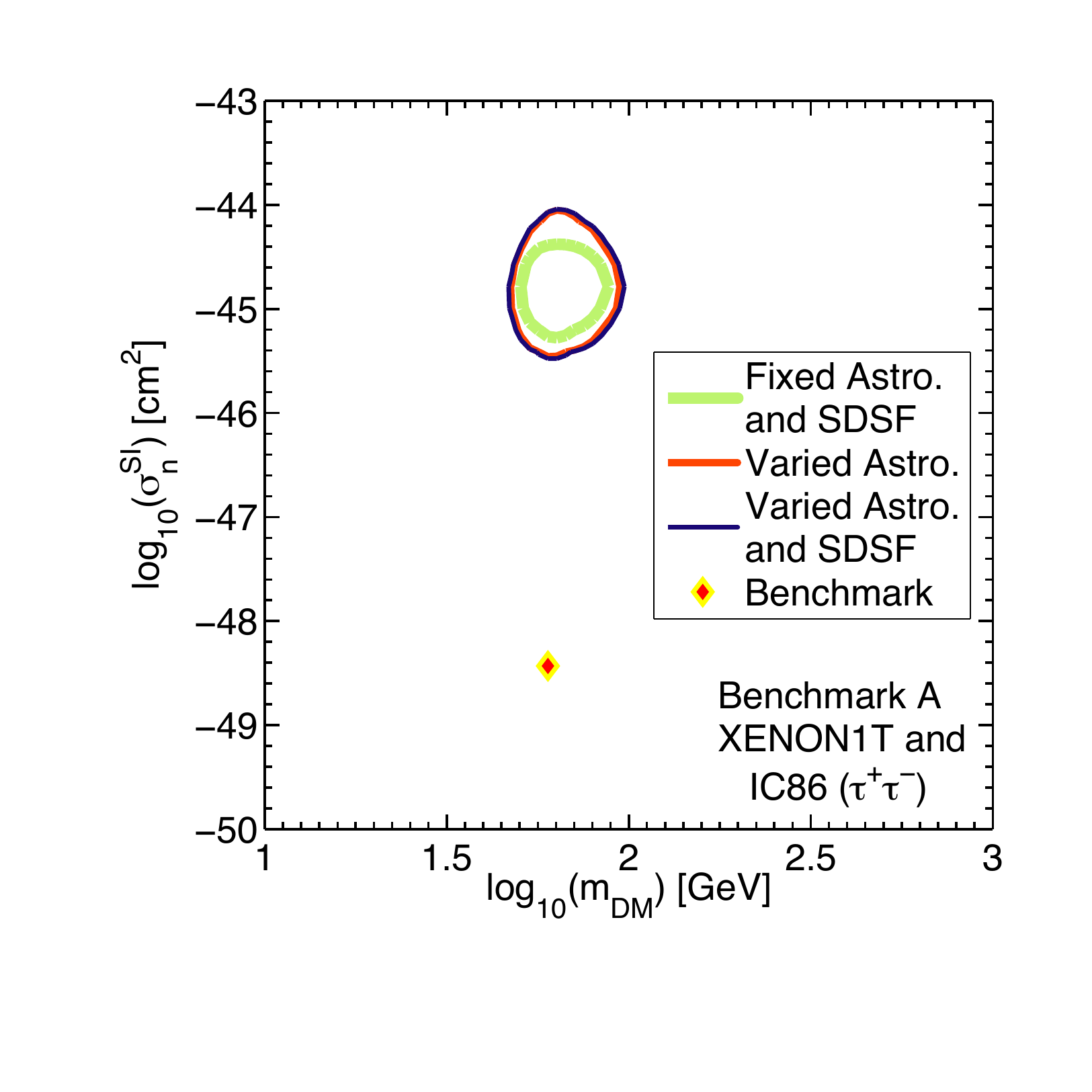}
\end{minipage}
\begin{minipage}[t]{0.32\textwidth}
\centering
\includegraphics[width=1.\columnwidth,trim=10mm 25mm 12mm 12mm, clip]{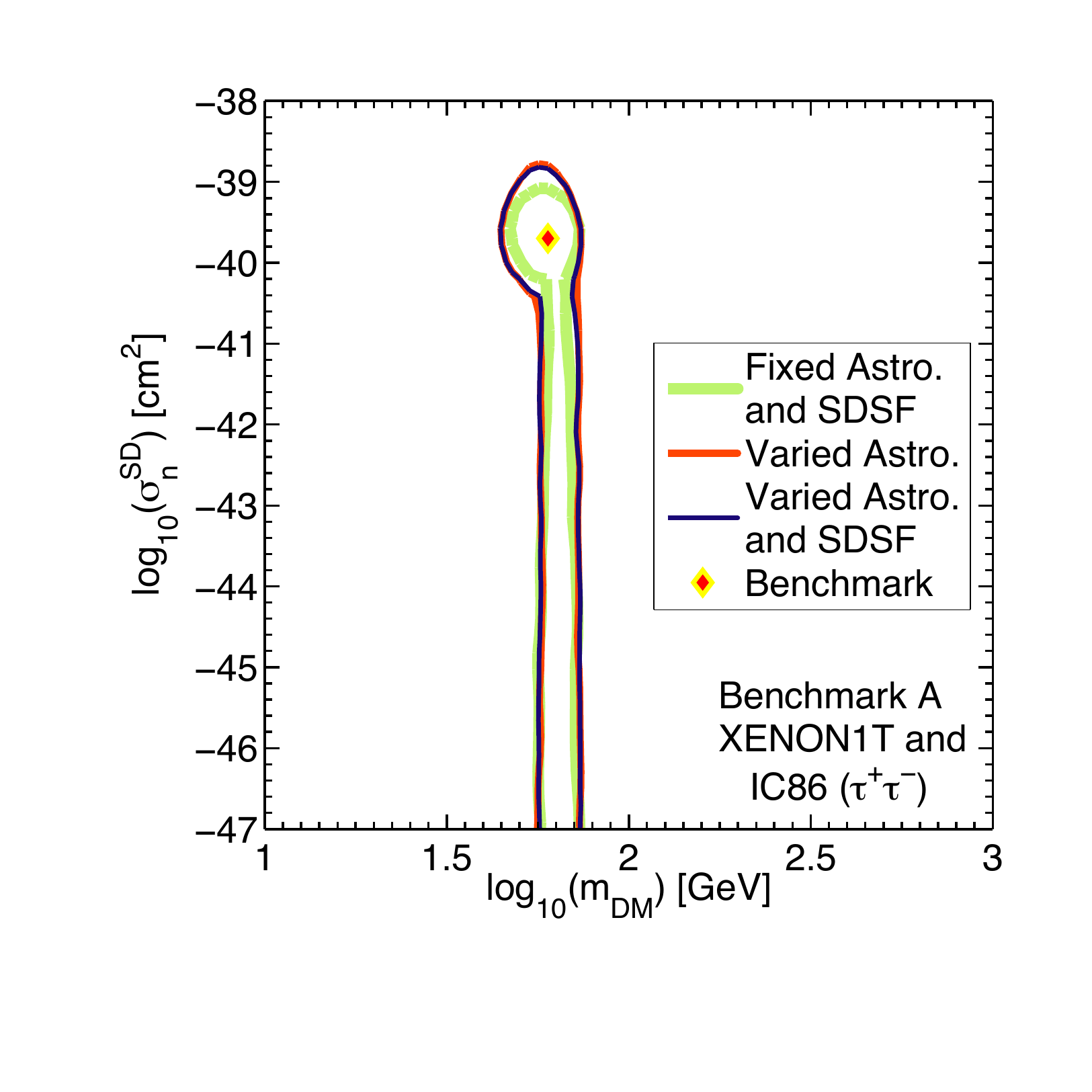}
\end{minipage}
\begin{minipage}[t]{0.32\textwidth}
\centering
\includegraphics[width=1.\columnwidth,trim=10mm 25mm 12mm 12mm, clip]{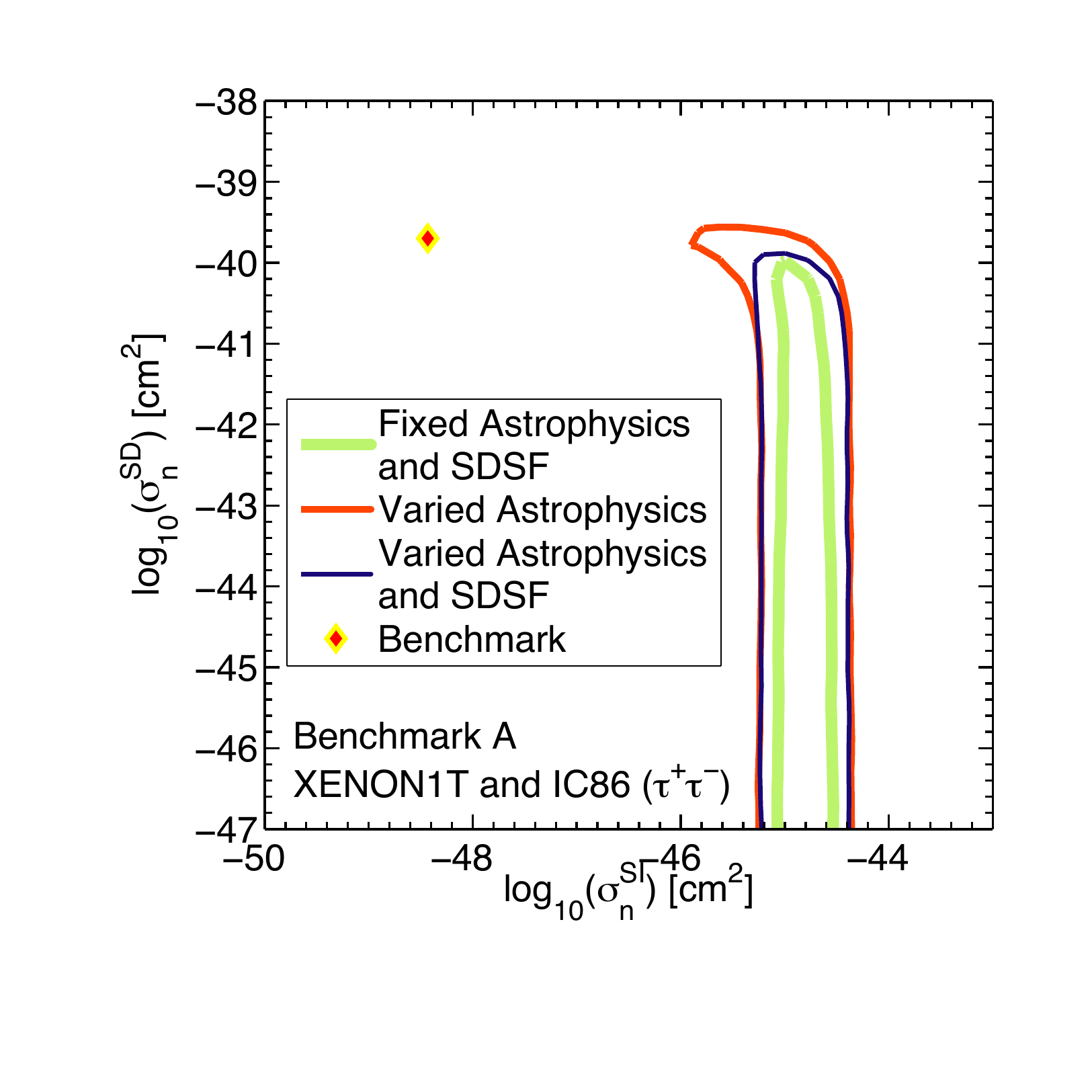}
\end{minipage}
\\
\begin{minipage}[t]{0.32\textwidth}
\centering
\includegraphics[width=1.\columnwidth,trim=10mm 25mm 12mm 12mm, clip]{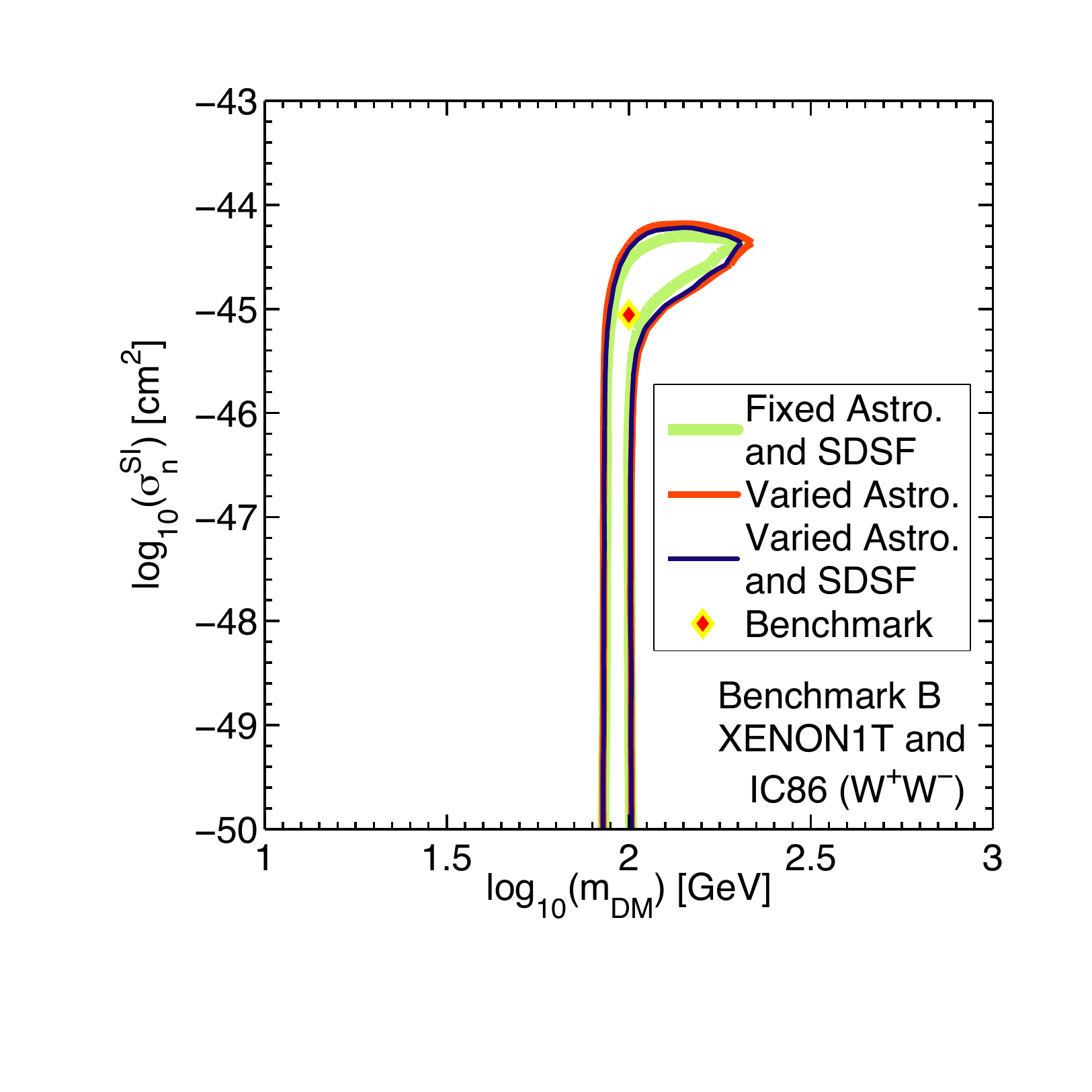}
\end{minipage}
\begin{minipage}[t]{0.32\textwidth}
\centering
\includegraphics[width=1.\columnwidth,trim=10mm 25mm 12mm 12mm, clip]{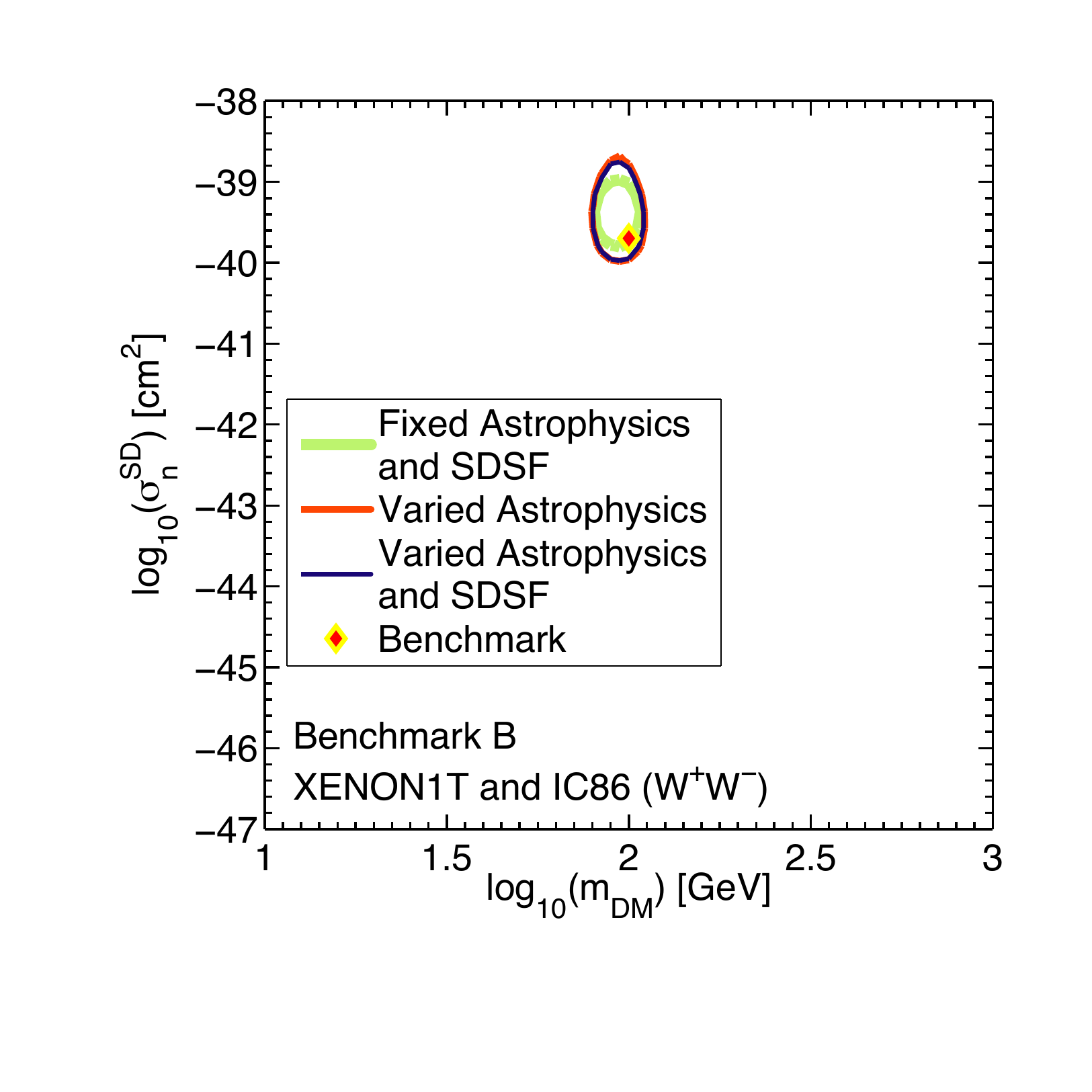}
\end{minipage}
\begin{minipage}[t]{0.32\textwidth}
\centering
\includegraphics[width=1.\columnwidth,trim=10mm 25mm 12mm 12mm, clip]{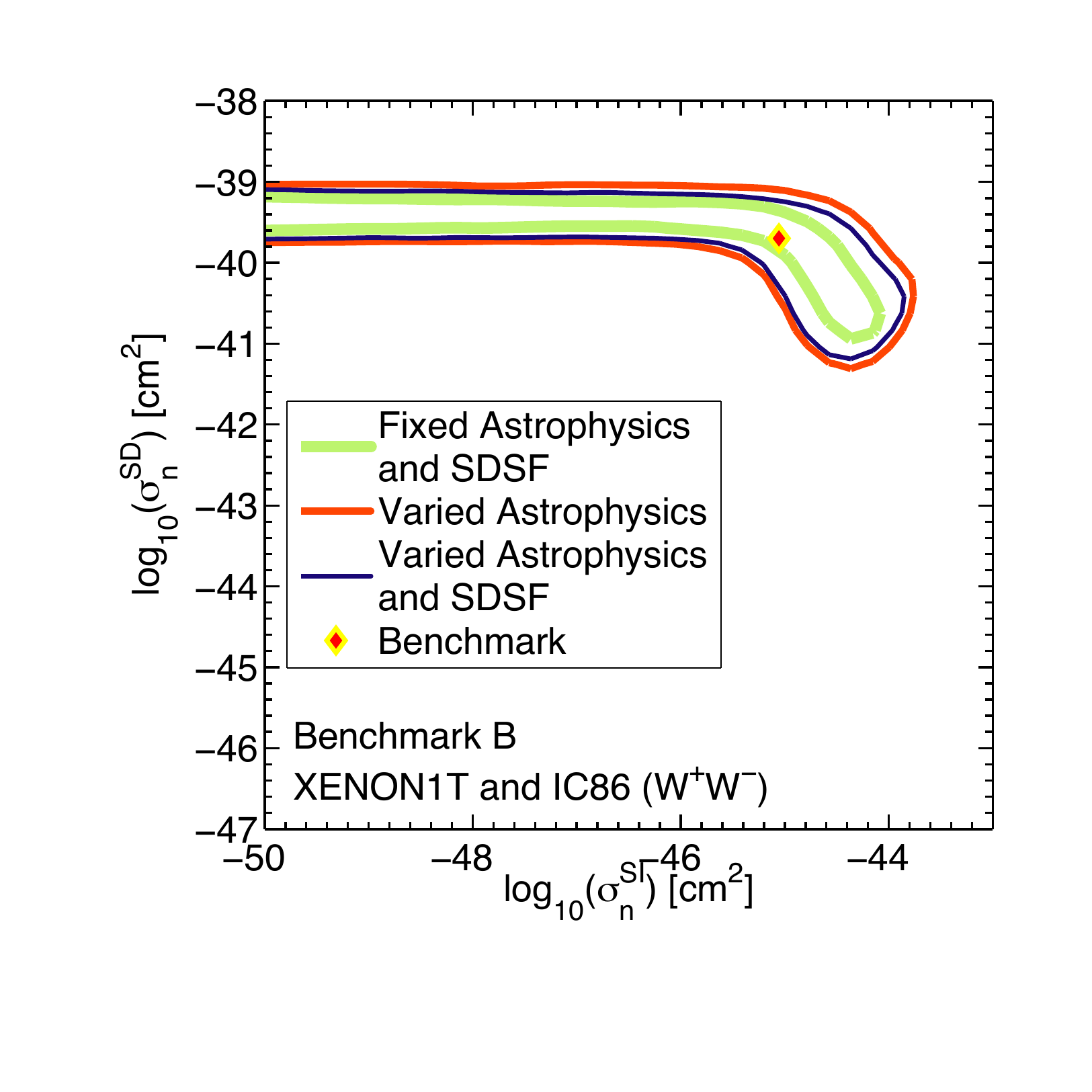}
\end{minipage}
\\
\begin{minipage}[t]{0.32\textwidth}
\centering
\includegraphics[width=1.\columnwidth,trim=10mm 25mm 12mm 12mm, clip]{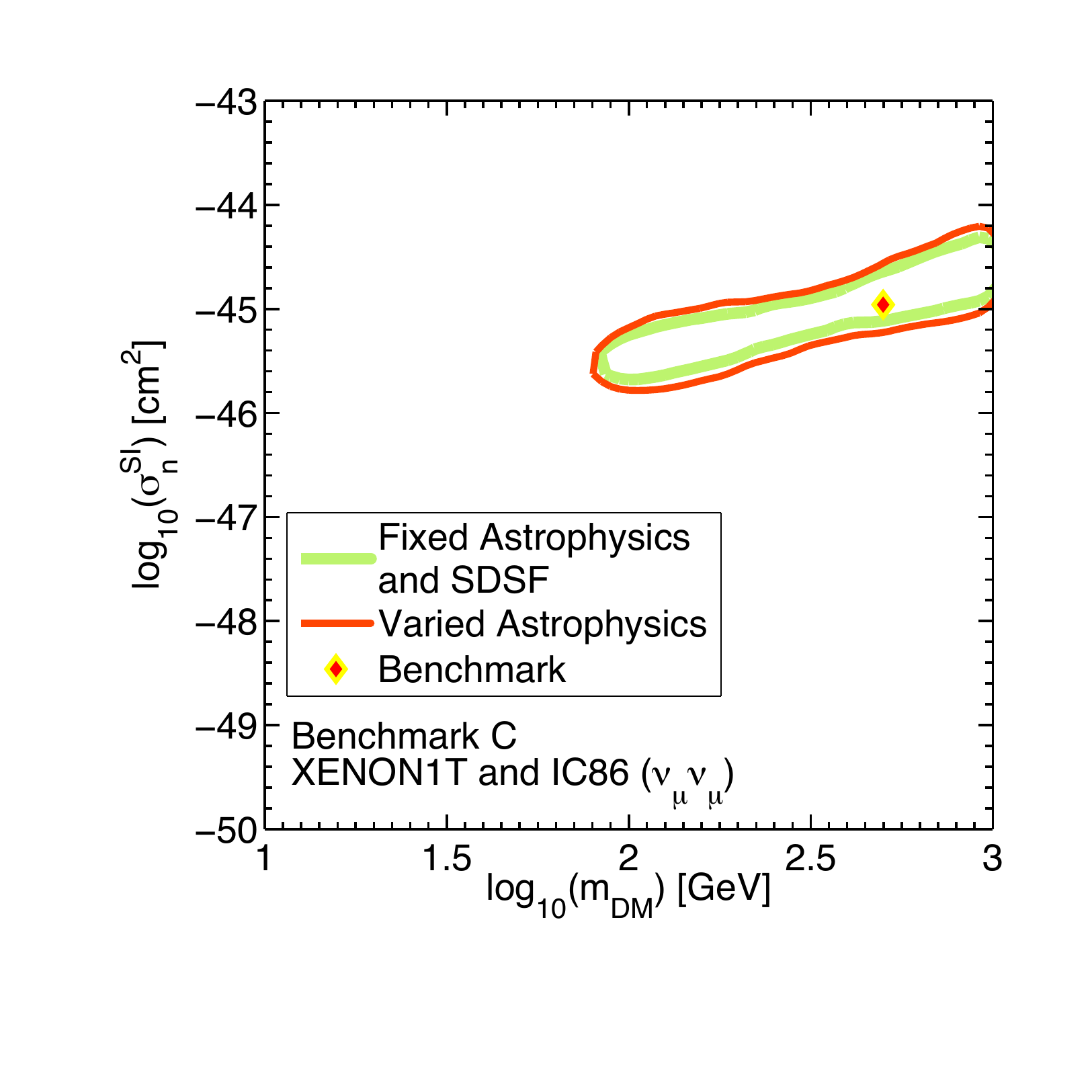}
\end{minipage}
\begin{minipage}[t]{0.32\textwidth}
\centering
\includegraphics[width=1.\columnwidth,trim=10mm 25mm 12mm 12mm, clip]{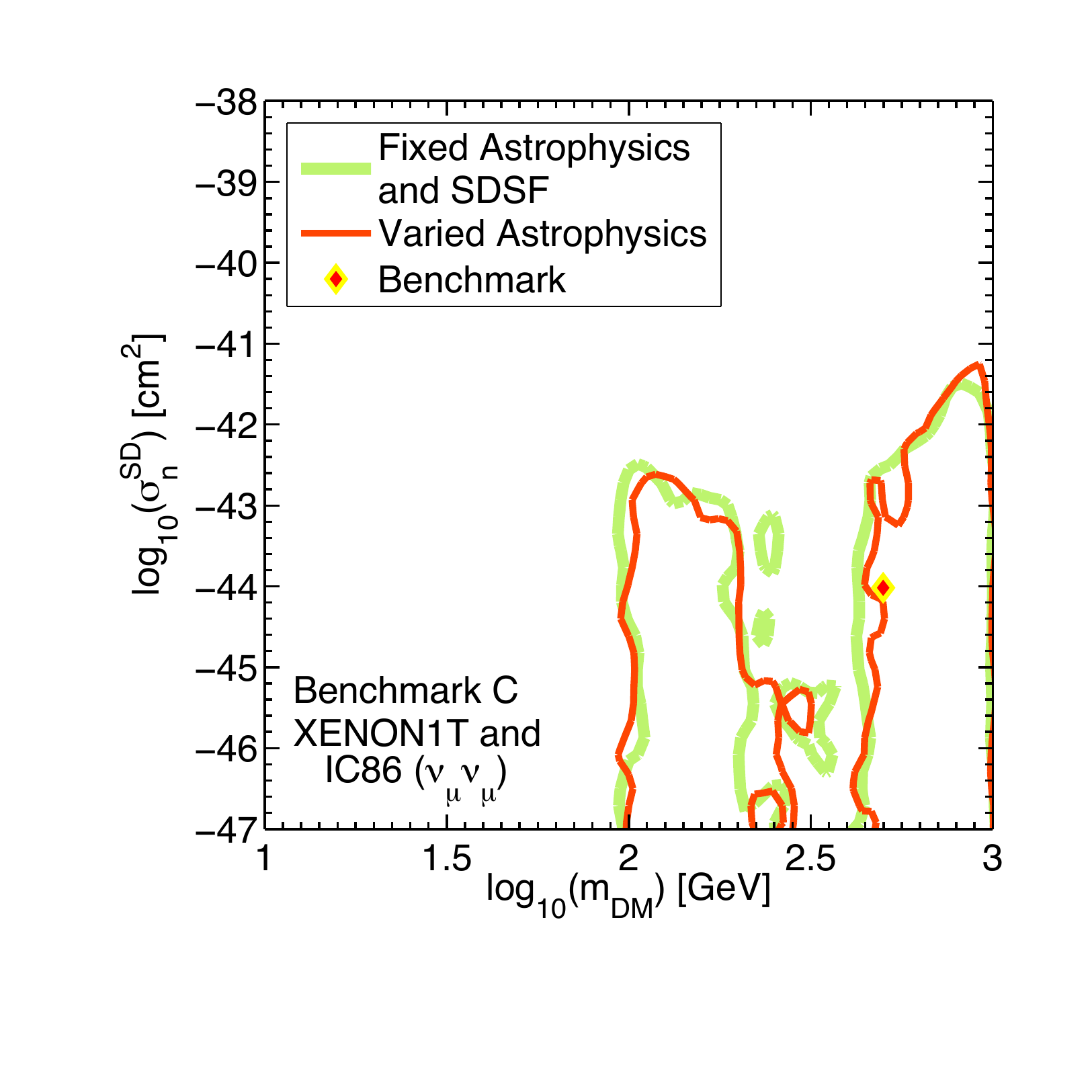}
\end{minipage}
\begin{minipage}[t]{0.32\textwidth}
\centering
\includegraphics[width=1.\columnwidth,trim=10mm 25mm 12mm 12mm, clip]{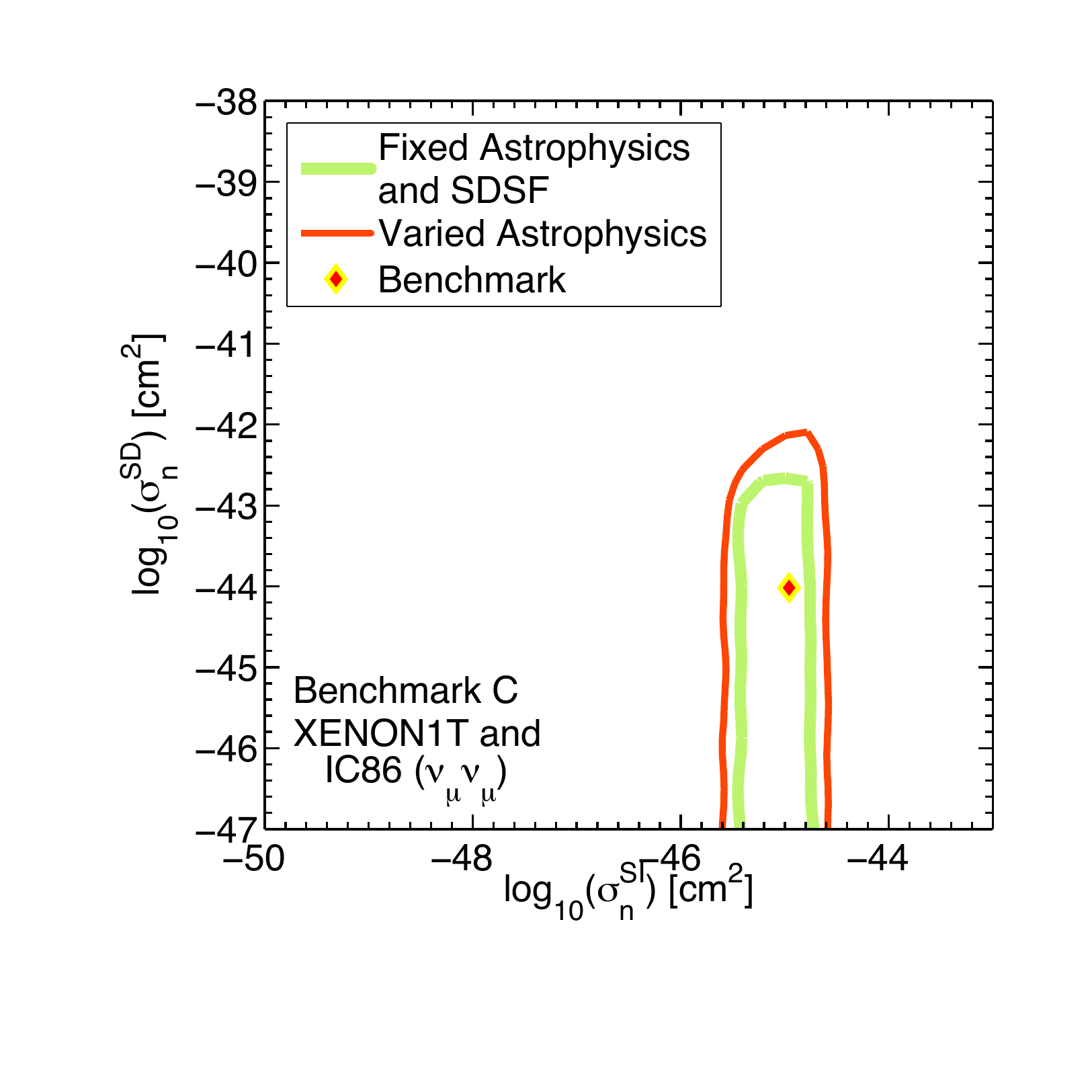}  
\end{minipage}
\caption{Reconstruction using XENON1T combined with IceCube (denoted IC86) for benchmark models $A$ with annihilation channel $\tau^+ \tau^-$ (top), $B$ with annihilation channel $W^+ W^-$ (middle), and $C$ with annihilation channel $\nu_\mu\bar{\nu}_\mu$ (bottom). The left, central and right plot shows the 2D marginal posterior pdf in the $\{m_{DM}, \sigma^{\rm SI}_n\}$-plane, $\{m_{\rm DM}, \sigma^{\rm SD}_n\}$-plane and $\{\sigma^{\rm SI}_n,\sigma^{\rm SD}_n\}$-plane respectively. The contours denote the 68\% credible region as labeled: the green line (light thick solid) stands for fixed astrophysics and fixed SD structure function (SDSF), the red curve (darker solid) denotes marginalized astrophysics, while the dark blue outer line stands for all nuisance marginalized over. The diamond point shows the parameter values of the benchmark model.}
\label{fig:XeICva}
\end{figure*}
The structure functions presented in Sec.~\ref{sec:theoxe} are important quantities to be computed. Regarding the xenon nuclei we are interested in, the variation of $S_{00}(q)$ leads to a different prediction of the events in the detector and hence, for example, to a huge uncertainty in the exclusion bound~\cite{:2013uj}. Considering our benchmark points, the difference in the counts for the two structure function is illustrated in Table \ref{tab:bench}. The NijmegenII structure function predicts fewer events in the detector with respect to the CEFT structure function: the variation is sizable for $A$ and $B$, where the events are affected by a factor of 3 of uncertainty while it is completely negligible for case $C$. The reconstruction in XENON1T of a benchmark model is therefore affected by the assumption of the structure function: if the reconstruction of $\sigma^{\rm SD}_n$ parameter is done assuming a different form factor with respect that used for the mock data a systematic offset is generated. If the reconstruction is done assuming CEFT and the mock data are generated with NijmegenII, this leads to an underestimation of the SD cross-section to account for the largest number of observed events; vice versa the SD cross-section will be overestimated. 

To account for the nuclear uncertainties in our Bayesian framework we follow~\cite{Cerdeno:2012ix}, which proposes a formula with three free parameters
\begin{equation}
S_{00}^{j}(u) = N_j \left( (1-\beta_j) e^{-\alpha_j u}+ \beta_j\right)\,,
\label{eq:fits00}
\end{equation}
where $j=129$ or $131$, depending on the xenon isotope, $N_j$ is a overall normalization factor, $\alpha_j$ drives the slope in the low momentum regime, while $\beta_j$ controls the height of a possible tail at large momentum. The variation of all parameters encompasses the whole region contained between the CEFT and NijmegenII parametrizations but it is not meant to be a proper fit of the nuclear structure functions. We take a flat prior for all these additional nuisance parameters $\psi_{\rm nucl}$, over their allowed range, which is: 
\begin{eqnarray}
& N_{129}  =   0.045 \to 0.070 \, \, \, {\rm and} \, \, \, N_{131} = 0.025 \to 0.052\,,\nonumber\\ 
& \alpha_{129}  =  3.8\to 4.0\, \, \, {\rm and} \, \, \, \alpha_{131} = 3.8. \to 4.0\,, \nonumber\\
& \beta_{129} = 0.013 \to 0.029\, \, \, {\rm and} \, \, \, \beta_{131} = 0.10 \to 0.12\,. \nonumber
\end{eqnarray}
We therefore generate for each benchmark model new mock data based on the mean value of these parameters and, as before, we assess the impact on the reconstruction of the benchmark points by letting them vary within their prior range, in combination with the astrophysical uncertainties. The new likelihood is given by
\begin{equation}
\label{eq:totExpLikelihood}
\mathcal{L}  =  \mathcal{L}(\psi_{\rm Tot}) \times \mathcal{L}_{\rm XENON1T}(\Theta,\psi_{\rm Tot}) \times \mathcal{L}_{\rm IC86}(\Theta,\psi_{\rm astro})\,,
\end{equation}
where $\psi_{\rm Tot}$ accounts for both astrophysical and nuclear uncertainties. Notice that the latter do not affect the reconstruction in IceCube: in solar capture the dominant contribution for SD arises from the hydrogen atom, which does not need a coherence factor for describing its single proton. Thus IceCube potentially has the capability of resolving the nuclear uncertainties when combined with XENON1T. 

In Fig.~\ref{fig:XeICva} the results for the 2D posterior pdf marginalized over all nuisance parameters are denoted by the dark blue solid line at 68\% CL. This contour is not shown for Benchmark $C$ as the nuclear uncertainties are not considered because of the negligible $\sigma^{\rm SD}_n$ cross-section. The main result is that the nuclear uncertainties are smaller or at most comparable with the astrophysical ones. This can be explained by the fact that we are considering the $S_{00}(q)$ nuclear structure function, which is known to be the better determined nuclear structure function, hence minimizing the role of nuclear uncertainties. Our findings are compatible with the analysis illustrated in~\cite{Cerdeno:2012ix,Cerdeno:2013gqa}.

\begin{figure*}[t]
\begin{minipage}[t]{0.32\textwidth}
\centering
\includegraphics[width=1.\columnwidth,trim=10mm 25mm 12mm 12mm, clip]{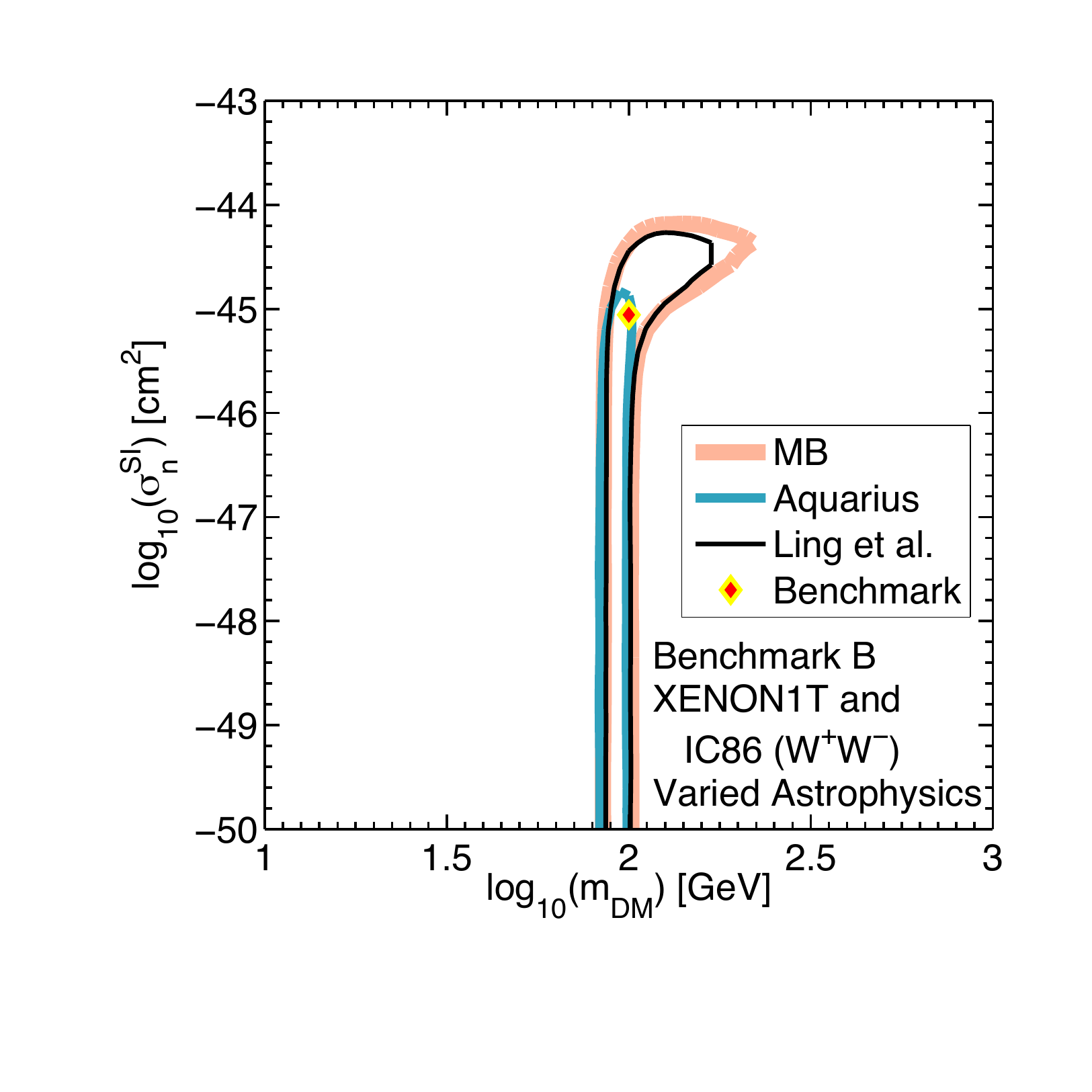}
\end{minipage}
\begin{minipage}[t]{0.32\textwidth}
\centering
\includegraphics[width=1.\columnwidth,trim=10mm 25mm 12mm 12mm, clip]{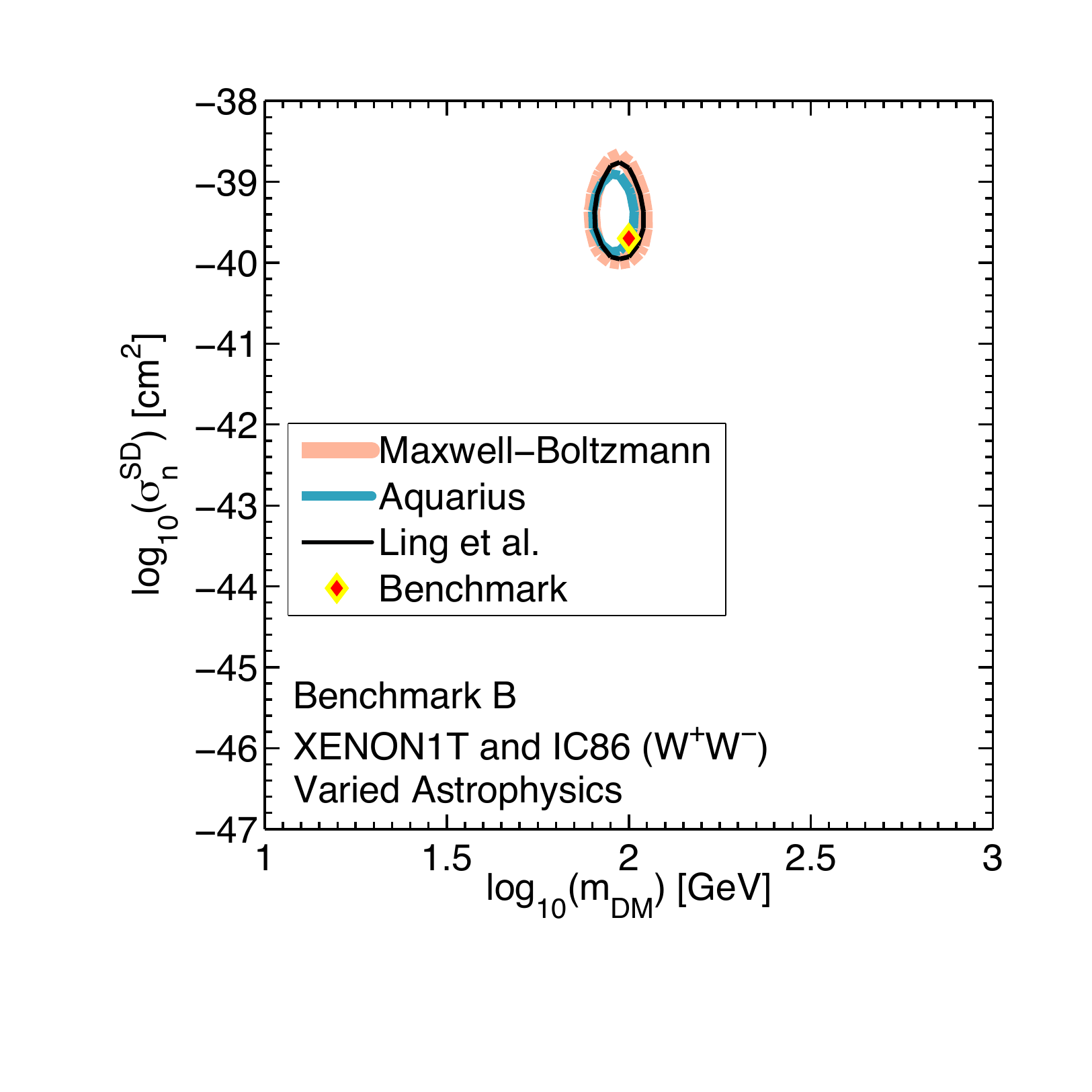}
\end{minipage}
\begin{minipage}[t]{0.32\textwidth}
\centering
\includegraphics[width=1.\columnwidth,trim=10mm 25mm 12mm 12mm, clip]{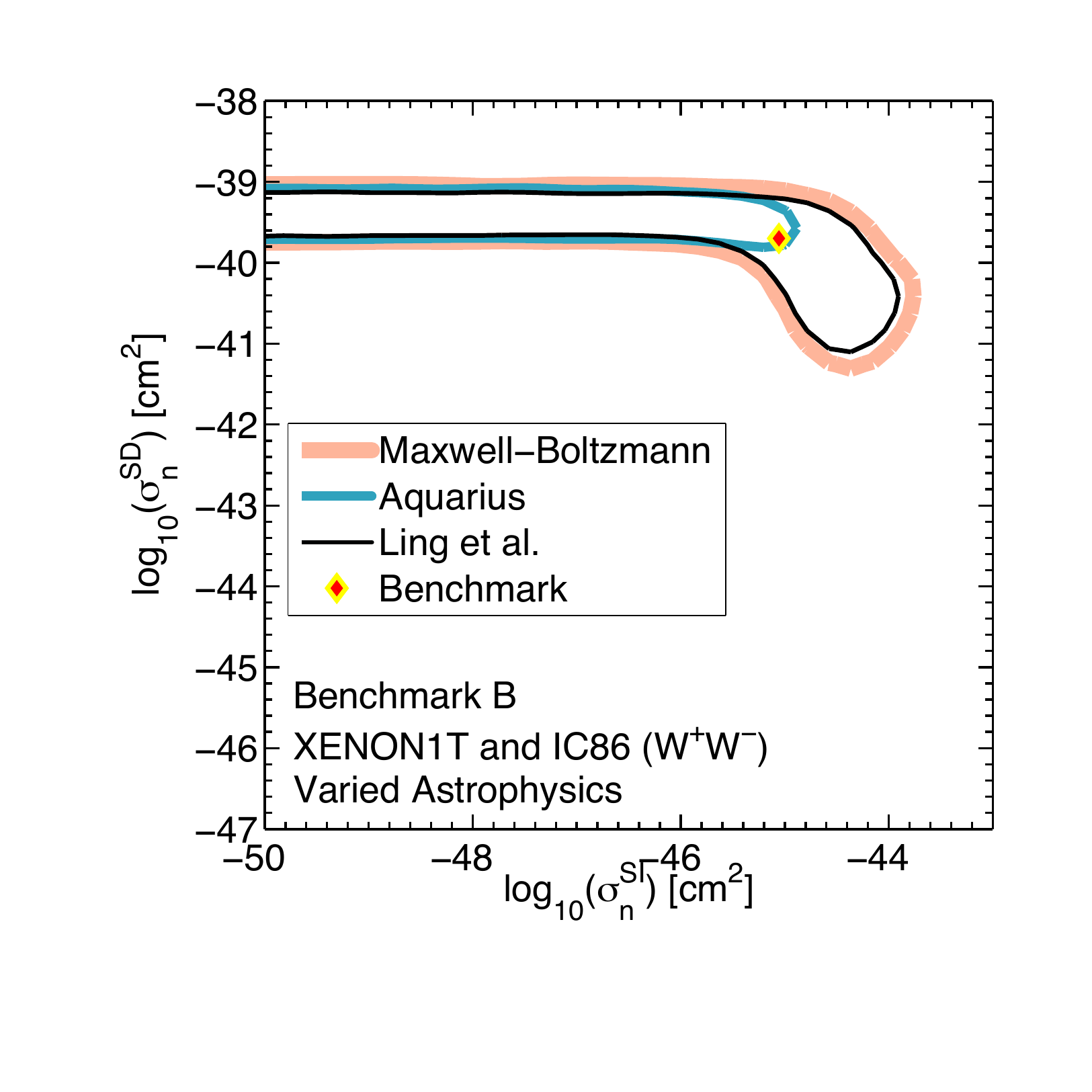}
\end{minipage}
\caption{Bias due to parametrization of DM velocity distribution for Benchmark $B$ with varied astrophysics. Each contour denotes the 68\% CL for the benchmark point generated with an N-body simulation or with the MB distribution, as labeled, and reconstructed assuming a standard halo model.}
\label{fig:Nbody}
\end{figure*}
\subsection{Effect of the shape of the velocity distribution}\label{sec:results}
The shape of the DM velocity distribution affects the direct detection and IceCube signal rates in different ways.  The nuclear recoil threshold for DM detection in XENON1T is in the range of 10 keV, which implies that only particles with the largest velocities have sufficient energy to produce a detectable nuclear recoil above the threshold, as given by the $v_{\rm min}$ definition in Eq.~(\ref{eq:vmin}). For a 100 GeV WIMP, as we have for Benchmark $B$, the typical $v_{\rm min}$ to produce a nuclear recoil is of the order of $v_{\rm min}\sim 300\, {\rm km \ s^{-1}}$. This means that it is crucial to accurately describe the high velocity tail of the DM velocity distribution in the Galactic halo. On the other hand, as explained in Sec.~\ref{sec:theoic}, the solar capture rate is sensitive to the low velocity tail of the DM velocity distribution, with the maximum velocity of a WIMP that can be captured being given by Eq.~(\ref{eq:uMax}). For a 100 GeV WIMP the maximum velocity is of the order of $u_{\rm max}\sim 200\,  {\rm km \ s ^{-1}}$. Thus it is important to describe the low tail of the velocity distribution as well. This is another physical quantity, besides the DM density in our galaxy, biasing the direct detection and indirect detection rate in different ways.

Several N-body simulations of Milky Way-like galaxies suggest that the velocity distribution deviates from the standard MB halo. To deal with the discrepancy in the high velocity tail for direct detection experiments Ref.~\cite{Lisanti:2010qx} has proposed a velocity distribution modeled as a function of one parameter $k$ and based on the these simulations.  

To assess if the shape of the velocity distribution is relevant for our analysis, we follow a different approach, which allows us to consider the effect of both tails of the velocity distribution. We consider the Aquarius N-body simulation~\cite{Vogelsberger:2008qb} with only DM, and the simulation by Ling \textit{et al.}~\cite{Ling:2009eh}, based on the RAMSES code, which includes baryons as well. 

First, in order to quantify the degree of discrepancy, we fit the mean of the N-body simulation Aquarius with a MB distribution, keeping the escape velocity fixed to $565\, {\rm km \ s^{-1}}$ as given by the simulation but letting $v_0$ vary: the best fit point underestimates both tails of the velocity distribution. For Benchmark $B$ the difference in the number of events is of the order of 17\% in XENON1T and of 15\% in IceCube. By allowing a similar procedure with $v_{\rm esc} = 520 \, {\rm km \ s^{-1}}$ in the case of Ling \textit{et al.} the velocity distribution is much closer to the MB shape, and the discrepancy in the number of predicted events is only of $~4\%$. 

Subsequently, the mean of these numerical velocity distributions is used to generate new mock data for our benchmarks in the case of XENON1T and rescaled for IceCube. These new benchmark models are then reconstructed assuming as usual a MB velocity distribution. We applied the procedure for Benchmark $B$, which is the point that takes full advantage of the combination of XENON1T and IceCube.

Figure~\ref{fig:Nbody} shows that our analysis is robust against variation of the shape of the velocity distribution, when both experiments are combined. From left to right the 2D marginalized posterior pdf for Benchmark $B$ is shown, in the $\{m_{\rm DM}, \sigma^{\rm SI}_n\}$-plane, $\{m_{\rm DM}, \sigma^{\rm SD}_n\}$-plane and $\{\sigma^{\rm SI}_n, \sigma^{\rm SD}_n\}$-plane. The middle thickness blue curve is the 68\% CL for a reconstruction of the DM physical parameters assuming that the true velocity distribution is given by the Aquarius simulation: the discrepancy with the MB curve (thick light orange) is very limited and affects mostly the SI cross-section with a more stringent upper bound, closer to the true value. The discrepancy with the Ling \textit{et al.} distribution is very small, as the 68\% contour level (thin black line) follows closely the MB contour. In other words the shape of the DM velocity distribution will likely have a limited impact in the reconstruction of WIMP physical parameters, in case of a positive detection.

\section{Conclusions}\label{sec:concl}

We have discussed the reconstruction capabilities for the physical WIMP parameters with a combination of data from XENON1T and IceCube. 
We have focused our attention on the degeneracy between the reconstruction of the SD and SI contributions, and we have shown the degeneracy between these two contributions arising from the analysis of the two experiments separately can be lifted from combining the two data sets. 

To illustrate and quantify the degree of complementarity, we have focused on three benchmark models: Benchmark $A$, characterized by a light DM mass and large SD contribution; Benchmark $B$ has intermediate mass and both sizable SI and SD cross-sections, and Benchmark $C$, with a large mass and dominant SI contribution.

The combined analysis for Benchmark $A$ allows us to illustrate the fact that even in cases where the parameters are such that no signal can be observed with IceCube, taking this fact into account in the global likelihood does improve the reconstruction of the DM parameters, as can be seen from the top row of Fig.~\ref{fig:Xeonly}.  The reconstruction of $\sigma^{\rm SI}_n$ and $\sigma^{\rm SD}_n$ in this case exhibits the same trend as the case of XENON1T only, except that now the 68\% CL for SD cross-section in the right panel is upper bounded, because of  the IceCube upper bound. The mass determination remains unchanged, demonstrating the good reconstruction capability of XENON1T and the reduced sensitivity of IceCube.

Benchmark model $B$ (second row of Fig.~\ref{fig:XeICfa}) fully exploits the properties of complementarity between DM search strategies and is the principal illustration of the main point of our analysis, as it can be detected by both experiments. Comparing with Figs.~\ref{fig:Xeonly} and~\ref{fig:ICfa} we can see that the combination of the two experiments allows for the SD cross-section (middle panel) to be well determined at the 68\% CL. The mass degeneracy is also significantly reduced and it is well determined at the 68\% CL. The contours for the SI cross-section as a function of $m_{\rm DM}$ have been contracted sensibly with respect to the separate experiments, but at $68\%$ CL the SI scattering cross-section is only upper bounded, as can be seen as well from the third panel: the SD is determined within one order of magnitude at most and contains the true value, while the SI cannot be larger than $10^{-44}{\rm cm^2}$. Therefore if a WIMP is detected by both experiments, meaning that it has sizable cross-sections, the SD cross-section and the mass can be well reconstructed, while a precise reconstruction of SI interaction would require the use of a further experiment; in this case a direct detection experiment with different target material would be the optimal choice~\cite{Cerdeno:2013gqa}.

Benchmark $C$, finally, shows remarkably the constraining power of IceCube, even with only 8 expected events from high energy neutrinos. The most dramatic improvement is in the determination of $\sigma^{\rm SI}_n$: where previously the 68\% and 95\% CLs extended from the upper bound all the way down to the bottom of our prior range, they are now bounded from below as well. As expected, only an upper limit can be set on $\sigma^{\rm SD}_n$ : the 95\% CL is clearly bounded from above at values of $\sigma^{\rm SD}_n \approx 10^{-42} {\rm cm^2}$. The effect of the SD upper bound is much more striking than for Benchmark $A$, signaling the fact the IceCube is more likely to be sensitive to high WIMP masses. 
In the absence of a publicly available energy response function for the 86-string configuration of IceCube our analysis could not utilize the energy information of the neutrinos observed by IceCube, limiting its mass determination capability. Including this spectral information would certainly improve the mass determination, one of the main limitations of the XENON1T-only reconstruction, and thus substantially reduce the degeneracy in the DM particle parameter space.

The inclusion of uncertainties on astrophysical parameters and nuclear structure functions only slightly affects the quality of the reconstruction of the 68\% CLs, the degradation being worst in the case of less constrained scenarios, such as Benchmark $C$, while $B$ appears to be fairly robust against variation of astrophysics and nuclear physics parameters. In other words the combination of XENON1T and IceCube in the case of a positive signal could also be a way of constraining the DM velocity distribution in the Galactic halo, once all the subtle effect of the astrophysical parameters entering into the capture and direct detection rates are consistently taken into account.

The strategy of combining direct and indirect searches for WIMPs allows for a better reconstruction of WIMP parameters, but the identification of the DM candidate will remain a difficult task. To understand the nature of DM it is necessary to complement the direct or indirect search strategies with LHC searches, as discussed {\it e.g.} in ~\cite{Bertone:2010rv,Bertone:2011pq}.

\begin{acknowledgments}
We gratefully acknowledge D. Cerde\~{n}o, M. Peir\'{o} and P. Scott  for interesting discussions and M. Lisanti and L. Strigari for providing us the data from N-body simulation based on $\Lambda$CDM cosmology. We acknowledge the support of the European Research Council through the ERC starting grant {\it WIMPs Kairos}, P.I. G. Bertone.
\end{acknowledgments}

\bibliographystyle{apsrev4-1}
\bibliography{biblio_chiara}

\begin{thebibliography}{81}%
\makeatletter
\providecommand \@ifxundefined [1]{%
 \@ifx{#1\undefined}
}%
\providecommand \@ifnum [1]{%
 \ifnum #1\expandafter \@firstoftwo
 \else \expandafter \@secondoftwo
 \fi
}%
\providecommand \@ifx [1]{%
 \ifx #1\expandafter \@firstoftwo
 \else \expandafter \@secondoftwo
 \fi
}%
\providecommand \natexlab [1]{#1}%
\providecommand \enquote  [1]{``#1''}%
\providecommand \bibnamefont  [1]{#1}%
\providecommand \bibfnamefont [1]{#1}%
\providecommand \citenamefont [1]{#1}%
\providecommand \href@noop [0]{\@secondoftwo}%
\providecommand \href [0]{\begingroup \@sanitize@url \@href}%
\providecommand \@href[1]{\@@startlink{#1}\@@href}%
\providecommand \@@href[1]{\endgroup#1\@@endlink}%
\providecommand \@sanitize@url [0]{\catcode `\\12\catcode `\$12\catcode
  `\&12\catcode `\#12\catcode `\^12\catcode `\_12\catcode `\%12\relax}%
\providecommand \@@startlink[1]{}%
\providecommand \@@endlink[0]{}%
\providecommand \url  [0]{\begingroup\@sanitize@url \@url }%
\providecommand \@url [1]{\endgroup\@href {#1}{\urlprefix }}%
\providecommand \urlprefix  [0]{URL }%
\providecommand \Eprint [0]{\href }%
\providecommand \doibase [0]{http://dx.doi.org/}%
\providecommand \selectlanguage [0]{\@gobble}%
\providecommand \bibinfo  [0]{\@secondoftwo}%
\providecommand \bibfield  [0]{\@secondoftwo}%
\providecommand \translation [1]{[#1]}%
\providecommand \BibitemOpen [0]{}%
\providecommand \bibitemStop [0]{}%
\providecommand \bibitemNoStop [0]{.\EOS\space}%
\providecommand \EOS [0]{\spacefactor3000\relax}%
\providecommand \BibitemShut  [1]{\csname bibitem#1\endcsname}%
\let\auto@bib@innerbib\@empty
\bibitem [{Ber(2010)}]{BertoneBook}%
  \BibitemOpen
  \href@noop {} {\emph {\bibinfo {title} {{Particle Dark Matter: Observations,
  Models and Searches}}}}\ (\bibinfo  {publisher} {Cambridge, UK: Cambridge
  University Press},\ \bibinfo {year} {2010})\ \bibinfo {note} {edited by
  Gianfranco Bertone}\BibitemShut {NoStop}%
\bibitem [{\citenamefont {Jungman}\ \emph {et~al.}(1996)\citenamefont
  {Jungman}, \citenamefont {Kamionkowski},\ and\ \citenamefont
  {Griest}}]{Jungman}%
  \BibitemOpen
  \bibfield  {author} {\bibinfo {author} {\bibfnamefont {G.}~\bibnamefont
  {Jungman}}, \bibinfo {author} {\bibfnamefont {M.}~\bibnamefont
  {Kamionkowski}}, \ and\ \bibinfo {author} {\bibfnamefont {K.}~\bibnamefont
  {Griest}},\ }\href {\doibase 10.1016/0370-1573(95)00058-5} {\bibfield
  {journal} {\bibinfo  {journal} {Phys. Rept.}\ }\textbf {\bibinfo {volume}
  {267}},\ \bibinfo {pages} {195} (\bibinfo {year} {1996})},\ \Eprint
  {http://arxiv.org/abs/hep-ph/9506380} {arXiv:hep-ph/9506380} \BibitemShut
  {NoStop}%
\bibitem [{\citenamefont {Munoz}(2004)}]{Munoz:2003gx}%
  \BibitemOpen
  \bibfield  {author} {\bibinfo {author} {\bibfnamefont {C.}~\bibnamefont
  {Munoz}},\ }\href {\doibase 10.1142/S0217751X04018154} {\bibfield  {journal}
  {\bibinfo  {journal} {Int. J. Mod. Phys.}\ }\textbf {\bibinfo {volume}
  {A19}},\ \bibinfo {pages} {3093} (\bibinfo {year} {2004})},\ \Eprint
  {http://arxiv.org/abs/hep-ph/0309346} {arXiv:hep-ph/0309346} \BibitemShut
  {NoStop}%
\bibitem [{\citenamefont {Bergstr{\"o}m}(2000)}]{Bergstrom}%
  \BibitemOpen
  \bibfield  {author} {\bibinfo {author} {\bibfnamefont {L.}~\bibnamefont
  {Bergstr{\"o}m}},\ }\href {\doibase 10.1088/0034-4885/63/5/2r3} {\bibfield
  {journal} {\bibinfo  {journal} {Rept. Prog. Phys.}\ }\textbf {\bibinfo
  {volume} {63}},\ \bibinfo {pages} {793} (\bibinfo {year} {2000})},\ \Eprint
  {http://arxiv.org/abs/hep-ph/0002126} {arXiv:hep-ph/0002126} \BibitemShut
  {NoStop}%
\bibitem [{\citenamefont {Bertone}\ \emph {et~al.}(2005)\citenamefont
  {Bertone}, \citenamefont {Hooper},\ and\ \citenamefont
  {Silk}}]{Bertone:2004pz}%
  \BibitemOpen
  \bibfield  {author} {\bibinfo {author} {\bibfnamefont {G.}~\bibnamefont
  {Bertone}}, \bibinfo {author} {\bibfnamefont {D.}~\bibnamefont {Hooper}}, \
  and\ \bibinfo {author} {\bibfnamefont {J.}~\bibnamefont {Silk}},\ }\href
  {\doibase 10.1016/j.physrep.2004.08.031} {\bibfield  {journal} {\bibinfo
  {journal} {Phys. Rept.}\ }\textbf {\bibinfo {volume} {405}},\ \bibinfo
  {pages} {279} (\bibinfo {year} {2005})},\ \Eprint
  {http://arxiv.org/abs/hep-ph/0404175} {arXiv:hep-ph/0404175} \BibitemShut
  {NoStop}%
\bibitem [{\citenamefont {Goodman}\ and\ \citenamefont
  {Witten}(1985)}]{Goodman:1984dc}%
  \BibitemOpen
  \bibfield  {author} {\bibinfo {author} {\bibfnamefont {M.~W.}\ \bibnamefont
  {Goodman}}\ and\ \bibinfo {author} {\bibfnamefont {E.}~\bibnamefont
  {Witten}},\ }\href {\doibase 10.1103/PhysRevD.31.3059} {\bibfield  {journal}
  {\bibinfo  {journal} {Phys. Rev.}\ }\textbf {\bibinfo {volume} {D31}},\
  \bibinfo {pages} {3059} (\bibinfo {year} {1985})}\BibitemShut {NoStop}%
\bibitem [{\citenamefont {Cerdeno}\ and\ \citenamefont
  {Green}(2010)}]{CerdenoGreen}%
  \BibitemOpen
  \bibfield  {author} {\bibinfo {author} {\bibfnamefont {D.~G.}\ \bibnamefont
  {Cerdeno}}\ and\ \bibinfo {author} {\bibfnamefont {A.~M.}\ \bibnamefont
  {Green}},\ }\href@noop {} {\  (\bibinfo {year} {2010})},\ \Eprint
  {http://arxiv.org/abs/1002.1912} {arXiv:1002.1912 [astro-ph.CO]} \BibitemShut
  {NoStop}%
\bibitem [{\citenamefont {Agnese}\ \emph {et~al.}(2013)\citenamefont {Agnese}
  \emph {et~al.}}]{Agnese:2013mga}%
  \BibitemOpen
  \bibfield  {author} {\bibinfo {author} {\bibfnamefont {R.}~\bibnamefont
  {Agnese}} \emph {et~al.} (\bibinfo {collaboration} {CDMS Collaboration}),\
  }\href@noop {} {\  (\bibinfo {year} {2013})},\ \Eprint
  {http://arxiv.org/abs/1304.4279} {arXiv:1304.4279 [hep-ex]} \BibitemShut
  {NoStop}%
\bibitem [{\citenamefont {Angloher}\ \emph {et~al.}(2012)\citenamefont
  {Angloher}, \citenamefont {Bauer}, \citenamefont {Bavykina}, \citenamefont
  {Bento}, \citenamefont {Bucci} \emph {et~al.}}]{Angloher:2011uu}%
  \BibitemOpen
  \bibfield  {author} {\bibinfo {author} {\bibfnamefont {G.}~\bibnamefont
  {Angloher}}, \bibinfo {author} {\bibfnamefont {M.}~\bibnamefont {Bauer}},
  \bibinfo {author} {\bibfnamefont {I.}~\bibnamefont {Bavykina}}, \bibinfo
  {author} {\bibfnamefont {A.}~\bibnamefont {Bento}}, \bibinfo {author}
  {\bibfnamefont {C.}~\bibnamefont {Bucci}},  \emph {et~al.},\ }\href@noop {}
  {\bibfield  {journal} {\bibinfo  {journal} {Eur.Phys.J.}\ }\textbf {\bibinfo
  {volume} {C72}},\ \bibinfo {pages} {1971} (\bibinfo {year} {2012})},\ \Eprint
  {http://arxiv.org/abs/1109.0702} {arXiv:1109.0702 [astro-ph.CO]} \BibitemShut
  {NoStop}%
\bibitem [{\citenamefont {Aalseth}\ \emph {et~al.}(2011)\citenamefont
  {Aalseth}, \citenamefont {Barbeau}, \citenamefont {Colaresi}, \citenamefont
  {Collar}, \citenamefont {Diaz~Leon} \emph {et~al.}}]{Aalseth:2011wp}%
  \BibitemOpen
  \bibfield  {author} {\bibinfo {author} {\bibfnamefont {C.}~\bibnamefont
  {Aalseth}}, \bibinfo {author} {\bibfnamefont {P.}~\bibnamefont {Barbeau}},
  \bibinfo {author} {\bibfnamefont {J.}~\bibnamefont {Colaresi}}, \bibinfo
  {author} {\bibfnamefont {J.}~\bibnamefont {Collar}}, \bibinfo {author}
  {\bibfnamefont {J.}~\bibnamefont {Diaz~Leon}},  \emph {et~al.},\ }\href
  {\doibase 10.1103/PhysRevLett.107.141301} {\bibfield  {journal} {\bibinfo
  {journal} {Phys.Rev.Lett.}\ }\textbf {\bibinfo {volume} {107}},\ \bibinfo
  {pages} {141301} (\bibinfo {year} {2011})},\ \Eprint
  {http://arxiv.org/abs/1106.0650} {arXiv:1106.0650 [astro-ph.CO]} \BibitemShut
  {NoStop}%
\bibitem [{\citenamefont {Bernabei}\ \emph {et~al.}(2010)\citenamefont
  {Bernabei}, \citenamefont {Belli}, \citenamefont {Cappella}, \citenamefont
  {Cerulli}, \citenamefont {Dai} \emph {et~al.}}]{Bernabei:2010mq}%
  \BibitemOpen
  \bibfield  {author} {\bibinfo {author} {\bibfnamefont {R.}~\bibnamefont
  {Bernabei}}, \bibinfo {author} {\bibfnamefont {P.}~\bibnamefont {Belli}},
  \bibinfo {author} {\bibfnamefont {F.}~\bibnamefont {Cappella}}, \bibinfo
  {author} {\bibfnamefont {R.}~\bibnamefont {Cerulli}}, \bibinfo {author}
  {\bibfnamefont {C.}~\bibnamefont {Dai}},  \emph {et~al.},\ }\href {\doibase
  10.1140/epjc/s10052-010-1303-9} {\bibfield  {journal} {\bibinfo  {journal}
  {Eur.Phys.J.}\ }\textbf {\bibinfo {volume} {C67}},\ \bibinfo {pages} {39}
  (\bibinfo {year} {2010})},\ \Eprint {http://arxiv.org/abs/1002.1028}
  {arXiv:1002.1028 [astro-ph.GA]} \BibitemShut {NoStop}%
\bibitem [{\citenamefont {Arina}(2012)}]{Arina:2012dr}%
  \BibitemOpen
  \bibfield  {author} {\bibinfo {author} {\bibfnamefont {C.}~\bibnamefont
  {Arina}},\ }\href {\doibase 10.1103/PhysRevD.86.123527} {\bibfield  {journal}
  {\bibinfo  {journal} {Phys.Rev.}\ }\textbf {\bibinfo {volume} {D86}},\
  \bibinfo {pages} {123527} (\bibinfo {year} {2012})},\ \Eprint
  {http://arxiv.org/abs/1210.4011} {arXiv:1210.4011 [hep-ph]} \BibitemShut
  {NoStop}%
\bibitem [{\citenamefont {Kopp}\ \emph {et~al.}(2012)\citenamefont {Kopp},
  \citenamefont {Schwetz},\ and\ \citenamefont {Zupan}}]{Kopp:2011yr}%
  \BibitemOpen
  \bibfield  {author} {\bibinfo {author} {\bibfnamefont {J.}~\bibnamefont
  {Kopp}}, \bibinfo {author} {\bibfnamefont {T.}~\bibnamefont {Schwetz}}, \
  and\ \bibinfo {author} {\bibfnamefont {J.}~\bibnamefont {Zupan}},\ }\href
  {\doibase 10.1088/1475-7516/2012/03/001} {\bibfield  {journal} {\bibinfo
  {journal} {JCAP}\ }\textbf {\bibinfo {volume} {1203}},\ \bibinfo {pages}
  {001} (\bibinfo {year} {2012})},\ \Eprint {http://arxiv.org/abs/1110.2721}
  {arXiv:1110.2721 [hep-ph]} \BibitemShut {NoStop}%
\bibitem [{\citenamefont {Aprile}\ \emph {et~al.}(2012)\citenamefont {Aprile}
  \emph {et~al.}}]{Aprile:2012nq}%
  \BibitemOpen
  \bibfield  {author} {\bibinfo {author} {\bibfnamefont {E.}~\bibnamefont
  {Aprile}} \emph {et~al.} (\bibinfo {collaboration} {XENON100
  Collaboration}),\ }\href {\doibase 10.1103/PhysRevLett.109.181301} {\bibfield
   {journal} {\bibinfo  {journal} {Phys.Rev.Lett.}\ }\textbf {\bibinfo {volume}
  {109}},\ \bibinfo {pages} {181301} (\bibinfo {year} {2012})},\ \Eprint
  {http://arxiv.org/abs/1207.5988} {1207.5988 [astro-ph.CO]} \BibitemShut
  {NoStop}%
\bibitem [{\citenamefont {Brown}\ \emph {et~al.}(2012)\citenamefont {Brown},
  \citenamefont {Henry}, \citenamefont {Kraus},\ and\ \citenamefont
  {McCabe}}]{Brown:2011dp}%
  \BibitemOpen
  \bibfield  {author} {\bibinfo {author} {\bibfnamefont {A.}~\bibnamefont
  {Brown}}, \bibinfo {author} {\bibfnamefont {S.}~\bibnamefont {Henry}},
  \bibinfo {author} {\bibfnamefont {H.}~\bibnamefont {Kraus}}, \ and\ \bibinfo
  {author} {\bibfnamefont {C.}~\bibnamefont {McCabe}},\ }\href {\doibase
  10.1103/PhysRevD.85.021301} {\bibfield  {journal} {\bibinfo  {journal}
  {Phys.Rev.}\ }\textbf {\bibinfo {volume} {D85}},\ \bibinfo {pages} {021301}
  (\bibinfo {year} {2012})},\ \Eprint {http://arxiv.org/abs/1109.2589}
  {arXiv:1109.2589 [astro-ph.CO]} \BibitemShut {NoStop}%
\bibitem [{\citenamefont {Ahmed}\ \emph {et~al.}(2011)\citenamefont {Ahmed}
  \emph {et~al.}}]{Ahmed:2010wy}%
  \BibitemOpen
  \bibfield  {author} {\bibinfo {author} {\bibfnamefont {Z.}~\bibnamefont
  {Ahmed}} \emph {et~al.} (\bibinfo {collaboration} {CDMS-II Collaboration}),\
  }\href {\doibase 10.1103/PhysRevLett.106.131302} {\bibfield  {journal}
  {\bibinfo  {journal} {Phys.Rev.Lett.}\ }\textbf {\bibinfo {volume} {106}},\
  \bibinfo {pages} {131302} (\bibinfo {year} {2011})},\ \Eprint
  {http://arxiv.org/abs/1011.2482} {arXiv:1011.2482 [astro-ph.CO]} \BibitemShut
  {NoStop}%
\bibitem [{\citenamefont {Ahmed}\ \emph {et~al.}(2012)\citenamefont {Ahmed}
  \emph {et~al.}}]{Ahmed:2012vq}%
  \BibitemOpen
  \bibfield  {author} {\bibinfo {author} {\bibfnamefont {Z.}~\bibnamefont
  {Ahmed}} \emph {et~al.} (\bibinfo {collaboration} {CDMS Collaboration}),\
  }\href@noop {} {\  (\bibinfo {year} {2012})},\ \Eprint
  {http://arxiv.org/abs/1203.1309} {arXiv:1203.1309 [astro-ph.CO]} \BibitemShut
  {NoStop}%
\bibitem [{\citenamefont {Ling}\ \emph {et~al.}(2010)\citenamefont {Ling},
  \citenamefont {Nezri}, \citenamefont {Athanassoula},\ and\ \citenamefont
  {Teyssier}}]{Ling:2009eh}%
  \BibitemOpen
  \bibfield  {author} {\bibinfo {author} {\bibfnamefont {F.}~\bibnamefont
  {Ling}}, \bibinfo {author} {\bibfnamefont {E.}~\bibnamefont {Nezri}},
  \bibinfo {author} {\bibfnamefont {E.}~\bibnamefont {Athanassoula}}, \ and\
  \bibinfo {author} {\bibfnamefont {R.}~\bibnamefont {Teyssier}},\ }\href
  {\doibase 10.1088/1475-7516/2010/02/012} {\bibfield  {journal} {\bibinfo
  {journal} {JCAP}\ }\textbf {\bibinfo {volume} {1002}},\ \bibinfo {pages}
  {012} (\bibinfo {year} {2010})},\ \Eprint {http://arxiv.org/abs/0909.2028}
  {arXiv:0909.2028 [astro-ph.GA]} \BibitemShut {NoStop}%
\bibitem [{\citenamefont {Kuhlen}\ \emph {et~al.}(2010)\citenamefont {Kuhlen},
  \citenamefont {Weiner}, \citenamefont {Diemand}, \citenamefont {Madau},
  \citenamefont {Moore}, \citenamefont {Potter}, \citenamefont {Stadel},\ and\
  \citenamefont {Zemp}}]{Kuhlen:2009vh}%
  \BibitemOpen
  \bibfield  {author} {\bibinfo {author} {\bibfnamefont {M.}~\bibnamefont
  {Kuhlen}}, \bibinfo {author} {\bibfnamefont {N.}~\bibnamefont {Weiner}},
  \bibinfo {author} {\bibfnamefont {J.}~\bibnamefont {Diemand}}, \bibinfo
  {author} {\bibfnamefont {P.}~\bibnamefont {Madau}}, \bibinfo {author}
  {\bibfnamefont {B.}~\bibnamefont {Moore}}, \bibinfo {author} {\bibfnamefont
  {D.}~\bibnamefont {Potter}}, \bibinfo {author} {\bibfnamefont
  {J.}~\bibnamefont {Stadel}}, \ and\ \bibinfo {author} {\bibfnamefont
  {M.}~\bibnamefont {Zemp}},\ }\href {\doibase 10.1088/1475-7516/2010/02/030}
  {\bibfield  {journal} {\bibinfo  {journal} {JCAP}\ }\textbf {\bibinfo
  {volume} {1002}},\ \bibinfo {pages} {030} (\bibinfo {year} {2010})},\ \Eprint
  {http://arxiv.org/abs/0912.2358} {arXiv:0912.2358 [astro-ph.GA]} \BibitemShut
  {NoStop}%
\bibitem [{\citenamefont {Vogelsberger}\ \emph {et~al.}(2009)\citenamefont
  {Vogelsberger}, \citenamefont {Helmi}, \citenamefont {Springel},
  \citenamefont {White}, \citenamefont {Wang}, \citenamefont {Frenk},
  \citenamefont {Jenkins}, \citenamefont {Ludlow},\ and\ \citenamefont
  {Navarro}}]{Vogelsberger:2008qb}%
  \BibitemOpen
  \bibfield  {author} {\bibinfo {author} {\bibfnamefont {M.}~\bibnamefont
  {Vogelsberger}}, \bibinfo {author} {\bibfnamefont {A.}~\bibnamefont {Helmi}},
  \bibinfo {author} {\bibfnamefont {V.}~\bibnamefont {Springel}}, \bibinfo
  {author} {\bibfnamefont {S.~D.}\ \bibnamefont {White}}, \bibinfo {author}
  {\bibfnamefont {J.}~\bibnamefont {Wang}}, \bibinfo {author} {\bibfnamefont
  {C.~S.}\ \bibnamefont {Frenk}}, \bibinfo {author} {\bibfnamefont
  {A.}~\bibnamefont {Jenkins}}, \bibinfo {author} {\bibfnamefont {A.~D.}\
  \bibnamefont {Ludlow}}, \ and\ \bibinfo {author} {\bibfnamefont {J.~F.}\
  \bibnamefont {Navarro}},\ }\href {\doibase 10.1111/j.1365-2966.2009.14630.x}
  {\bibfield  {journal} {\bibinfo  {journal} {Mon.Not.Roy.Astron.Soc.}\
  }\textbf {\bibinfo {volume} {395}},\ \bibinfo {pages} {797} (\bibinfo {year}
  {2009})},\ \Eprint {http://arxiv.org/abs/0812.0362} {arXiv:0812.0362
  [astro-ph]} \BibitemShut {NoStop}%
\bibitem [{\citenamefont {Green}(2010)}]{Green:2010gw}%
  \BibitemOpen
  \bibfield  {author} {\bibinfo {author} {\bibfnamefont {A.~M.}\ \bibnamefont
  {Green}},\ }\href {\doibase 10.1088/1475-7516/2010/10/034} {\bibfield
  {journal} {\bibinfo  {journal} {JCAP}\ }\textbf {\bibinfo {volume} {1010}},\
  \bibinfo {pages} {034} (\bibinfo {year} {2010})},\ \Eprint
  {http://arxiv.org/abs/1009.0916} {arXiv:1009.0916 [astro-ph.CO]} \BibitemShut
  {NoStop}%
\bibitem [{\citenamefont {McCabe}(2010)}]{McCabe:2010zh}%
  \BibitemOpen
  \bibfield  {author} {\bibinfo {author} {\bibfnamefont {C.}~\bibnamefont
  {McCabe}},\ }\href {\doibase 10.1103/PhysRevD.82.023530} {\bibfield
  {journal} {\bibinfo  {journal} {Phys.Rev.}\ }\textbf {\bibinfo {volume}
  {D82}},\ \bibinfo {pages} {023530} (\bibinfo {year} {2010})},\ \Eprint
  {http://arxiv.org/abs/1005.0579} {arXiv:1005.0579 [hep-ph]} \BibitemShut
  {NoStop}%
\bibitem [{\citenamefont {Pato}\ \emph {et~al.}(2013)\citenamefont {Pato},
  \citenamefont {Strigari}, \citenamefont {Trotta},\ and\ \citenamefont
  {Bertone}}]{Pato:2012fw}%
  \BibitemOpen
  \bibfield  {author} {\bibinfo {author} {\bibfnamefont {M.}~\bibnamefont
  {Pato}}, \bibinfo {author} {\bibfnamefont {L.~E.}\ \bibnamefont {Strigari}},
  \bibinfo {author} {\bibfnamefont {R.}~\bibnamefont {Trotta}}, \ and\ \bibinfo
  {author} {\bibfnamefont {G.}~\bibnamefont {Bertone}},\ }\href {\doibase
  10.1088/1475-7516/2013/02/041} {\bibfield  {journal} {\bibinfo  {journal}
  {JCAP}\ }\textbf {\bibinfo {volume} {1302}},\ \bibinfo {pages} {041}
  (\bibinfo {year} {2013})},\ \Eprint {http://arxiv.org/abs/1211.7063}
  {arXiv:1211.7063 [astro-ph.CO]} \BibitemShut {NoStop}%
\bibitem [{\citenamefont {Strigari}\ and\ \citenamefont
  {Trotta}(2009)}]{Strigari:2009zb}%
  \BibitemOpen
  \bibfield  {author} {\bibinfo {author} {\bibfnamefont {L.~E.}\ \bibnamefont
  {Strigari}}\ and\ \bibinfo {author} {\bibfnamefont {R.}~\bibnamefont
  {Trotta}},\ }\href {\doibase 10.1088/1475-7516/2009/11/019} {\bibfield
  {journal} {\bibinfo  {journal} {JCAP}\ }\textbf {\bibinfo {volume} {0911}},\
  \bibinfo {pages} {019} (\bibinfo {year} {2009})},\ \Eprint
  {http://arxiv.org/abs/0906.5361} {arXiv:0906.5361 [astro-ph.HE]} \BibitemShut
  {NoStop}%
\bibitem [{\citenamefont {Pato}\ \emph {et~al.}(2011)\citenamefont {Pato},
  \citenamefont {Baudis}, \citenamefont {Bertone}, \citenamefont {Ruiz~de
  Austri}, \citenamefont {Strigari},\ and\ \citenamefont
  {Trotta}}]{Pato:2010zk}%
  \BibitemOpen
  \bibfield  {author} {\bibinfo {author} {\bibfnamefont {M.}~\bibnamefont
  {Pato}}, \bibinfo {author} {\bibfnamefont {L.}~\bibnamefont {Baudis}},
  \bibinfo {author} {\bibfnamefont {G.}~\bibnamefont {Bertone}}, \bibinfo
  {author} {\bibfnamefont {R.}~\bibnamefont {Ruiz~de Austri}}, \bibinfo
  {author} {\bibfnamefont {L.~E.}\ \bibnamefont {Strigari}}, \ and\ \bibinfo
  {author} {\bibfnamefont {R.}~\bibnamefont {Trotta}},\ }\href {\doibase
  10.1103/PhysRevD.83.083505} {\bibfield  {journal} {\bibinfo  {journal}
  {Phys.Rev.}\ }\textbf {\bibinfo {volume} {D83}},\ \bibinfo {pages} {083505}
  (\bibinfo {year} {2011})},\ \Eprint {http://arxiv.org/abs/1012.3458}
  {arXiv:1012.3458 [astro-ph.CO]} \BibitemShut {NoStop}%
\bibitem [{\citenamefont {Fox}\ \emph {et~al.}(2011)\citenamefont {Fox},
  \citenamefont {Liu},\ and\ \citenamefont {Weiner}}]{Fox:2010bz}%
  \BibitemOpen
  \bibfield  {author} {\bibinfo {author} {\bibfnamefont {P.~J.}\ \bibnamefont
  {Fox}}, \bibinfo {author} {\bibfnamefont {J.}~\bibnamefont {Liu}}, \ and\
  \bibinfo {author} {\bibfnamefont {N.}~\bibnamefont {Weiner}},\ }\href
  {\doibase 10.1103/PhysRevD.83.103514} {\bibfield  {journal} {\bibinfo
  {journal} {Phys.Rev.}\ }\textbf {\bibinfo {volume} {D83}},\ \bibinfo {pages}
  {103514} (\bibinfo {year} {2011})},\ \Eprint {http://arxiv.org/abs/1011.1915}
  {arXiv:1011.1915 [hep-ph]} \BibitemShut {NoStop}%
\bibitem [{\citenamefont {Strege}\ \emph {et~al.}(2012)\citenamefont {Strege},
  \citenamefont {Trotta}, \citenamefont {Bertone}, \citenamefont {Peter},\ and\
  \citenamefont {Scott}}]{Strege:2012kv}%
  \BibitemOpen
  \bibfield  {author} {\bibinfo {author} {\bibfnamefont {C.}~\bibnamefont
  {Strege}}, \bibinfo {author} {\bibfnamefont {R.}~\bibnamefont {Trotta}},
  \bibinfo {author} {\bibfnamefont {G.}~\bibnamefont {Bertone}}, \bibinfo
  {author} {\bibfnamefont {A.~H.}\ \bibnamefont {Peter}}, \ and\ \bibinfo
  {author} {\bibfnamefont {P.}~\bibnamefont {Scott}},\ }\href {\doibase
  10.1103/PhysRevD.86.023507} {\bibfield  {journal} {\bibinfo  {journal}
  {Phys.Rev.}\ }\textbf {\bibinfo {volume} {D86}},\ \bibinfo {pages} {023507}
  (\bibinfo {year} {2012})},\ \Eprint {http://arxiv.org/abs/1201.3631}
  {arXiv:1201.3631 [hep-ph]} \BibitemShut {NoStop}%
\bibitem [{\citenamefont {Kavanagh}\ and\ \citenamefont
  {Green}(2012)}]{Kavanagh:2012nr}%
  \BibitemOpen
  \bibfield  {author} {\bibinfo {author} {\bibfnamefont {B.~J.}\ \bibnamefont
  {Kavanagh}}\ and\ \bibinfo {author} {\bibfnamefont {A.~M.}\ \bibnamefont
  {Green}},\ }\href@noop {} {\  (\bibinfo {year} {2012})},\ \Eprint
  {http://arxiv.org/abs/1207.2039} {arXiv:1207.2039 [astro-ph.CO]} \BibitemShut
  {NoStop}%
\bibitem [{\citenamefont {Cerdeno}\ \emph
  {et~al.}(2013{\natexlab{a}})\citenamefont {Cerdeno}, \citenamefont {Fornasa},
  \citenamefont {Huh},\ and\ \citenamefont {Peiro}}]{Cerdeno:2012ix}%
  \BibitemOpen
  \bibfield  {author} {\bibinfo {author} {\bibfnamefont {D.}~\bibnamefont
  {Cerdeno}}, \bibinfo {author} {\bibfnamefont {M.}~\bibnamefont {Fornasa}},
  \bibinfo {author} {\bibfnamefont {J.}~\bibnamefont {Huh}}, \ and\ \bibinfo
  {author} {\bibfnamefont {M.}~\bibnamefont {Peiro}},\ }\href {\doibase
  10.1103/PhysRevD.87.023512} {\bibfield  {journal} {\bibinfo  {journal}
  {Phys.Rev.}\ }\textbf {\bibinfo {volume} {D87}},\ \bibinfo {pages} {023512}
  (\bibinfo {year} {2013}{\natexlab{a}})},\ \Eprint
  {http://arxiv.org/abs/1208.6426} {arXiv:1208.6426 [hep-ph]} \BibitemShut
  {NoStop}%
\bibitem [{\citenamefont {Aprile}\ \emph {et~al.}(2013)\citenamefont {Aprile}
  \emph {et~al.}}]{:2013uj}%
  \BibitemOpen
  \bibfield  {author} {\bibinfo {author} {\bibfnamefont {E.}~\bibnamefont
  {Aprile}} \emph {et~al.} (\bibinfo {collaboration} {XENON100
  collaboration}),\ }\href@noop {} {\  (\bibinfo {year} {2013})},\ \Eprint
  {http://arxiv.org/abs/1301.6620} {arXiv:1301.6620 [astro-ph.CO]} \BibitemShut
  {NoStop}%
\bibitem [{\citenamefont {Bertone}\ \emph {et~al.}(2007)\citenamefont
  {Bertone}, \citenamefont {Cerdeno}, \citenamefont {Collar},\ and\
  \citenamefont {Odom}}]{Bertone:2007xj}%
  \BibitemOpen
  \bibfield  {author} {\bibinfo {author} {\bibfnamefont {G.}~\bibnamefont
  {Bertone}}, \bibinfo {author} {\bibfnamefont {D.~G.}\ \bibnamefont
  {Cerdeno}}, \bibinfo {author} {\bibfnamefont {J.}~\bibnamefont {Collar}}, \
  and\ \bibinfo {author} {\bibfnamefont {B.~C.}\ \bibnamefont {Odom}},\ }\href
  {\doibase 10.1103/PhysRevLett.99.151301} {\bibfield  {journal} {\bibinfo
  {journal} {Phys.Rev.Lett.}\ }\textbf {\bibinfo {volume} {99}},\ \bibinfo
  {pages} {151301} (\bibinfo {year} {2007})},\ \Eprint
  {http://arxiv.org/abs/0705.2502} {arXiv:0705.2502 [astro-ph]} \BibitemShut
  {NoStop}%
\bibitem [{\citenamefont {Pato}(2011)}]{Pato:2011de}%
  \BibitemOpen
  \bibfield  {author} {\bibinfo {author} {\bibfnamefont {M.}~\bibnamefont
  {Pato}},\ }\href {\doibase 10.1088/1475-7516/2011/10/035} {\bibfield
  {journal} {\bibinfo  {journal} {JCAP}\ }\textbf {\bibinfo {volume} {1110}},\
  \bibinfo {pages} {035} (\bibinfo {year} {2011})},\ \Eprint
  {http://arxiv.org/abs/1106.0743} {arXiv:1106.0743 [astro-ph.CO]} \BibitemShut
  {NoStop}%
\bibitem [{\citenamefont {Cerdeno}\ \emph
  {et~al.}(2013{\natexlab{b}})\citenamefont {Cerdeno}, \citenamefont {Cuesta},
  \citenamefont {Fornasa}, \citenamefont {Garcia}, \citenamefont {Ginestra}
  \emph {et~al.}}]{Cerdeno:2013gqa}%
  \BibitemOpen
  \bibfield  {author} {\bibinfo {author} {\bibfnamefont {D.}~\bibnamefont
  {Cerdeno}}, \bibinfo {author} {\bibfnamefont {C.}~\bibnamefont {Cuesta}},
  \bibinfo {author} {\bibfnamefont {M.}~\bibnamefont {Fornasa}}, \bibinfo
  {author} {\bibfnamefont {E.}~\bibnamefont {Garcia}}, \bibinfo {author}
  {\bibfnamefont {C.}~\bibnamefont {Ginestra}},  \emph {et~al.},\ }\href@noop
  {} {\  (\bibinfo {year} {2013}{\natexlab{b}})},\ \Eprint
  {http://arxiv.org/abs/1304.1758} {arXiv:1304.1758 [hep-ph]} \BibitemShut
  {NoStop}%
\bibitem [{\citenamefont {Bertone}\ \emph {et~al.}(2010)\citenamefont
  {Bertone}, \citenamefont {Cerdeno}, \citenamefont {Fornasa}, \citenamefont
  {de~Austri},\ and\ \citenamefont {Trotta}}]{Bertone:2010rv}%
  \BibitemOpen
  \bibfield  {author} {\bibinfo {author} {\bibfnamefont {G.}~\bibnamefont
  {Bertone}}, \bibinfo {author} {\bibfnamefont {D.~G.}\ \bibnamefont
  {Cerdeno}}, \bibinfo {author} {\bibfnamefont {M.}~\bibnamefont {Fornasa}},
  \bibinfo {author} {\bibfnamefont {R.~R.}\ \bibnamefont {de~Austri}}, \ and\
  \bibinfo {author} {\bibfnamefont {R.}~\bibnamefont {Trotta}},\ }\href
  {\doibase 10.1103/PhysRevD.82.055008} {\bibfield  {journal} {\bibinfo
  {journal} {Phys.Rev.}\ }\textbf {\bibinfo {volume} {D82}},\ \bibinfo {pages}
  {055008} (\bibinfo {year} {2010})},\ \Eprint {http://arxiv.org/abs/1005.4280}
  {arXiv:1005.4280 [hep-ph]} \BibitemShut {NoStop}%
\bibitem [{\citenamefont {Bernal}\ \emph {et~al.}(2009)\citenamefont {Bernal},
  \citenamefont {Goudelis}, \citenamefont {Mambrini},\ and\ \citenamefont
  {Munoz}}]{Bernal:2008zk}%
  \BibitemOpen
  \bibfield  {author} {\bibinfo {author} {\bibfnamefont {N.}~\bibnamefont
  {Bernal}}, \bibinfo {author} {\bibfnamefont {A.}~\bibnamefont {Goudelis}},
  \bibinfo {author} {\bibfnamefont {Y.}~\bibnamefont {Mambrini}}, \ and\
  \bibinfo {author} {\bibfnamefont {C.}~\bibnamefont {Munoz}},\ }\href
  {\doibase 10.1088/1475-7516/2009/01/046} {\bibfield  {journal} {\bibinfo
  {journal} {JCAP}\ }\textbf {\bibinfo {volume} {0901}},\ \bibinfo {pages}
  {046} (\bibinfo {year} {2009})},\ \Eprint {http://arxiv.org/abs/0804.1976}
  {arXiv:0804.1976 [hep-ph]} \BibitemShut {NoStop}%
\bibitem [{\citenamefont {Aartsen}\ \emph {et~al.}(2013)\citenamefont {Aartsen}
  \emph {et~al.}}]{Aartsen:2012kia}%
  \BibitemOpen
  \bibfield  {author} {\bibinfo {author} {\bibfnamefont {M.}~\bibnamefont
  {Aartsen}} \emph {et~al.} (\bibinfo {collaboration} {IceCube
  collaboration}),\ }\href {\doibase 10.1103/PhysRevLett.110.131302} {\bibfield
   {journal} {\bibinfo  {journal} {Phys.Rev.Lett.}\ }\textbf {\bibinfo {volume}
  {110}},\ \bibinfo {pages} {131302} (\bibinfo {year} {2013})},\ \Eprint
  {http://arxiv.org/abs/1212.4097} {arXiv:1212.4097 [astro-ph.HE]} \BibitemShut
  {NoStop}%
\bibitem [{\citenamefont {Hooper}\ and\ \citenamefont
  {Taylor}(2007)}]{Hooper:2006wv}%
  \BibitemOpen
  \bibfield  {author} {\bibinfo {author} {\bibfnamefont {D.}~\bibnamefont
  {Hooper}}\ and\ \bibinfo {author} {\bibfnamefont {A.~M.}\ \bibnamefont
  {Taylor}},\ }\href {\doibase 10.1088/1475-7516/2007/03/017} {\bibfield
  {journal} {\bibinfo  {journal} {JCAP}\ }\textbf {\bibinfo {volume} {0703}},\
  \bibinfo {pages} {017} (\bibinfo {year} {2007})},\ \Eprint
  {http://arxiv.org/abs/hep-ph/0607086} {arXiv:hep-ph/0607086 [hep-ph]}
  \BibitemShut {NoStop}%
\bibitem [{\citenamefont {Niro}\ \emph {et~al.}(2009)\citenamefont {Niro},
  \citenamefont {Bottino}, \citenamefont {Fornengo},\ and\ \citenamefont
  {Scopel}}]{PhysRevD.80.095019}%
  \BibitemOpen
  \bibfield  {author} {\bibinfo {author} {\bibfnamefont {V.}~\bibnamefont
  {Niro}}, \bibinfo {author} {\bibfnamefont {A.}~\bibnamefont {Bottino}},
  \bibinfo {author} {\bibfnamefont {N.}~\bibnamefont {Fornengo}}, \ and\
  \bibinfo {author} {\bibfnamefont {S.}~\bibnamefont {Scopel}},\ }\href
  {\doibase 10.1103/PhysRevD.80.095019} {\bibfield  {journal} {\bibinfo
  {journal} {Phys. Rev. D}\ }\textbf {\bibinfo {volume} {80}},\ \bibinfo
  {pages} {095019} (\bibinfo {year} {2009})}\BibitemShut {NoStop}%
\bibitem [{\citenamefont {Aprile}(2012)}]{Aprile:2012zx}%
  \BibitemOpen
  \bibfield  {author} {\bibinfo {author} {\bibfnamefont {E.}~\bibnamefont
  {Aprile}} (\bibinfo {collaboration} {XENON1T collaboration}),\ }\href@noop {}
  {\  (\bibinfo {year} {2012})},\ \Eprint {http://arxiv.org/abs/1206.6288}
  {arXiv:1206.6288 [astro-ph.IM]} \BibitemShut {NoStop}%
\bibitem [{\citenamefont {Abbasi}\ \emph
  {et~al.}(2012{\natexlab{a}})\citenamefont {Abbasi} \emph
  {et~al.}}]{IceCube:2011aj}%
  \BibitemOpen
  \bibfield  {author} {\bibinfo {author} {\bibfnamefont {R.}~\bibnamefont
  {Abbasi}} \emph {et~al.} (\bibinfo {collaboration} {IceCube Collaboration}),\
  }\href {\doibase 10.1103/PhysRevD.85.042002} {\bibfield  {journal} {\bibinfo
  {journal} {Phys.Rev.}\ }\textbf {\bibinfo {volume} {D85}},\ \bibinfo {pages}
  {042002} (\bibinfo {year} {2012}{\natexlab{a}})},\ \Eprint
  {http://arxiv.org/abs/1112.1840} {arXiv:1112.1840 [astro-ph.HE]} \BibitemShut
  {NoStop}%
\bibitem [{\citenamefont {Smith}\ \emph {et~al.}(1996)\citenamefont {Smith},
  \citenamefont {Arnison}, \citenamefont {Homer}, \citenamefont {Lewin},
  \citenamefont {Alner} \emph {et~al.}}]{Smith:1996fu}%
  \BibitemOpen
  \bibfield  {author} {\bibinfo {author} {\bibfnamefont {P.}~\bibnamefont
  {Smith}}, \bibinfo {author} {\bibfnamefont {G.}~\bibnamefont {Arnison}},
  \bibinfo {author} {\bibfnamefont {G.}~\bibnamefont {Homer}}, \bibinfo
  {author} {\bibfnamefont {J.}~\bibnamefont {Lewin}}, \bibinfo {author}
  {\bibfnamefont {G.}~\bibnamefont {Alner}},  \emph {et~al.},\ }\href {\doibase
  10.1016/0370-2693(96)00350-4} {\bibfield  {journal} {\bibinfo  {journal}
  {Phys.Lett.}\ }\textbf {\bibinfo {volume} {B379}},\ \bibinfo {pages} {299}
  (\bibinfo {year} {1996})}\BibitemShut {NoStop}%
\bibitem [{\citenamefont {Savage}\ \emph {et~al.}(2009)\citenamefont {Savage},
  \citenamefont {Gelmini}, \citenamefont {Gondolo},\ and\ \citenamefont
  {Freese}}]{Savage:2008er}%
  \BibitemOpen
  \bibfield  {author} {\bibinfo {author} {\bibfnamefont {C.}~\bibnamefont
  {Savage}}, \bibinfo {author} {\bibfnamefont {G.}~\bibnamefont {Gelmini}},
  \bibinfo {author} {\bibfnamefont {P.}~\bibnamefont {Gondolo}}, \ and\
  \bibinfo {author} {\bibfnamefont {K.}~\bibnamefont {Freese}},\ }\href
  {\doibase 10.1088/1475-7516/2009/04/010} {\bibfield  {journal} {\bibinfo
  {journal} {JCAP}\ }\textbf {\bibinfo {volume} {0904}},\ \bibinfo {pages}
  {010} (\bibinfo {year} {2009})},\ \Eprint {http://arxiv.org/abs/0808.3607}
  {arXiv:0808.3607 [astro-ph]} \BibitemShut {NoStop}%
\bibitem [{\citenamefont {Helm}(1956)}]{Helm:1956zz}%
  \BibitemOpen
  \bibfield  {author} {\bibinfo {author} {\bibfnamefont {R.~H.}\ \bibnamefont
  {Helm}},\ }\href {\doibase 10.1103/PhysRev.104.1466} {\bibfield  {journal}
  {\bibinfo  {journal} {Phys. Rev.}\ }\textbf {\bibinfo {volume} {104}},\
  \bibinfo {pages} {1466} (\bibinfo {year} {1956})}\BibitemShut {NoStop}%
\bibitem [{\citenamefont {Lewin}\ and\ \citenamefont
  {Smith}(1996)}]{Lewin:1995rx}%
  \BibitemOpen
  \bibfield  {author} {\bibinfo {author} {\bibfnamefont {J.~D.}\ \bibnamefont
  {Lewin}}\ and\ \bibinfo {author} {\bibfnamefont {P.~F.}\ \bibnamefont
  {Smith}},\ }\href {\doibase 10.1016/S0927-6505(96)00047-3} {\bibfield
  {journal} {\bibinfo  {journal} {Astropart. Phys.}\ }\textbf {\bibinfo
  {volume} {6}},\ \bibinfo {pages} {87} (\bibinfo {year} {1996})}\BibitemShut
  {NoStop}%
\bibitem [{\citenamefont {Duda}\ \emph {et~al.}(2007)\citenamefont {Duda},
  \citenamefont {Kemper},\ and\ \citenamefont {Gondolo}}]{Duda:2006uk}%
  \BibitemOpen
  \bibfield  {author} {\bibinfo {author} {\bibfnamefont {G.}~\bibnamefont
  {Duda}}, \bibinfo {author} {\bibfnamefont {A.}~\bibnamefont {Kemper}}, \ and\
  \bibinfo {author} {\bibfnamefont {P.}~\bibnamefont {Gondolo}},\ }\href@noop
  {} {\bibfield  {journal} {\bibinfo  {journal} {JCAP}\ }\textbf {\bibinfo
  {volume} {0704}},\ \bibinfo {pages} {012} (\bibinfo {year} {2007})},\ \Eprint
  {http://arxiv.org/abs/hep-ph/0608035} {arXiv:hep-ph/0608035} \BibitemShut
  {NoStop}%
\bibitem [{\citenamefont {Belanger}\ \emph {et~al.}(2009)\citenamefont
  {Belanger}, \citenamefont {Nezri},\ and\ \citenamefont
  {Pukhov}}]{Belanger:2008gy}%
  \BibitemOpen
  \bibfield  {author} {\bibinfo {author} {\bibfnamefont {G.}~\bibnamefont
  {Belanger}}, \bibinfo {author} {\bibfnamefont {E.}~\bibnamefont {Nezri}}, \
  and\ \bibinfo {author} {\bibfnamefont {A.}~\bibnamefont {Pukhov}},\ }\href
  {\doibase 10.1103/PhysRevD.79.015008} {\bibfield  {journal} {\bibinfo
  {journal} {Phys.Rev.}\ }\textbf {\bibinfo {volume} {D79}},\ \bibinfo {pages}
  {015008} (\bibinfo {year} {2009})},\ \Eprint {http://arxiv.org/abs/0810.1362}
  {arXiv:0810.1362 [hep-ph]} \BibitemShut {NoStop}%
\bibitem [{\citenamefont {Ressell}\ and\ \citenamefont
  {Dean}(1997)}]{Ressell:1997kx}%
  \BibitemOpen
  \bibfield  {author} {\bibinfo {author} {\bibfnamefont {M.}~\bibnamefont
  {Ressell}}\ and\ \bibinfo {author} {\bibfnamefont {D.}~\bibnamefont {Dean}},\
  }\href {\doibase 10.1103/PhysRevC.56.535} {\bibfield  {journal} {\bibinfo
  {journal} {Phys.Rev.}\ }\textbf {\bibinfo {volume} {C56}},\ \bibinfo {pages}
  {535} (\bibinfo {year} {1997})},\ \Eprint
  {http://arxiv.org/abs/hep-ph/9702290} {arXiv:hep-ph/9702290 [hep-ph]}
  \BibitemShut {NoStop}%
\bibitem [{\citenamefont {Menendez}\ \emph {et~al.}(2012)\citenamefont
  {Menendez}, \citenamefont {Gazit},\ and\ \citenamefont
  {Schwenk}}]{Menendez:2012tm}%
  \BibitemOpen
  \bibfield  {author} {\bibinfo {author} {\bibfnamefont {J.}~\bibnamefont
  {Menendez}}, \bibinfo {author} {\bibfnamefont {D.}~\bibnamefont {Gazit}}, \
  and\ \bibinfo {author} {\bibfnamefont {A.}~\bibnamefont {Schwenk}},\ }\href
  {\doibase 10.1103/PhysRevD.86.103511} {\bibfield  {journal} {\bibinfo
  {journal} {Phys.Rev.}\ }\textbf {\bibinfo {volume} {D86}},\ \bibinfo {pages}
  {103511} (\bibinfo {year} {2012})},\ \Eprint {http://arxiv.org/abs/1208.1094}
  {arXiv:1208.1094 [astro-ph.CO]} \BibitemShut {NoStop}%
\bibitem [{\citenamefont {Gondolo}\ \emph {et~al.}(2004)\citenamefont
  {Gondolo}, \citenamefont {Edsj{\"o}}, \citenamefont {Ullio}, \citenamefont
  {Bergstr{\"o}m}, \citenamefont {Schelke},\ and\ \citenamefont
  {Baltz}}]{Gondolo2004}%
  \BibitemOpen
  \bibfield  {author} {\bibinfo {author} {\bibfnamefont {P.}~\bibnamefont
  {Gondolo}}, \bibinfo {author} {\bibfnamefont {J.}~\bibnamefont {Edsj{\"o}}},
  \bibinfo {author} {\bibfnamefont {P.}~\bibnamefont {Ullio}}, \bibinfo
  {author} {\bibfnamefont {L.}~\bibnamefont {Bergstr{\"o}m}}, \bibinfo {author}
  {\bibfnamefont {M.}~\bibnamefont {Schelke}}, \ and\ \bibinfo {author}
  {\bibfnamefont {E.~A.}\ \bibnamefont {Baltz}},\ }\href {\doibase
  10.1088/1475-7516/2004/07/008} {\bibfield  {journal} {\bibinfo  {journal}
  {JCAP}\ }\textbf {\bibinfo {volume} {0407}},\ \bibinfo {pages} {008}
  (\bibinfo {year} {2004})},\ \Eprint {http://arxiv.org/abs/astro-ph/0406204}
  {arXiv:astro-ph/0406204 [astro-ph]} \BibitemShut {NoStop}%
\bibitem [{\citenamefont {Gould}(1987)}]{Gould1987}%
  \BibitemOpen
  \bibfield  {author} {\bibinfo {author} {\bibfnamefont {A.}~\bibnamefont
  {Gould}},\ }\href {\doibase 10.1086/165653} {\bibfield  {journal} {\bibinfo
  {journal} {Astrophys.J.}\ }\textbf {\bibinfo {volume} {321}},\ \bibinfo
  {pages} {571} (\bibinfo {year} {1987})}\BibitemShut {NoStop}%
\bibitem [{\citenamefont {Sivertsson}\ and\ \citenamefont
  {Edsj{\"o}}(2012)}]{Sivertsson:2012qj}%
  \BibitemOpen
  \bibfield  {author} {\bibinfo {author} {\bibfnamefont {S.}~\bibnamefont
  {Sivertsson}}\ and\ \bibinfo {author} {\bibfnamefont {J.}~\bibnamefont
  {Edsj{\"o}}},\ }\href {\doibase 10.1103/PhysRevD.85.129905,
  10.1103/PhysRevD.85.123514} {\bibfield  {journal} {\bibinfo  {journal}
  {Phys.Rev.}\ }\textbf {\bibinfo {volume} {D85}},\ \bibinfo {pages} {123514}
  (\bibinfo {year} {2012})},\ \Eprint {http://arxiv.org/abs/1201.1895}
  {arXiv:1201.1895 [astro-ph.HE]} \BibitemShut {NoStop}%
\bibitem [{\citenamefont {Scott}\ \emph {et~al.}(2009)\citenamefont {Scott},
  \citenamefont {Fairbairn},\ and\ \citenamefont {Edsj{\"o}}}]{Scott:2008ns}%
  \BibitemOpen
  \bibfield  {author} {\bibinfo {author} {\bibfnamefont {P.}~\bibnamefont
  {Scott}}, \bibinfo {author} {\bibfnamefont {M.}~\bibnamefont {Fairbairn}}, \
  and\ \bibinfo {author} {\bibfnamefont {J.}~\bibnamefont {Edsj{\"o}}},\ }\href
  {\doibase 10.1111/j.1365-2966.2008.14282.x} {\bibfield  {journal} {\bibinfo
  {journal} {Mon.Not.Roy.Astron.Soc.}\ }\textbf {\bibinfo {volume} {394}},\
  \bibinfo {pages} {82} (\bibinfo {year} {2009})},\ \Eprint
  {http://arxiv.org/abs/0809.1871} {arXiv:0809.1871 [astro-ph]} \BibitemShut
  {NoStop}%
\bibitem [{\citenamefont {Serpico}\ and\ \citenamefont
  {Bertone}(2010)}]{Serpico:2010ae}%
  \BibitemOpen
  \bibfield  {author} {\bibinfo {author} {\bibfnamefont {P.~D.}\ \bibnamefont
  {Serpico}}\ and\ \bibinfo {author} {\bibfnamefont {G.}~\bibnamefont
  {Bertone}},\ }\href {\doibase 10.1103/PhysRevD.82.063505} {\bibfield
  {journal} {\bibinfo  {journal} {Phys.Rev.}\ }\textbf {\bibinfo {volume}
  {D82}},\ \bibinfo {pages} {063505} (\bibinfo {year} {2010})},\ \Eprint
  {http://arxiv.org/abs/1006.3268} {arXiv:1006.3268 [astro-ph.HE]} \BibitemShut
  {NoStop}%
\bibitem [{\citenamefont {Griest}\ and\ \citenamefont
  {Seckel}(1987)}]{Griest1987}%
  \BibitemOpen
  \bibfield  {author} {\bibinfo {author} {\bibfnamefont {K.}~\bibnamefont
  {Griest}}\ and\ \bibinfo {author} {\bibfnamefont {D.}~\bibnamefont
  {Seckel}},\ }\href {\doibase 10.1016/0550-3213(87)90293-8} {\bibfield
  {journal} {\bibinfo  {journal} {Nucl.Phys.}\ }\textbf {\bibinfo {volume}
  {B283}},\ \bibinfo {pages} {681} (\bibinfo {year} {1987})}\BibitemShut
  {NoStop}%
\bibitem [{\citenamefont {Halzen}(2006)}]{Halzen2006}%
  \BibitemOpen
  \bibfield  {author} {\bibinfo {author} {\bibfnamefont {F.}~\bibnamefont
  {Halzen}},\ }\href {\doibase 10.1140/epjc/s2006-02536-4} {\bibfield
  {journal} {\bibinfo  {journal} {Eur.Phys.J.}\ }\textbf {\bibinfo {volume}
  {C46}},\ \bibinfo {pages} {669} (\bibinfo {year} {2006})},\ \Eprint
  {http://arxiv.org/abs/astro-ph/0602132} {arXiv:astro-ph/0602132 [astro-ph]}
  \BibitemShut {NoStop}%
\bibitem [{\citenamefont {Silverwood}\ \emph {et~al.}(2013)\citenamefont
  {Silverwood}, \citenamefont {Scott}, \citenamefont {Danninger}, \citenamefont
  {Savage}, \citenamefont {Edsj{\"o}}, \citenamefont {Adams}, \citenamefont
  {Brown},\ and\ \citenamefont {Hultqvist}}]{Silverwood:2012tp}%
  \BibitemOpen
  \bibfield  {author} {\bibinfo {author} {\bibfnamefont {H.}~\bibnamefont
  {Silverwood}}, \bibinfo {author} {\bibfnamefont {P.}~\bibnamefont {Scott}},
  \bibinfo {author} {\bibfnamefont {M.}~\bibnamefont {Danninger}}, \bibinfo
  {author} {\bibfnamefont {C.}~\bibnamefont {Savage}}, \bibinfo {author}
  {\bibfnamefont {J.}~\bibnamefont {Edsj{\"o}}}, \bibinfo {author}
  {\bibfnamefont {J.}~\bibnamefont {Adams}}, \bibinfo {author} {\bibfnamefont
  {A.~M.}\ \bibnamefont {Brown}}, \ and\ \bibinfo {author} {\bibfnamefont
  {K.}~\bibnamefont {Hultqvist}},\ }\href {\doibase
  10.1088/1475-7516/2013/03/027} {\bibfield  {journal} {\bibinfo  {journal}
  {JCAP}\ }\textbf {\bibinfo {volume} {1303}},\ \bibinfo {pages} {027}
  (\bibinfo {year} {2013})},\ \Eprint {http://arxiv.org/abs/1210.0844}
  {arXiv:1210.0844 [hep-ph]} \BibitemShut {NoStop}%
\bibitem [{\citenamefont {Cowan}\ \emph {et~al.}(2011)\citenamefont {Cowan},
  \citenamefont {Cranmer}, \citenamefont {Gross},\ and\ \citenamefont
  {Vitells}}]{Cowan:2010js}%
  \BibitemOpen
  \bibfield  {author} {\bibinfo {author} {\bibfnamefont {G.}~\bibnamefont
  {Cowan}}, \bibinfo {author} {\bibfnamefont {K.}~\bibnamefont {Cranmer}},
  \bibinfo {author} {\bibfnamefont {E.}~\bibnamefont {Gross}}, \ and\ \bibinfo
  {author} {\bibfnamefont {O.}~\bibnamefont {Vitells}},\ }\href {\doibase
  10.1140/epjc/s10052-011-1554-0} {\bibfield  {journal} {\bibinfo  {journal}
  {Eur.Phys.J.}\ }\textbf {\bibinfo {volume} {C71}},\ \bibinfo {pages} {1554}
  (\bibinfo {year} {2011})},\ \Eprint {http://arxiv.org/abs/1007.1727}
  {arXiv:1007.1727 [physics.data-an]} \BibitemShut {NoStop}%
\bibitem [{\citenamefont {Feroz}\ and\ \citenamefont
  {Hobson}(2008)}]{Feroz:2007kg}%
  \BibitemOpen
  \bibfield  {author} {\bibinfo {author} {\bibfnamefont {F.}~\bibnamefont
  {Feroz}}\ and\ \bibinfo {author} {\bibfnamefont {M.}~\bibnamefont {Hobson}},\
  }\href {\doibase 10.1111/j.1365-2966.2007.12353.x} {\bibfield  {journal}
  {\bibinfo  {journal} {Mon.Not.Roy.Astron.Soc.}\ }\textbf {\bibinfo {volume}
  {384}},\ \bibinfo {pages} {449} (\bibinfo {year} {2008})},\ \Eprint
  {http://arxiv.org/abs/0704.3704} {arXiv:0704.3704 [astro-ph]} \BibitemShut
  {NoStop}%
\bibitem [{\citenamefont {Feroz}\ \emph {et~al.}(2009)\citenamefont {Feroz},
  \citenamefont {Hobson},\ and\ \citenamefont {Bridges}}]{Feroz:2008xx}%
  \BibitemOpen
  \bibfield  {author} {\bibinfo {author} {\bibfnamefont {F.}~\bibnamefont
  {Feroz}}, \bibinfo {author} {\bibfnamefont {M.}~\bibnamefont {Hobson}}, \
  and\ \bibinfo {author} {\bibfnamefont {M.}~\bibnamefont {Bridges}},\ }\href
  {\doibase DOI: 10.1111/j.1365-2966.2009.14548.x} {\bibfield  {journal}
  {\bibinfo  {journal} {Mon.Not.Roy.Astron.Soc.}\ }\textbf {\bibinfo {volume}
  {398}},\ \bibinfo {pages} {1601} (\bibinfo {year} {2009})},\ \Eprint
  {http://arxiv.org/abs/0809.3437} {arXiv:0809.3437 [astro-ph]} \BibitemShut
  {NoStop}%
\bibitem [{\citenamefont {Feroz}\ \emph {et~al.}(2011)\citenamefont {Feroz},
  \citenamefont {Cranmer}, \citenamefont {Hobson}, \citenamefont {Ruiz~de
  Austri},\ and\ \citenamefont {Trotta}}]{Feroz:2011bj}%
  \BibitemOpen
  \bibfield  {author} {\bibinfo {author} {\bibfnamefont {F.}~\bibnamefont
  {Feroz}}, \bibinfo {author} {\bibfnamefont {K.}~\bibnamefont {Cranmer}},
  \bibinfo {author} {\bibfnamefont {M.}~\bibnamefont {Hobson}}, \bibinfo
  {author} {\bibfnamefont {R.}~\bibnamefont {Ruiz~de Austri}}, \ and\ \bibinfo
  {author} {\bibfnamefont {R.}~\bibnamefont {Trotta}},\ }\href {\doibase
  10.1007/JHEP06(2011)042} {\bibfield  {journal} {\bibinfo  {journal} {JHEP}\
  }\textbf {\bibinfo {volume} {1106}},\ \bibinfo {pages} {042} (\bibinfo {year}
  {2011})},\ \Eprint {http://arxiv.org/abs/1101.3296} {arXiv:1101.3296
  [hep-ph]} \BibitemShut {NoStop}%
\bibitem [{\citenamefont {de~Austri}\ \emph {et~al.}()\citenamefont
  {de~Austri}, \citenamefont {Trotta},\ and\ \citenamefont
  {Feroz}}]{superbayes}%
  \BibitemOpen
  \bibfield  {author} {\bibinfo {author} {\bibfnamefont {R.~R.}\ \bibnamefont
  {de~Austri}}, \bibinfo {author} {\bibfnamefont {R.}~\bibnamefont {Trotta}}, \
  and\ \bibinfo {author} {\bibfnamefont {F.}~\bibnamefont {Feroz}},\
  }\href@noop {} {\enquote {\bibinfo {title} {{\texttt{SuperBayeS} package}},}\
  }\bibinfo {note} {\url{http://www.superbayes.org/}}\BibitemShut {NoStop}%
\bibitem [{\citenamefont {Trotta}\ \emph {et~al.}(2008)\citenamefont {Trotta},
  \citenamefont {Feroz}, \citenamefont {Hobson}, \citenamefont {Roszkowski},\
  and\ \citenamefont {Ruiz~de Austri}}]{Trotta:2008bp}%
  \BibitemOpen
  \bibfield  {author} {\bibinfo {author} {\bibfnamefont {R.}~\bibnamefont
  {Trotta}}, \bibinfo {author} {\bibfnamefont {F.}~\bibnamefont {Feroz}},
  \bibinfo {author} {\bibfnamefont {M.~P.}\ \bibnamefont {Hobson}}, \bibinfo
  {author} {\bibfnamefont {L.}~\bibnamefont {Roszkowski}}, \ and\ \bibinfo
  {author} {\bibfnamefont {R.}~\bibnamefont {Ruiz~de Austri}},\ }\href
  {\doibase 10.1088/1126-6708/2008/12/024} {\bibfield  {journal} {\bibinfo
  {journal} {JHEP}\ }\textbf {\bibinfo {volume} {0812}},\ \bibinfo {pages}
  {024} (\bibinfo {year} {2008})},\ \Eprint {http://arxiv.org/abs/0809.3792}
  {arXiv:0809.3792 [hep-ph]} \BibitemShut {NoStop}%
\bibitem [{\citenamefont {Bovy}\ \emph {et~al.}(2012)\citenamefont {Bovy},
  \citenamefont {Prieto}, \citenamefont {Beers}, \citenamefont {Bizyaev},
  \citenamefont {da~Costa} \emph {et~al.}}]{Bovy:2012ba}%
  \BibitemOpen
  \bibfield  {author} {\bibinfo {author} {\bibfnamefont {J.}~\bibnamefont
  {Bovy}}, \bibinfo {author} {\bibfnamefont {C.~A.}\ \bibnamefont {Prieto}},
  \bibinfo {author} {\bibfnamefont {T.~C.}\ \bibnamefont {Beers}}, \bibinfo
  {author} {\bibfnamefont {D.}~\bibnamefont {Bizyaev}}, \bibinfo {author}
  {\bibfnamefont {L.~N.}\ \bibnamefont {da~Costa}},  \emph {et~al.},\ }\href
  {\doibase 10.1088/0004-637X/759/2/131} {\bibfield  {journal} {\bibinfo
  {journal} {Astrophys.J.}\ }\textbf {\bibinfo {volume} {759}},\ \bibinfo
  {pages} {131} (\bibinfo {year} {2012})},\ \Eprint
  {http://arxiv.org/abs/1209.0759} {arXiv:1209.0759 [astro-ph.GA]} \BibitemShut
  {NoStop}%
\bibitem [{\citenamefont {Reid}\ \emph {et~al.}(2009)\citenamefont {Reid},
  \citenamefont {Menten}, \citenamefont {Zheng}, \citenamefont {Brunthaler},
  \citenamefont {Moscadelli} \emph {et~al.}}]{Reid:2009nj}%
  \BibitemOpen
  \bibfield  {author} {\bibinfo {author} {\bibfnamefont {M.}~\bibnamefont
  {Reid}}, \bibinfo {author} {\bibfnamefont {K.}~\bibnamefont {Menten}},
  \bibinfo {author} {\bibfnamefont {X.}~\bibnamefont {Zheng}}, \bibinfo
  {author} {\bibfnamefont {A.}~\bibnamefont {Brunthaler}}, \bibinfo {author}
  {\bibfnamefont {L.}~\bibnamefont {Moscadelli}},  \emph {et~al.},\ }\href
  {\doibase 10.1088/0004-637X/700/1/137} {\bibfield  {journal} {\bibinfo
  {journal} {Astrophys.J.}\ }\textbf {\bibinfo {volume} {700}},\ \bibinfo
  {pages} {137} (\bibinfo {year} {2009})},\ \Eprint
  {http://arxiv.org/abs/0902.3913} {arXiv:0902.3913 [astro-ph.GA]} \BibitemShut
  {NoStop}%
\bibitem [{\citenamefont {Gillessen}\ \emph {et~al.}(2009)\citenamefont
  {Gillessen}, \citenamefont {Eisenhauer}, \citenamefont {Trippe},
  \citenamefont {Alexander}, \citenamefont {Genzel}, \citenamefont {Martins},\
  and\ \citenamefont {Ott}}]{Gillessen:2008qv}%
  \BibitemOpen
  \bibfield  {author} {\bibinfo {author} {\bibfnamefont {S.}~\bibnamefont
  {Gillessen}}, \bibinfo {author} {\bibfnamefont {F.}~\bibnamefont
  {Eisenhauer}}, \bibinfo {author} {\bibfnamefont {S.}~\bibnamefont {Trippe}},
  \bibinfo {author} {\bibfnamefont {T.}~\bibnamefont {Alexander}}, \bibinfo
  {author} {\bibfnamefont {R.}~\bibnamefont {Genzel}}, \bibinfo {author}
  {\bibfnamefont {F.}~\bibnamefont {Martins}}, \ and\ \bibinfo {author}
  {\bibfnamefont {T.}~\bibnamefont {Ott}},\ }\href {\doibase
  10.1088/0004-637X/692/2/1075} {\bibfield  {journal} {\bibinfo  {journal}
  {Astrophys.J.}\ }\textbf {\bibinfo {volume} {692}},\ \bibinfo {pages} {1075}
  (\bibinfo {year} {2009})},\ \Eprint {http://arxiv.org/abs/0810.4674}
  {arXiv:0810.4674 [astro-ph]} \BibitemShut {NoStop}%
\bibitem [{\citenamefont {Smith}\ \emph {et~al.}(2007)\citenamefont {Smith},
  \citenamefont {Ruchti}, \citenamefont {Helmi}, \citenamefont {Wyse},
  \citenamefont {Fulbright} \emph {et~al.}}]{Smith:2006ym}%
  \BibitemOpen
  \bibfield  {author} {\bibinfo {author} {\bibfnamefont {M.~C.}\ \bibnamefont
  {Smith}}, \bibinfo {author} {\bibfnamefont {G.}~\bibnamefont {Ruchti}},
  \bibinfo {author} {\bibfnamefont {A.}~\bibnamefont {Helmi}}, \bibinfo
  {author} {\bibfnamefont {R.}~\bibnamefont {Wyse}}, \bibinfo {author}
  {\bibfnamefont {J.}~\bibnamefont {Fulbright}},  \emph {et~al.},\ }\href
  {\doibase 10.1111/j.1365-2966.2007.11964.x} {\bibfield  {journal} {\bibinfo
  {journal} {Mon.Not.Roy.Astron.Soc.}\ }\textbf {\bibinfo {volume} {379}},\
  \bibinfo {pages} {755} (\bibinfo {year} {2007})},\ \Eprint
  {http://arxiv.org/abs/astro-ph/0611671} {arXiv:astro-ph/0611671 [astro-ph]}
  \BibitemShut {NoStop}%
\bibitem [{\citenamefont {Dehnen}\ and\ \citenamefont
  {Binney}(1998)}]{Dehnen:1997cq}%
  \BibitemOpen
  \bibfield  {author} {\bibinfo {author} {\bibfnamefont {W.}~\bibnamefont
  {Dehnen}}\ and\ \bibinfo {author} {\bibfnamefont {J.}~\bibnamefont
  {Binney}},\ }\href@noop {} {\bibfield  {journal} {\bibinfo  {journal} {Mon.
  Not. Roy. Astron. Soc.}\ }\textbf {\bibinfo {volume} {298}},\ \bibinfo
  {pages} {387} (\bibinfo {year} {1998})},\ \Eprint
  {http://arxiv.org/abs/astro-ph/9710077} {arXiv:astro-ph/9710077} \BibitemShut
  {NoStop}%
\bibitem [{\citenamefont {Weber}\ and\ \citenamefont
  {de~Boer}(2010)}]{Weber:2009pt}%
  \BibitemOpen
  \bibfield  {author} {\bibinfo {author} {\bibfnamefont {M.}~\bibnamefont
  {Weber}}\ and\ \bibinfo {author} {\bibfnamefont {W.}~\bibnamefont
  {de~Boer}},\ }\href {\doibase 10.1051/0004-6361/200913381} {\bibfield
  {journal} {\bibinfo  {journal} {Astron.Astrophys.}\ }\textbf {\bibinfo
  {volume} {509}},\ \bibinfo {pages} {A25} (\bibinfo {year} {2010})},\ \Eprint
  {http://arxiv.org/abs/0910.4272} {arXiv:0910.4272 [astro-ph.CO]} \BibitemShut
  {NoStop}%
\bibitem [{\citenamefont {Salucci}\ \emph {et~al.}(2010)\citenamefont
  {Salucci}, \citenamefont {Nesti}, \citenamefont {Gentile},\ and\
  \citenamefont {Martins}}]{Salucci:2010qr}%
  \BibitemOpen
  \bibfield  {author} {\bibinfo {author} {\bibfnamefont {P.}~\bibnamefont
  {Salucci}}, \bibinfo {author} {\bibfnamefont {F.}~\bibnamefont {Nesti}},
  \bibinfo {author} {\bibfnamefont {G.}~\bibnamefont {Gentile}}, \ and\
  \bibinfo {author} {\bibfnamefont {C.}~\bibnamefont {Martins}},\ }\href
  {\doibase 10.1051/0004-6361/201014385} {\bibfield  {journal} {\bibinfo
  {journal} {Astron.Astrophys.}\ }\textbf {\bibinfo {volume} {523}},\ \bibinfo
  {pages} {A83} (\bibinfo {year} {2010})},\ \Eprint
  {http://arxiv.org/abs/1003.3101} {arXiv:1003.3101 [astro-ph.GA]} \BibitemShut
  {NoStop}%
\bibitem [{\citenamefont {Bovy}\ and\ \citenamefont
  {Tremaine}(2012)}]{Bovy:2012tw}%
  \BibitemOpen
  \bibfield  {author} {\bibinfo {author} {\bibfnamefont {J.}~\bibnamefont
  {Bovy}}\ and\ \bibinfo {author} {\bibfnamefont {S.}~\bibnamefont
  {Tremaine}},\ }\href {\doibase 10.1088/0004-637X/756/1/89} {\bibfield
  {journal} {\bibinfo  {journal} {Astrophys.J.}\ }\textbf {\bibinfo {volume}
  {756}},\ \bibinfo {pages} {89} (\bibinfo {year} {2012})},\ \Eprint
  {http://arxiv.org/abs/1205.4033} {arXiv:1205.4033 [astro-ph.GA]} \BibitemShut
  {NoStop}%
\bibitem [{\citenamefont {Halzen}\ and\ \citenamefont
  {Klein}(2010)}]{HalzenKlein2010}%
  \BibitemOpen
  \bibfield  {author} {\bibinfo {author} {\bibfnamefont {F.}~\bibnamefont
  {Halzen}}\ and\ \bibinfo {author} {\bibfnamefont {S.~R.}\ \bibnamefont
  {Klein}},\ }\href {\doibase 10.1063/1.3480478} {\bibfield  {journal}
  {\bibinfo  {journal} {Rev.Sci.Instrum.}\ }\textbf {\bibinfo {volume} {81}},\
  \bibinfo {pages} {081101} (\bibinfo {year} {2010})},\ \Eprint
  {http://arxiv.org/abs/1007.1247} {arXiv:1007.1247 [astro-ph.HE]} \BibitemShut
  {NoStop}%
\bibitem [{\citenamefont {Abbasi}\ \emph
  {et~al.}(2012{\natexlab{b}})\citenamefont {Abbasi} \emph
  {et~al.}}]{IceCubeDeepCore2012}%
  \BibitemOpen
  \bibfield  {author} {\bibinfo {author} {\bibfnamefont {R.}~\bibnamefont
  {Abbasi}} \emph {et~al.} (\bibinfo {collaboration} {IceCube Collaboration}),\
  }\href {\doibase 10.1016/j.astropartphys.2012.01.004} {\bibfield  {journal}
  {\bibinfo  {journal} {Astropart.Phys.}\ }\textbf {\bibinfo {volume} {35}},\
  \bibinfo {pages} {615} (\bibinfo {year} {2012}{\natexlab{b}})},\ \Eprint
  {http://arxiv.org/abs/1109.6096} {arXiv:1109.6096 [astro-ph.IM]} \BibitemShut
  {NoStop}%
\bibitem [{\citenamefont {Goldberg}(1983)}]{Goldberg1983}%
  \BibitemOpen
  \bibfield  {author} {\bibinfo {author} {\bibfnamefont {H.}~\bibnamefont
  {Goldberg}},\ }\href {\doibase 10.1103/PhysRevLett.50.1419} {\bibfield
  {journal} {\bibinfo  {journal} {Phys.Rev.Lett.}\ }\textbf {\bibinfo {volume}
  {50}},\ \bibinfo {pages} {1419} (\bibinfo {year} {1983})}\BibitemShut
  {NoStop}%
\bibitem [{\citenamefont {Hooper}\ and\ \citenamefont
  {Silk}(2004)}]{HooperSilk2004}%
  \BibitemOpen
  \bibfield  {author} {\bibinfo {author} {\bibfnamefont {D.}~\bibnamefont
  {Hooper}}\ and\ \bibinfo {author} {\bibfnamefont {J.}~\bibnamefont {Silk}},\
  }\href {\doibase 10.1088/1367-2630/6/1/023} {\bibfield  {journal} {\bibinfo
  {journal} {New J.Phys.}\ }\textbf {\bibinfo {volume} {6}},\ \bibinfo {pages}
  {23} (\bibinfo {year} {2004})},\ \Eprint
  {http://arxiv.org/abs/hep-ph/0311367} {arXiv:hep-ph/0311367 [hep-ph]}
  \BibitemShut {NoStop}%
\bibitem [{\citenamefont {Gondolo}\ \emph {et~al.}(2009)\citenamefont
  {Gondolo}, \citenamefont {Edsj{\"o}}, \citenamefont {Bergstr{\"o}m},
  \citenamefont {Ullio}, \citenamefont {Schelke}, \citenamefont {Baltz},
  \citenamefont {Bringmann},\ and\ \citenamefont {Duda}}]{DarkSUSY2009}%
  \BibitemOpen
  \bibfield  {author} {\bibinfo {author} {\bibfnamefont {P.}~\bibnamefont
  {Gondolo}}, \bibinfo {author} {\bibfnamefont {J.}~\bibnamefont {Edsj{\"o}}},
  \bibinfo {author} {\bibfnamefont {L.}~\bibnamefont {Bergstr{\"o}m}}, \bibinfo
  {author} {\bibfnamefont {P.}~\bibnamefont {Ullio}}, \bibinfo {author}
  {\bibfnamefont {M.}~\bibnamefont {Schelke}}, \bibinfo {author} {\bibfnamefont
  {E.~A.}\ \bibnamefont {Baltz}}, \bibinfo {author} {\bibfnamefont
  {T.}~\bibnamefont {Bringmann}}, \ and\ \bibinfo {author} {\bibfnamefont
  {G.}~\bibnamefont {Duda}},\ }\href@noop {} {\enquote {\bibinfo {title}
  {Darksusy: Manual and long description of routines},}\ }\bibinfo
  {howpublished} {\url{http://www.fysik.su.se/~edsjo/darksusy/docs/Manual.pdf}}
  (\bibinfo {year} {2009})\BibitemShut {NoStop}%
\bibitem [{\citenamefont {Scott}\ \emph {et~al.}(2012)\citenamefont {Scott},
  \citenamefont {Savage},\ and\ \citenamefont {Edsj{\"o}}}]{Scott:2012mq}%
  \BibitemOpen
  \bibfield  {author} {\bibinfo {author} {\bibfnamefont {P.}~\bibnamefont
  {Scott}}, \bibinfo {author} {\bibfnamefont {C.}~\bibnamefont {Savage}}, \
  and\ \bibinfo {author} {\bibfnamefont {J.}~\bibnamefont {Edsj{\"o}}}
  (\bibinfo {collaboration} {IceCube Collaboration}),\ }\href {\doibase
  10.1088/1475-7516/2012/11/057} {\bibfield  {journal} {\bibinfo  {journal}
  {JCAP}\ }\textbf {\bibinfo {volume} {1211}},\ \bibinfo {pages} {057}
  (\bibinfo {year} {2012})},\ \Eprint {http://arxiv.org/abs/1207.0810}
  {arXiv:1207.0810 [hep-ph]} \BibitemShut {NoStop}%
\bibitem [{\citenamefont {Danninger}(2012)}]{Danninger2011}%
  \BibitemOpen
  \bibfield  {author} {\bibinfo {author} {\bibfnamefont {M.}~\bibnamefont
  {Danninger}} (\bibinfo {collaboration} {IceCube Collaboration}),\ }\href
  {\doibase 10.1088/1742-6596/375/1/012038} {\bibfield  {journal} {\bibinfo
  {journal} {J.Phys.Conf.Ser.}\ }\textbf {\bibinfo {volume} {375}},\ \bibinfo
  {pages} {012038} (\bibinfo {year} {2012})}\BibitemShut {NoStop}%
\bibitem [{\citenamefont {Montaruli}(2011)}]{MontaruliIceCube2011}%
  \BibitemOpen
  \bibfield  {author} {\bibinfo {author} {\bibfnamefont {T.}~\bibnamefont
  {Montaruli}} (\bibinfo {collaboration} {IceCube Collaboration}),\ }\href
  {\doibase 10.1016/j.nuclphysbps.2011.03.014} {\bibfield  {journal} {\bibinfo
  {journal} {Nucl.Phys.Proc.Suppl.}\ }\textbf {\bibinfo {volume} {212-213}},\
  \bibinfo {pages} {99} (\bibinfo {year} {2011})},\ \Eprint
  {http://arxiv.org/abs/1012.0881} {arXiv:1012.0881 [astro-ph.HE]} \BibitemShut
  {NoStop}%
\bibitem [{\citenamefont {Pato}\ \emph {et~al.}(2010)\citenamefont {Pato},
  \citenamefont {Agertz}, \citenamefont {Bertone}, \citenamefont {Moore},\ and\
  \citenamefont {Teyssier}}]{Pato:2010yq}%
  \BibitemOpen
  \bibfield  {author} {\bibinfo {author} {\bibfnamefont {M.}~\bibnamefont
  {Pato}}, \bibinfo {author} {\bibfnamefont {O.}~\bibnamefont {Agertz}},
  \bibinfo {author} {\bibfnamefont {G.}~\bibnamefont {Bertone}}, \bibinfo
  {author} {\bibfnamefont {B.}~\bibnamefont {Moore}}, \ and\ \bibinfo {author}
  {\bibfnamefont {R.}~\bibnamefont {Teyssier}},\ }\href {\doibase
  10.1103/PhysRevD.82.023531} {\bibfield  {journal} {\bibinfo  {journal}
  {Phys.Rev.}\ }\textbf {\bibinfo {volume} {D82}},\ \bibinfo {pages} {023531}
  (\bibinfo {year} {2010})},\ \Eprint {http://arxiv.org/abs/1006.1322}
  {arXiv:1006.1322 [astro-ph.HE]} \BibitemShut {NoStop}%
\bibitem [{\citenamefont {Lisanti}\ \emph {et~al.}(2011)\citenamefont
  {Lisanti}, \citenamefont {Strigari}, \citenamefont {Wacker},\ and\
  \citenamefont {Wechsler}}]{Lisanti:2010qx}%
  \BibitemOpen
  \bibfield  {author} {\bibinfo {author} {\bibfnamefont {M.}~\bibnamefont
  {Lisanti}}, \bibinfo {author} {\bibfnamefont {L.~E.}\ \bibnamefont
  {Strigari}}, \bibinfo {author} {\bibfnamefont {J.~G.}\ \bibnamefont
  {Wacker}}, \ and\ \bibinfo {author} {\bibfnamefont {R.~H.}\ \bibnamefont
  {Wechsler}},\ }\href {\doibase 10.1103/PhysRevD.83.023519} {\bibfield
  {journal} {\bibinfo  {journal} {Phys.Rev.}\ }\textbf {\bibinfo {volume}
  {D83}},\ \bibinfo {pages} {023519} (\bibinfo {year} {2011})},\ \Eprint
  {http://arxiv.org/abs/1010.4300} {arXiv:1010.4300 [astro-ph.CO]} \BibitemShut
  {NoStop}%
\bibitem [{\citenamefont {Bertone}\ \emph {et~al.}(2012)\citenamefont
  {Bertone}, \citenamefont {Cerdeno}, \citenamefont {Fornasa}, \citenamefont
  {Pieri}, \citenamefont {Ruiz~de Austri},\ and\ \citenamefont
  {Trotta}}]{Bertone:2011pq}%
  \BibitemOpen
  \bibfield  {author} {\bibinfo {author} {\bibfnamefont {G.}~\bibnamefont
  {Bertone}}, \bibinfo {author} {\bibfnamefont {D.}~\bibnamefont {Cerdeno}},
  \bibinfo {author} {\bibfnamefont {M.}~\bibnamefont {Fornasa}}, \bibinfo
  {author} {\bibfnamefont {L.}~\bibnamefont {Pieri}}, \bibinfo {author}
  {\bibfnamefont {R.}~\bibnamefont {Ruiz~de Austri}}, \ and\ \bibinfo {author}
  {\bibfnamefont {R.}~\bibnamefont {Trotta}},\ }\href {\doibase
  10.1103/PhysRevD.85.055014} {\bibfield  {journal} {\bibinfo  {journal}
  {Phys.Rev.}\ }\textbf {\bibinfo {volume} {D85}},\ \bibinfo {pages} {055014}
  (\bibinfo {year} {2012})},\ \Eprint {http://arxiv.org/abs/1111.2607}
  {arXiv:1111.2607 [astro-ph.HE]} \BibitemShut {NoStop}%
\end{thebibliography}%

\end{document}